\documentclass{aa}
\usepackage[dvipsnames]{xcolor}

\usepackage[colorlinks,linkcolor={blue},citecolor={blue},urlcolor={blue}]{hyperref} 
\usepackage{graphicx}
\usepackage{txfonts}
\usepackage{natbib}

\usepackage[dvipsnames]{xcolor}

\colorlet{avic}{blue}

\begin{document}

\title{Supermassive black holes in six triaxial galaxies: \newline Insights from SINFONI and MUSE observations}

   \author{Sabine Thater\inst{1}  \and Avinash Chaturvedi\inst{2} \and Davor Krajnovi\'{c}\inst{2} \and  Michele Cappellari\inst{3} \and Sadegh Khochfar\inst{4}  \and  \newline  Thorsten Naab\inst{5}  \and Marc Sarzi\inst{6} \and
   Glenn van de Ven\inst{1} 
}

   \institute{Department of Astrophysics, University of Vienna, T\"urkenschanzstraße 17, 1180 Vienna, Austria\\
              \email{sabine.thater@univie.ac.at}
         \and
   Leibniz-Institute for Astrophysics Potsdam (AIP), An der Sternwarte 16, D-14482 Potsdam, Germany
   \and
   Sub-department of Astrophysics, Department of Physics, University of Oxford, Denys Wilkinson Building, Keble Road, Oxford OX1 3RH
   \and
    Institute for Astronomy, University of Edinburgh, Royal Observatory, Edinburgh EH9 3HJ, UK
    \and
    Max-Planck-Institut für Astrophysik, Karl-Schwarzschild-Str. 1, 85741 Garching, Germany
    \and
    Armagh Observatory and Planetarium, College Hill, Armagh BT61 9DG, UK
    }

   \date{Received: 29 October 2025/ Accepted: 11 May 2026}

 \abstract
{Dynamical modelling can be used to constrain the masses of central black holes; however, modelling massive galaxies is challenging due to their complexity. In this work, we report six new supermassive black hole mass measurements of massive early-type galaxies from stellar kinematics, which were extracted from adaptive optics-assisted SINFONI and MUSE observations. We combine the stellar kinematics with HST photometry to build \texttt{DYNAMITE} triaxial Schwarzschild orbit-superposition models. Our Schwarzschild models can recover the complex triaxial features of the galaxies and constrain the black hole masses of all six galaxies. We find that strong triaxial kinematic features can bias the mass measurements and correct for this effect. The derived black hole masses are $(1.14^{+0.41}_{-0.63}) \times 10^9$ M$_{\odot}$ for NGC~3706, $(1.19^{+1.34}_{-0.80}) \times 10^9$ M$_{\odot}$ for NGC~3923, $(1.14^{+1.08}_{-0.95}) \times 10^9$ M$_{\odot}$ for NGC~4261, $(4.68^{+2.99}_{-4.26}) \times 10^8$ M$_{\odot}$ for NGC~4636, $(3.51^{+3.37}_{-2.57}) \times 10^9$ M$_{\odot}$ for IC~4296, and $(2.43^{+1.53}_{-1.65}) \times 10^9$ M$_{\odot}$ for IC~4329 at 3$\sigma$ confidence level. We compare our measurements with published results from axisymmetric Schwarzschild modelling and with our Jeans Anisotropic Models (JAM), and obtain mostly consistent black hole masses. Most of our black hole mass estimates can be well constrained using only MUSE observations. All of our mass measurements are in agreement with local black hole scaling relations. 
}


\titlerunning{MBHs in triaxial galaxies with SINFONI and MUSE}
   
\authorrunning{Thater et al.}

\maketitle

\nolinenumbers

\section{Introduction}

Supermassive black holes (SMBHs) are known to be hosted by almost all massive galaxies in the local universe and up to large redshifts. Their role in the evolution of galaxies is not yet fully understood, but the existence of tight scaling relations between SMBH mass ($M_{\rm  BH}$) and different galaxy properties implies a co-evolution between the SMBHs and their host galaxies \citep[e.g., reviews by][]{Kormendy2013, Graham2016, Saglia2016, Mezcua2017}. The tightest scaling relations are found between $M_{\rm  BH}$ and the global velocity dispersion ($\sigma_{\rm e}$) or the stellar mass ($M_{*}$) of the host galaxy or its bulge. With nearly 200 SMBH masses estimated dynamically \citep[e.g. see compilation by][]{Bosch2016}, an increased scatter has become apparent for both low-mass \citep{Graham2015, Nguyen2017, Nguyen2018} and high-mass galaxies \citep{Dullo2019, Sahu2020, Nicola2024, Nicola2025}, suggesting a non-universal origin of the scaling relations. The origin of the scatter might be driven by evolutionary scenarios \citep{Krajnovic2018b, Graham2022} or by black hole seeding mechanisms \citep[e.g.][]{Agarwal2013, Latif2022}. The key is that initial seed masses are above the current scaling relation, but the self-regulated feedback from black holes and mergers drives them towards the local scaling relations \citep[e.g.][]{Hirschmann2010}. 

 Photometric observations have shown that massive early-type galaxies (ETGs) can be distinguished in core and power-law galaxies depending on the shape of their nuclear profiles \citep{Ferrarese1994, Lauer1985, Lauer1995, Faber1997, Rest2001}. Instead of a continuously steep power-law surface brightness cusp, core galaxies have a shallow central light profile, depleted of stars with respect to the extrapolation of the outer surface brightness distribution \citep{Graham2003}. While core scouring by merging black holes after one or more dissipationless galaxy mergers  \citep{Milosavljevic2002, Khochfar2004, Rantala2018, Rantala2019, Frigo2021, Mannerkoski2021, Rawlings2024}  is a popular scenario to explain this relation, the formation of depleted cores and the role of the SMBHs are still debated. This becomes apparent when one compares predictions of $M_{\rm BH}$ from the scaling relations based on $\sigma_{\rm e}$ and the core size ($R_{\rm b}$): $M_{\rm BH}$ in large-core galaxies estimated from the ${M_{\rm BH} - \sigma_{\rm e}}$ relation are under-massive, by up to a factor of 40, relative to expectations from their large $R_{\rm b}$ values \citep{Dullo2019}.

 \begin{table*}
   \caption{The galaxy sample}
   \label{t:properties}
   \centering 
  \begin{tabular}{lccccccccccccl}
    \hline
    \noalign{\smallskip}
Galaxy &  Type &  Distance & Scale & M$_{\rm{K}}$ & $R_{\rm{b}}$ & $R_{\rm{e}}$  & $\sigma_{\rm{e}}$ &  $i$ & Large Scale\\

       &       &  (Mpc)    & (pc arcsec$^{-1}$)&  (mag)  & (arcsec)     & (arcsec)     & (km s$^{-1}$)& $(^{\circ})$ &  \\
    (1)    &  (2)    &   (3)        &    (4)     &   (5)             &   (6)   & (7)      & (8) & (9)& (10) \\
\hline
NGC  3706 &   E   &   36.0 
& 175  &  -25.10      & 0.11 &   30.6     & 270  &    79 &  SINFONI \\ 
NGC 3923 &  E4-5   &  21.0 & 102  &   -25.09      & 1.68 & 86.4   &  240  &   48 &    MUSE\\ 
NGC 4261 &   E2-3   &  30.8 & 149 &   -25.18  &  1.18 &  38.0     &  265  &   89 &   MUSE\\ 
NGC 4636 &  E0-1   &   14.3 & 69   &   -24.36  & 1.52  & 89.1     &  181 &   89 &  MUSE \\ 
IC 4296 &   E   &   49.8 & 241 &   -25.99   & 1.44  &   68.4     &  330  &   57 &  MUSE\\ 
IC 4329 &   S0   &   56.9   &  276 & -25.74  & 0.79  &     120.0     &  296   &   67 &   MUSE\\ 
       \noalign{\smallskip}
    \hline
  \end{tabular}

\tablefoot{Column 1: Galaxy name. Column 2: Morphological type \citep{deVaucouleurs1991rc3}. For NGC 3706, we adopt the classification by \cite{Dullo2013}. Column 3: Distance to the galaxy taken from \cite{Cappellari2011} for NGC 4261 and NGC 4636, and NED for the remaining galaxies. Column 4: Linear scale derived from the distance.  Column 5: 2MASS total K-band magnitude \citep{Jarrett2000}. Column 6: Break radius from a  core-S\'ersic fit to the surface brightness profiles. Values are taken from \cite{Dullo2014}, \cite{Chaturvedi2025}, \cite{Rusli2013a}, \cite{Chaturvedi2025},  \cite{Capetti2005} and  \cite{Lauer2005}. Column 7: Effective radius derived from SDSS r-band \citep{Cappellari2013b} or CGS B-band \citep{Ho2011} imaging data. Column 8: Effective velocity dispersion derived by co-adding the spectra of the large-scale optical IFU data in elliptical annuli of the size of the effective radius. For NGC 4261 and NGC 4636 taken from \cite{Cappellari2013b}.
  Column 9: Inclination from \cite{Cappellari2013b} for NGC 4261 and NGC 4636 and \cite{Ho2011} for the remaining galaxies.  Column 10: Large-scale kinematics data used for the dynamical Schwarzschild models.}
\end{table*}

The evolution of SMBHs might also differ for low- and high-mass galaxies \citep[e.g.,][]{Krajnovic2018b}. While accretion of gas is the main process for growing SMBHs, high-mass SMBHs can also acquire a significant amount of mass via SMBH merging during gas-poor major mergers \citep{Schawinski2006, Volonteri2010, Rantala2024}. This effect can be up to a factor of two for the most massive galaxies \citep{Yoo2007, Kulier2015}. Furthermore, it is unclear whether the different evolutionary paths of the SMBHs or the host galaxies are the main driver for the large scatter at the high-mass end. 

Massive galaxies often exhibit triaxial features \citep{Binney1978a, Binney1978b, Cappellari2016} that cannot be described by axisymmetric models, which may lead to biased or uncertain estimates of the SMBH mass. NGC3379 was the first galaxy for which axisymmetric and triaxial $M_{\rm BH}$ estimates were presented. This galaxy has only a minor twist in the photometry, has very regular stellar kinematics, and overall shows weak signs of triaxiality. However, the two mass estimates differ by a factor of two \citep{vandenBosch2010}, albeit within the modelling uncertainties. A truly triaxial galaxy, PGC046832, with rotations around both the major and minor axes, provided a very different picture \citep{denBrok2021}: triaxial models were not able to constrain the lower limit to $M_{\rm BH}$. In contrast, the axisymmetric models estimated $M_{\rm BH} = 6\times 10^9 M_\odot$. This value, which is consistent with the prediction from the $M_{\rm BH} - R_{\rm b}$ relation \citep{Dullo2019b}, is also a factor of three larger than the upper mass limit provided by the triaxial models. The reason for the discrepancy remains unknown. 
The triaxial models can reproduce the complex velocity map well, featuring multiple spin reversals and the rotation around the major axis, while the axisymmetric modelling, lacking relevant orbital types, cannot. However, the insensitivity of triaxial models to the SMBH mass in some galaxies is puzzling and warrants further investigation \citep[e.g.][]{Mehrgan2019, Neureiter2022, Liepold2025}.

To properly understand the underlying reasons for the high scatter observed at the high-mass end of SMBH mass measurements, it is important to obtain more $M_{\rm BH}$ estimates in massive galaxies and examine their evolution in detail. Furthermore, it is essential to investigate how SMBH mass measurements based on axisymmetric and triaxial dynamical models compare for galaxies that are evidently triaxial in nature. In this work, we present new black hole mass measurements of six massive galaxies using integral field unit (IFU) data obtained with the SINFONI instrument in combination with adaptive optics (AO).
Spherical Jeans Anisotropic modelling (JAM) and axisymmetric Schwarzschild modelling were performed for these galaxies in \cite{Thater2019b}. A key difference of this work is the usage of new high-quality MUSE IFU observations and the construction of triaxial Schwarzschild models. The new MUSE data enable the detection of long-axis tube rotation, showing that these galaxies are indeed triaxial. This is the first in a series of publications on these galaxies, presenting $M_{\rm BH}$ derived from triaxial Schwarzschild models. The second paper will discuss the effect of including varying stellar mass-to-light ratios ($M_*/L$) on $M_{\rm BH}$ \citep{Chaturvedi2025}, and in the third paper, we will investigate the internal orbital structure (Chaturvedi et al., in prep.).

The structure of this paper is as follows. We describe the sample selection and our sample galaxies in Section~2. The different IFU observations and their data reduction are described in Section~3. In Section~4, we explain our kinematics extraction and present the kinematic maps. The kinematic information is combined with the luminous mass models to construct dynamical triaxial Schwarzschild models in Section~5. We discuss our results and give an outlook on future work in Section~6.

\section{Galaxy sample}

The galaxy sample (Table~\ref{t:properties}) presented in this paper is related to the sample of \cite{Krajnovic2018} and \cite{Thater2019}. Our six galaxies have early-type morphology like the galaxies in the previous work. However, they differ in three critical properties: they are more massive, have surface brightness cores, and can be classified as slow-rotators \citep{Emsellem2011}. 

Our sample galaxies were selected prioritising the determination of the SMBH mass. They needed a resolvable sphere of influence (SoI), high-resolution Hubble Space Telescope (HST) imaging, and observations with large-scale IFUs. The targets were required to be observable from Paranal, with the Spectrograph for INtegral Field Observations in the Near Infrared (SINFONI) in the "seeing enhancer" mode, where the adaptive optics works with only the laser guide star (and no tip-tilt star). This last condition required a steep central surface brightness and good observing conditions (good natural seeing). 

The size of the SoI is defined as $R_{\rm SoI} = G M_{\rm BH}/\sigma_{\rm e}^{2}$, where $G$ is the gravitational constant and $\sigma_{\rm e}$ is the velocity dispersion of the galaxy within one effective radius $R_{\rm e}$. At the time of the preparation of the observations, we calculated $R_{\rm SoI}$ using $M_{\rm BH}$ based on the scaling relation by \cite{Tremaine2002} and the velocity dispersions from the ATLAS$^{\textrm{3D}}$ Survey \citep{Cappellari2011} reported in \cite{Cappellari2013}. If velocity dispersions were not available, the central velocity dispersions from Hyperleda were used and extrapolated to the effective radius. 

High-resolution imaging is required for two reasons. Firstly, a robust $M_{\rm BH}$ measurement requires detailed knowledge of the light distribution in the galaxy centre, preferably at a higher resolution than the available kinematics. Furthermore, as the point-spread function (PSF) of the seeing enhancer mode is difficult to estimate based on the telescope telemetry only, high-resolution imaging is necessary to determine the actual PSF of the kinematics data (see Section~\ref{ss:SINFONI PSF} and Table~\ref{t:spatialres}). 

The large-scale kinematics are necessary to constrain the dynamical models, specifically the contributions of the stellar and dark matter components \citep{Krajnovic2005, Krajnovic2009}, and in this way break the degeneracy with the gravitational potential of the SMBH. At the time of the sample selection, this was primarily based on the SAURON \citep{Bacon2001} data from the ATLAS$^{\rm 3D}$ Survey \citep{Cappellari2011} and observations with VIMOS-IFU. In the meantime, almost all galaxies presented here also have higher-quality MUSE data.

\section{Observations \& data reduction}

In this section, we present the characteristics of the imaging data obtained with the HST and the spectroscopic observations collected with SINFONI and MUSE.

\subsection{SINFONI IFU data}
The high-resolution IFU observations of our six sample galaxies were performed with the Spectrograph for INtegral Field Observations in the Near Infrared \citep[SINFONI;][]{Eisenhauer2003, Bonnet2004} instrument mounted on UT4 (Yepun) of the Very Large Telescope (VLT). Between May 2009 and May 2013, we observed each galaxy at K-band grating (1.94 - 2.45 $\mu$m) providing a spectral resolution of $R \sim 4000$ and a rectangular pixel scaling of $50 \times 100$ mas ($125 \times 250$ mas) leading to a total field-of-view (FOV) on the sky of about $3.2 \arcsec \times 3.2 \arcsec$ ($8 \arcsec \times 8 \arcsec$) per pointing. The observations were performed using an object-sky-object nodding scheme. At the beginning and end of each observing block, a standard star was observed at a similar airmass and with the same instrumental setup to perform the telluric correction with similar atmospheric and instrumental conditions. In addition, we made use of the adaptive optics mode, either using a natural guide star or an artificial sodium laser guide star to correct for ground-layer turbulence and optimise the spatial resolution. Details of the observing runs for each galaxy are provided in Table 1 of the supplementary material. Our observations yield typical spatial resolutions of around $0.2''$ (FWHM) and Strehl ratios of $10\%$ (see Table~\ref{t:spatialres} and Section~\ref{ss:SINFONI PSF}.

We used the SINFONI data reduction pipeline \citep[version 2.4.8;][]{Modigliani2007} provided by ESO to reduce the SINFONI data. It handles the bias-correction, dark-subtraction, flat-fielding, non-linearity correction, distortion correction, and wavelength calibration, and finally creates data cubes of the single exposures. In addition to the steps of the pipeline, we applied our external routine to correct telluric features using standard stars observed on the same night. We refer the reader to \citet{Thater2019} for details about the data reduction. Finally, we merged the individual data frames by recentering the isophotes of the reconstructed images and Voronoi-binned \citep[Python version 3.1.0\footnote{\url{https://pypi.org/project/vorbin/}};][]{Cappellari2003} the generated data cube. The target signal-to-noise (S/N) of the Voronoi-binning was chosen to be between 50 and 70, balancing the desire to keep the central spaxels unbinned and ensure a sufficiently high resolution in the outer region. We thus obtained bin diameter sizes of around $0.2''$ in the centre and $0.7-1.0''$ at radii larger than $2''$ for the observations at 250mas, and bin diameter sizes of $<0.1''$ in the centre and $0.3-0.4''$ at radii larger than $1''$ for the observation of NGC~3706 at 100mas.

\begin{table}
\caption{SINFONI spatial resolution}
\centering
\begin{tabular}{lcccc}
\hline\hline
Galaxy  & FWHM$_{N}$ & FWHM$_{B}$ & f$_{N}$ & Strehl\\
        & (arcsec)    &  (arcsec) &         & \\
(1)   & (2) & (3) & (4) & (5)    \\
\hline
NGC 3706 & 0.28 $\pm$ 0.05 & 1.02 & 0.94 &  12 \%\\
NGC 3923 & 0.20 $\pm$ 0.04 & 0.78 & 0.20 &  14 \%\\
NGC 4261 & 0.19 $\pm$ 0.02 & 0.56  &  0.41 & 12 \%\\
NGC 4636 & 0.24 $\pm$ 0.04 & 0.62 & 0.10 & 15 \%\\
IC 4296 & 0.18 $\pm$ 0.05 & 0.58 & 0.47 & 9 \%\\
IC 4329 & 0.23 $\pm$ 0.03  & 0.68 & 0.02 & 10 \%\\
\hline
\\
\end{tabular}
\tablefoot{The SINFONI PSF of the data was parametrised by a double Gaussian with a narrow and a broad component. The parameters are given in the following columns. Column 1: Galaxy name. Column 2: FWHM of the narrow Gaussian component. Column 3: FWHM of the broad Gaussian component. Column 4: Relative intensity of the narrow component. Column 5: Strehl ratio of the data.} 
\label{t:spatialres}
\end{table}

\subsection{MUSE IFU data}

All except one galaxy (NGC 3706) have archival Multi-Unit Spectroscopic Explorer \citep[MUSE;][]{Bacon2010} data covering large spatial scales. They were observed in the wide-field, non-AO mode within different programmes: NGC 3923 PID: 0110.A-4397 (PI: Olivares), NGC 4261 PID: 097.A-0366(A) (PI: Hammer), NGC 4636 PID: 0105.B-0598 (PI: Bian) and IC 4296 PID: 094.B-0298 (PI: Walcher) and PID: 097.B-0776 (PI: Emsellem), while IC 4392 was observed during MUSE commissioning. We used all available data, resulting in 2700s, 4680s, 8960s and 2780s on source for NGC3923, NGC4261, NGC 4636 and IC 4296, respectively. For IC 4329, when we used only the February 2014 data set, totalling 3600s on source.

\begin{figure*}
  \centering
    \includegraphics[width=0.9\textwidth]{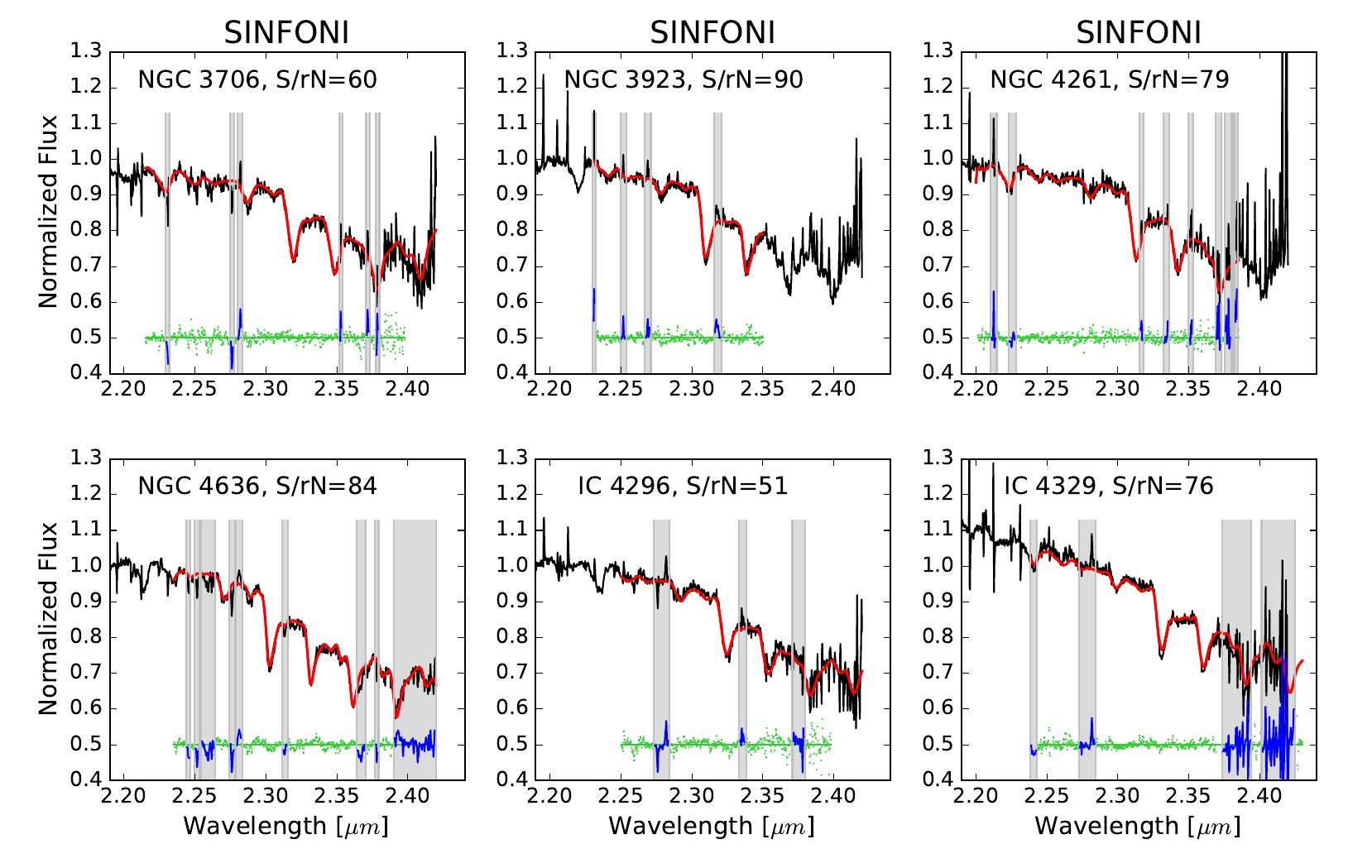}

      \caption{Integrated SINFONI spectra and \texttt{pPXF} fits of our target galaxies. The integrated spectra (black solid lines) were obtained by summing up all spectra of the IFU data cubes and fitted using the \texttt{pPXF} routine (red lines) in order to derive an optimal template. The fitting residuals between the spectra and the best-fitting pPXF models are shown as green dots and are shifted up by 0.5. Regions which were masked in the fit (often due to emission lines or insufficient sky subtraction) are indicated as grey shaded regions, and their residuals are indicated in blue.}
      \label{ff:ppxf_overview}
\end{figure*}

We reduced the data using the standard MUSE data reduction pipeline\footnote{\url{https://www.eso.org/sci/software/pipelines/muse}} \citep{Weilbacher2015}, version 1.6. It includes bias and sky subtraction, flat-field correction and wavelength calibration for each on-source observation. In addition to the on-source exposures, a separate sky field and standard star were reduced to extract a sky spectrum, the flux response curve and the telluric correction curve. These were used to correct the on-source exposures. After the data reduction of individual exposures, we estimated and applied the respective offsets and then merged on-source exposures using the MUSE pipeline recipes. The final data cubes cover approximately 1 arcmin$^2$ (IC 4296 about 1.5 arcmin$^2$), sampled with $0.2''\times 0.2''$ spaxels, covering a spectral range between $4800\,\AA$ and $9300\,\AA$, with a spectral sampling of 1.25$\AA$. We Voronoi-binned the MUSE observations to a target S/N of 60, resulting in bin diameter sizes of $0.5''$ in the centre and $3''$ at $R > 10''$. For the extraction of kinematics (see Section~\ref{s:kin}), we used only the blue part of the spectra until 700 nm, to avoid the red part of the spectral range with lower quality sky subtraction. However, we tested the kinematics extraction in the Calcium-Triplet region, and the results were consistent. Emission lines within the spectral range were also masked.

\subsection{Imaging data}
A detailed galaxy mass model requires both high-resolution imaging of the centre and large-scale deep imaging covering ideally at least twice the kinematic data of the galaxy. For the high-resolution imaging of the centre, we downloaded reduced and calibrated HST archival data from the ESA Hubble Science Archive. Depending on the availability, we used Wide-Field Planetary Camera \citep[WFPC2; ][]{Holtzman1995} or Advanced Camera for Survey \citep[ACS; ][]{Ford1998} images. In Table 3 of the supplementary material, we give information about the HST filter 
used for each galaxy. For NGC 4261, we used the luminous mass model from \cite{Boizelle2021} and therefore did not work with the actual images. After downloading the HST images, cosmic rays were removed from the calibrated images via Laplacian edge detection using the python implementation\footnote{The python implementation was written by Malte Tewes. A newer version can be downloaded from \url{https://github.com/cmccully/lacosmicx}.} of L.A.Cosmic \citep{vanDokkum2001}. In addition to the HST images, we obtained Sloan Digital Sky Survey (SDSS) mosaics from the Science Archive Server of DR12 \citep{Alam2015} for NGC 4636, and images of the Carnegie-Irvine Galaxy Survey \citep[CGS; ][]{Ho2011, Li2011, Huang2013} for the remaining four galaxies. The large-scale images cover a square FOV of $500''$ to $600''$.

\begin{figure*}
  \centering
    \includegraphics[width=0.86\textwidth]{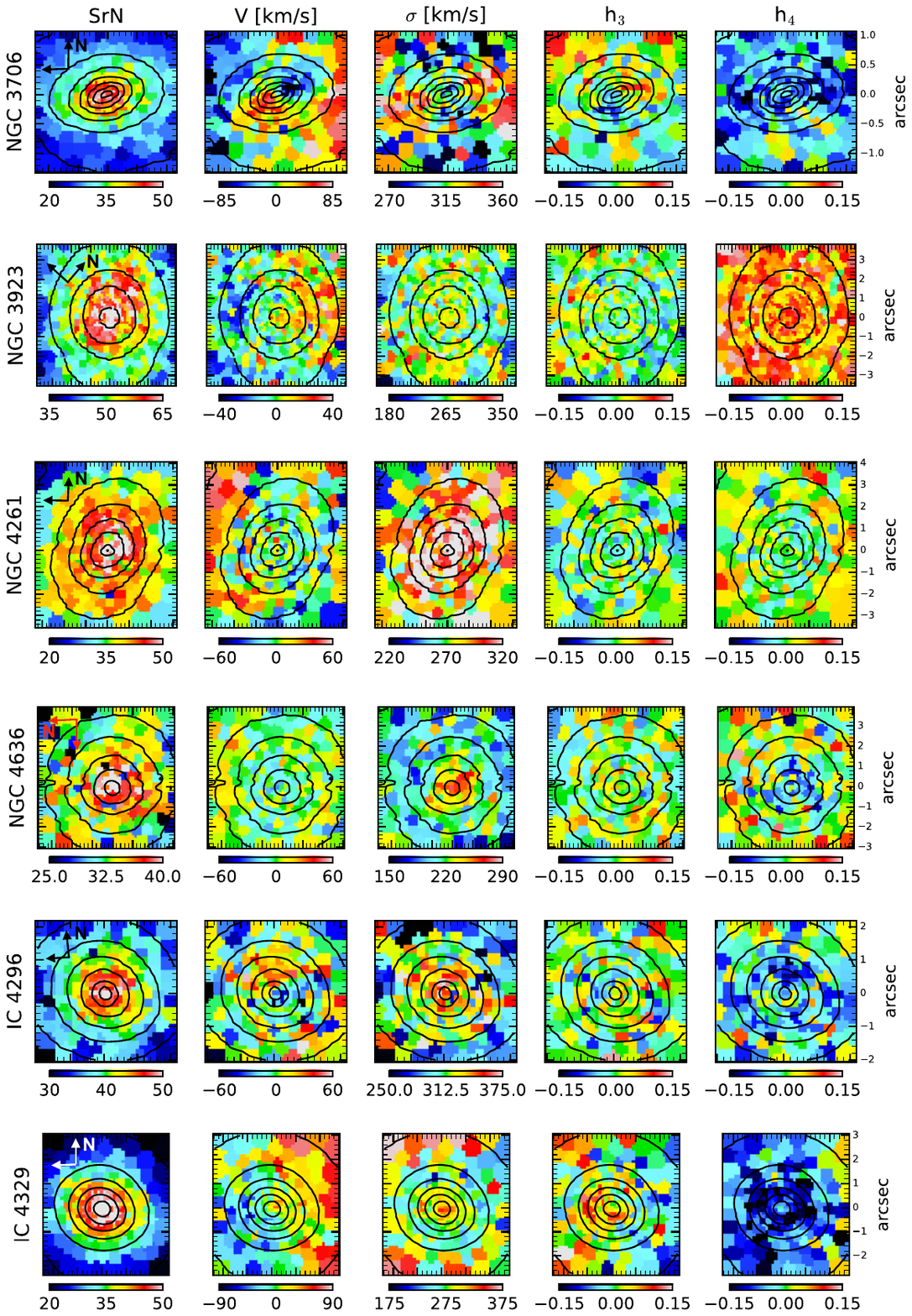}
      \caption{SINFONI stellar kinematics (derived from CO bandhead spectroscopy) of our target galaxies (from top to bottom) NGC 3706, NGC 3923, NGC 4261, NGC 4636, IC 4296 and IC 4329. From left to right, the panels show maps of signal-to-residual noise (S/N), mean velocity (V), velocity dispersion ($\sigma$) and the Gauss-Hermite moments $h_{3}$ and $h_4$. The black contours indicate the galaxy surface brightness from the collapsed data cube. The data orientation is indicated by the arrows in the SrN maps, pointing to north and east.
      }
      \label{ff:sinfoni_kinematics}
\end{figure*}

\section{Stellar kinematics}
\label{s:kin}

We measured the line-of-sight velocity distribution (LOSVD) for each Voronoi bin of the IFU observations using the Python implementation of the penalised Pixel Fitting method\footnote{\url{https://pypi.org/project/ppxf/}} \citep[\texttt{pPXF},][]{Cappellari2004, Cappellari2017, Cappellari2023}. The LOSVD is recovered by fitting an optimal template to the observed galaxy spectrum. The optimal template is a non-negative linear combination of a set of stellar templates. They were taken from the Gemini Spectral Library of Near-IR Late-Type stellar template library \citep{Winge2009} for our near-infrared SINFONI data and from the MILES stellar template library \citep{Sanchez-Blazquez2006} for the optical MUSE data. The emission lines were masked during the fit because we showed in \cite{Thater2022} that fitting emission lines simultaneously to the stellar absorption lines did not significantly improve the stellar kinematics fits. Furthermore, we improved our kinematic fits by clipping the $3\sigma$ outliers in all binned spectra. The $3\sigma$ clipping particularly helped to mask the remaining skyline residuals.

For each dataset and each galaxy, we performed two sets of stellar kinematics extractions, parametrising the LOSVD as simple Gaussian (V, $\sigma$) or as Gauss-Hermite polynomials \citep[V, $\sigma$, h$_3$, h$_4$;][]{Gerhard1993, vanderMarel1993}. The first set was used for dynamical models based on the Jeans equations (see Appendix~\ref{ss:jeans}), while all four moments were required to construct robust Schwarzschild models (see Section~\ref{ss:schwarzschild}). We carefully tested the effects of using different wavelength ranges, masking and unmasking insufficiently reduced sky- and telluric lines, and varying the additive polynomial to mitigate template mismatch effects in our stellar kinematics extraction and obtain a qualitative sense of the systematic uncertainties. 

We used Monte Carlo simulations (500 realisations) to derive the statistical uncertainties of the kinematic measurements, as explained in \cite{Thater2019}. The kinematic errors derived from our SINFONI kinematics are relatively high and typically reach about 15 km s$^{-1}$ for the mean velocity (V) and about 20 km s$^{-1}$ for the velocity dispersion ($\sigma$). IC 4296 has a particularly low S/N, which translates to larger errors in the kinematic measurements (20 km s$^{-1}$ for V and  $\sim30$ km s$^{-1}$ for $\sigma$). The errors of the optical MUSE kinematics reach typically around 7 km s$^{-1}$ for V and about 10 km s$^{-1}$ for $\sigma$ because of its higher S/N. 

In Figure~\ref{ff:ppxf_overview}, we show the \texttt{pPXF} fits to the integrated SINFONI spectra for each galaxy. The key feature in the SINFONI K-band is the CO absorption band head around $2.3~\mu$m. Due to strong telluric residuals, it was not always possible to recover the third and fourth bands sufficiently, and we decided not to fit these lines for half of our sample. The panels also show the spectral regions that were masked during the fits due to emission lines or an insufficient sky subtraction. In Figure~\ref{ff:ppxf_overview_ls} we show a similar set of spectra for the large-scale data (MUSE or large plate SINFONI).

The extracted SINFONI stellar kinematics maps of our sample galaxies are shown in Fig.~\ref{ff:sinfoni_kinematics}, and the large-scale kinematics obtained from MUSE (and SINFONI for NGC 3706) observations are presented in Fig.~\ref{ff:kinematics_ls1}. Some of the massive galaxies in this study show very peculiar kinematic features. NGC\,3706, NGC\,4636 and IC\,4296 contain kinematically distinct cores (KDCs) in their nuclei. In the case of NGC 3706, the velocity dispersion map shows the "$2\sigma$-peak" characteristic for counter-rotating structures \citep{Krajnovic2011}. The velocity maps of NGC\,4261 and NGC\,4636 show almost no nuclear rotation, but in the case of NGC\,4261, the velocity rises towards the edge of the SINFONI FOV for radii $>2''$. The rotation is around the major axis, which we will refer to as {\it long-axis rotation}. This sense of rotation is rare in the general population of ETGs, but seems to be a common feature in massive galaxies \citep{Krajnovic2018, Ene2018, Krajnovic2026}. Simulations indicated that it originates from dissipation-less equal-mass mergers, predominantly of already massive systems \citep{Ebrova2015, Li2016,  Ebrova2017, Tsatsi2017}. 

The long-axis rotation in NGC\,4261 is much clearer in the large-scale MUSE data and was already reported in \cite{Krajnovic2011}.  Three further galaxies also have long-axis rotation: NGC\,3923, NGC\,4636 and IC\,4329. In the case of NGC\,3923, the long-axis rotation is visible already on the SINFONI data, while for NGC\,4636 one has to look at the large scales covered by MUSE data. NGC\,4636 is actually an example of an extremely rare kinematic case: MUSE data reveal that it has a central KDC, which is in a near counter-rotation with the outer kinematics, while both components rotate approximately around the major axis of the galaxy. Furthermore, its velocity dispersion map shows a ring of higher values of about 15\arcsec\, in radius, presumably indicating the region where the two counter-rotating components overlap the most. For IC\,4329, the SINFONI data reveal a more standard short-axis rotation (oblate rotation) in the central region, while the MUSE velocity map traces a large kinematic twist where the central short-axis rotation transitions into the large-scale long-axis rotation (prolate rotation).

While the maximum velocities of our sample galaxies are typically low, the measured velocity dispersion reaches up to 300 km s$^{-1}$ in the centre and for some galaxies stays at that level over a significant fraction of SINFONI FOVs. The low-rotation features are also imprinted in the h$_3$-maps, while $h_4$ maps often show negative values in the central 1-2\arcsec. The latter is indicative of tangentially anisotropic velocity ellipsoid \citep[e.g. flat topped LOSVD;][]{vanderMarel1993, Gerhard1993}. The tangentially biased h$_4$ moments thus provide a hint of possible previous core scouring by merging BHs \citep{Thomas2014}. This is supported by the fact that the only core-less galaxy of our sample (NGC3923) has a positive kurtosis in the SINFONI data. We caution that the Hermite moments strongly depend on the quality of the data and are less accurate for the fits of only two out of the four CO-bands (NGC\,3923, NGC\,4261 and IC\,4329).

A visual comparison of kinematics between the small and large spatial scales shows mostly consistent results. We highlight two differences: the nuclear regions of almost all large-scale $h_4$ kinematic maps have values around zero, and, in the case of IC 4296, the lower spatial resolution of the MUSE data does not allow the identification of the KDC seen with SINFONI (Fig.~\ref{ff:sinfoni_kinematics}). 

A quantitative comparison between the velocity dispersion and the h$_4$-moment of the small- and large-scale data is shown in Fig. S5 of the supplementary material. For the comparison, we averaged the kinematic bins within circular annuli around the kinematic centre. In some cases, the small- and large-scale data did not match within the errors (indicated by shaded regions), with differences at the 5-10\% level. We found the largest differences for IC 4296 and IC 4329, which were likely driven by the strong residual skylines in the SINFONI observations, leading to enhanced uncertainties in the higher moments. To correct for this mismatch, we multiplied the SINFONI velocity dispersion by a constant factor and/or added/subtracted a constant shift to the SINFONI h$_4$ moments to match them with the large-scale data. Because the large-scale data were observed at a different spatial resolution, we only matched the data at radii > 1 arcsec. The matching was done for NGC 3923, NGC 4636, IC 4296 and IC 4329. The respective factors and shifts for each galaxy are shown in Figure S5 of the supplementary material. After the shift, the even moments of the different datasets matched very well and showed the same trends. We discussed the influence of this shift on the black hole mass measurements in \cite{Krajnovic2018b} and \cite{Thater2019}, and concluded that it likely imposes an uncertainty of less than a factor of two\footnote{\cite{Krajnovic2018b} show that an upward shift in velocity dispersion by 8\%  increases the mass of the black hole by 80\%. To test the significance of the h$_4$ shift, we have also run a small set of DYNAMITE models without applying the shift and did not find a significant change in the derived radial anisotropy profiles. This means that the tangential signature of the core scouring, which is expected for core galaxies, is preserved.}.

\begin{figure}
  \centering
     
    \includegraphics[width=0.36\textwidth]{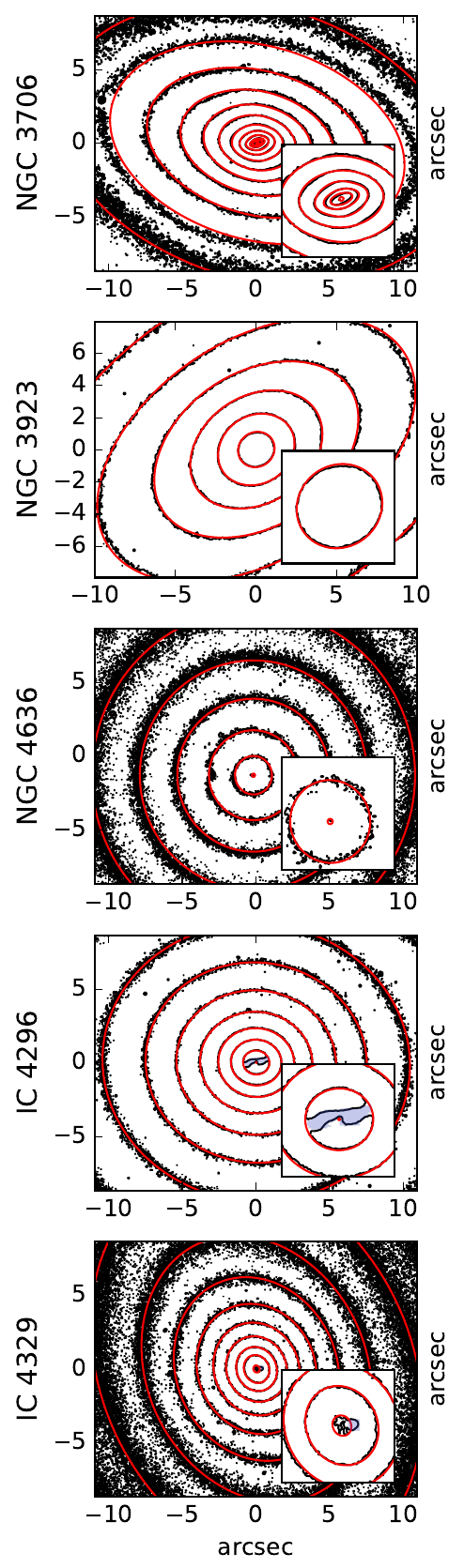}
      \caption{Isophotal maps of the WFPC2 and ACS images of our target galaxies within a FOV of $20\arcsec \times 20$\arcsec from the photometric centre. In the bottom right of each panel, we show a cutout of the central $3\arcsec \times 3$\arcsec. The contours of our best-fitting MGE model (red) are superimposed on the HST images (black). 
      We masked the central dust tori IC~4296 and the foreground star for NGC~4329 before MGE modelling their surface brightness (blue shaded area).}
      \label{ff:mge_overview}
\end{figure}

\section{Dynamical modelling}

Here, we present the triaxial Schwarzschild models of our six galaxies to derive the central black hole masses. The galaxies have previously been modelled by \cite{Thater2019b} using axisymmetric Schwarzschild modelling and spherical Jeans Anisotropic modelling. While we present the new results assuming triaxiality in the main section of this paper, we summarise the previous results in Sections~\ref{ss:jeans} and~\ref{ss:comparison_tri_axi} of the Appendix. The new results (using triaxial Schwarzschild modelling) differ from the previous results (using axisymmetric Schwarzschild modelling) in the shape of the gravitational potential (triaxial versus axisymmetric), the inclusion of dark matter and the usage of new MUSE observations for NGC 3923, NGC 4261 and NGC 4636 compared to ATLAS$^{\textrm{3D}}$ observations. Therefore, the comparison of the different methods in this paper should be taken with caution, and we will discuss the differences in Section~\ref{ss:comp}.

\subsection{Mass models}\label{ss:mass_models}
Detailed modelling of the distribution of mass and light within the galaxy is indispensable for dynamical modelling. We used the Multi-Gaussian Expansion \citep[MGE; ][]{Emsellem1994, Cappellari2002} to describe the surface brightness of our target galaxies analytically. In this method, the surface brightness is measured along equally spaced wedges and then parameterised by a sum of two-dimensional concentric Gaussians. The parametrised surface brightness is then deprojected to produce the luminosity density, and, assuming a constant mass-to-light ratio, then converted to a stellar mass distribution. During the process, the spatial resolution of the HST observations is taken into account. To derive the spatial resolution of our HST images, we created artificial point spread functions with TinyTim \citep{Krist2001} and modelled them with normalised concentric circular Gaussians using the MGE routine (see Table 2 of the supplementary material). 

Because \cite{Boizelle2021} published a detailed MGE model for NGC 4261, we did not repeat the MGE modelling here. For the other galaxies, we fitted both high-resolution HST and deep large-scale imaging data simultaneously, using the MgeFit Python package\footnote{\url{https://pypi.org/project/mgefit/}} Version 5.0 of \cite{Cappellari2002}. We aligned the surface brightness profiles by re-scaling the large-scale data to the central HST profiles and used the HST imaging for the photometric calibration. We carefully checked the position angles (PA) of our galaxies over wide radial ranges and found no change in any galaxy except for a photometric twist of about 30 degrees in the centre of NGC 3706. That is why we used the MGE routine with PA twist for NGC 3706 and kept the PA fixed for the remaining galaxies. During the fit, we carefully masked foreground stars, nearby galaxies and dust lanes. The dust masking is particularly important in the centre of our galaxies to assign the right amounts of mass to the luminous and dark components. Dust lanes were only visible in IC 4296, and we created a detailed dust mask as described in the Appendix of \cite{Thater2019}. Pixels were masked if they deviated significantly from the general galaxy surface brightness profile. We show this mask for IC 4296 in the Appendix \ref{app:A.1} of this work. After careful inspection, we noted that the dust features were only present in the very centre of IC 4296. We, therefore, decided not to perform a dust correction on the SDSS images. 

Figure~\ref{ff:mge_overview} shows the contours of our best-fitting MGE models, which are overlaid on the isophotal maps of the HST images. Most of the model contours reproduce the observed light contours very well. We show a similar plot covering the full FOV of the large-scale data in Fig. S1 of the supplementary material. Figures S2 and S3 of the supplementary material show a comparison of the count profiles of the small and large-scale images and the respective MGE fit. The percentage differences between the data and the model indicate a good quality of the MGE fits ($<10\%$ within $100''$). 

The final MGE models are described by a sum of 8 to 12 concentric Gaussians, which are listed in Table 3 of the supplementary material. We calibrated our models following the guidelines of the MGE readme, using the photometric zero points from \cite{Dolphin2009}, paying particular attention to the gain and filter of our observation, and the Vega magnitudes of the sun from \cite{Willmer2018}. One can obtain the density profile of the luminous mass by deprojecting the light model and multiplying it by the stellar mass-to-light ratio $M_*/L$ of the galaxy in the respective band mentioned in Table 3 of the supplementary material. 

These MGE models were used as inputs for all modelling types: JAM, axisymmetric and triaxial Schwarzschild. For NGC 3706, we kept the position angle constant in the dynamical modelling because JAM and the axisymmetric Schwarzschild modelling do not support twists, and for the triaxial Schwarzschild modelling, we could not find a deprojectable solution when including the changing position angles. Modelling the twisted MGE would require a modification of the MGE code (as done in \citealt{denBrok2021}), and we will further investigate this in future work. 

In \cite{Thater2019b} we used the MGE models together with a constant $M_*/L$. In this paper, we do the same, but add a separate parametrisation for dark matter, effectively changing the total mass model, while keeping the luminous mass distribution the same. In a companion paper by \cite{Chaturvedi2025}, we apply a varying $M_*/L$ based on variation in age, metallicity and initial mass function to describe the luminous mass distribution, and compare the results.

\begin{sidewaystable}
  \label{tt:results_LS}
\caption{Results of the iterative parameter search for the triaxial Schwarzschild models.}
\centering
\begin{tabular}{lccccccc|cccc|cc}
\hline\hline
Galaxy  & $\log (M_{\rm BH}/$M$_{\odot}$)  &  $M_*/L$ & $p_{\rm min}$ & $q_{\rm min}$ & $u_{\rm min}$ & $\log f_{\rm DM}$ &  $\chi^2$/d.o.f. & $\theta$ &  $\phi$ & $\psi$ & $R_{\rm s}$/kpc & $T_{\rm Re}$ & $\log (M_{*}/$M$_{\odot}$)\\
 (1) & (2) & (3) & (4) & (5) & (6) & (7)   & (8) & (9) & (10)& (11) & (12) & (13)  & (14)\\
\hline
NGC 3706 & $9.05\pm 0.15$ & 4.85$\pm 0.50$ & 0.80$\pm 0.05$ & 0.40$\pm 0.04$ & 0.98$\pm 0.02$ & 2.4$\pm 2.0$ & 1.24 & $79^{+5}_{-21}$ & $75^{+14}_{-15}$ & $91^{+7}_{-1}$ & 114 & 0.43$\pm 0.04$ & 11.31$\pm 0.04$\\ 

NGC 3923 & $9.25\pm 0.35$ & 4.85$\pm 0.85$ & 0.71$\pm 0.07$ & 0.53$\pm 0.10$ & 0.90$\pm 0.05$ & 2.4$\pm 1.0$ & 1.29 & $80^{+1}_{-12}$ & $51^{+5}_{-2}$ & $94^{+6}_{-1}$ & 132 & 0.67$\pm 0.06$ & 11.63$\pm 0.07$\\ 

NGC 4261 & $<9.20$ & 2.20$\pm 0.30$ & 0.82$\pm 0.08$ & 0.68$\pm 0.06$ & 0.97$\pm 0.03$ & $<2.6$ & 1.16 &  $70^{+15}_{-12}$ & $66^{+22}_{-9}$ & $95^{+3}_{-5}$ & 177 & 0.61$\pm 0.08$ & 11.70$\pm 0.04$ \\ 

NGC 4636 & $7.25\pm 0.50$ & 4.55$\pm 0.65$ & 0.70$\pm 0.10$ & 0.42$\pm 0.20$ & 0.80$\pm 0.10$ & 0.8$\pm 1.2$ & 3.42 &  $58^{+25}_{-36}$ & $27^{+26}_{-17}$ & $112^{+10}_{-20}$ & 27 & 0.62$\pm 0.04$ & 11.37$\pm 0.07$\\ 

IC 4296 & $9.50\pm 0.10$ & 4.42$\pm 0.22$ & 0.92$\pm 0.05$ & 0.80$\pm 0.04$ & 0.99$\pm 0.05$ & 0.4$\pm 0.3$ & 1.63 &  $25^{+8}_{-23}$ & $86^{+1}_{-34}$ & $93^{+34}_{-1}$ & 29 & 0.42$\pm 0.04$ & 12.19$\pm 0.02$\\ 

IC 4329 & $9.10\pm 0.50$ & 6.14$\pm 0.54$ & 0.68$\pm 0.08$ & 0.47$\pm 0.07$ & 0.90$\pm 0.06$ & 1.4$\pm 1.0$ & 1.18 &  $62^{+9}_{-10}$ & $50^{+10}_{-8}$ & $104^{+3}_{-4}$ & 149 & 0.68$\pm 0.04$ & 12.04$\pm 0.02$ \\ 
\hline
\end{tabular}
\\
\tablefoot{Columns 2 to 7 show the best-fit parameters derived from the iterative parameter search from the combination of the SINFONI and large-scale MUSE data. Uncertainties are given within $1\sigma$ confidence levels. Column 1: Name of the galaxy. Columns 2 and 3: Black hole mass $M_{\rm BH}$ and stellar mass-to-light ratio $M_*/L$ in the band given in Table 3. Columns 4 to 7: Parameters of the best-fitting models to constrain the global gravitational potential: $p_{\rm min}$ (intrinsic medium-to-major axis ratio), $q_{\rm min}$ (intrinsic minor-to-major axis ratio), $u_{\rm min}$ (ratio between projected and intrinsic major axis), and $f_{\rm DM} = M_{200}/M_*$, the dark matter fraction at $r_{200}$. Column 8: The $\chi^2$ over the degrees of freedom (d.o.f.). The degrees of freedom are the sum of kinematic constraints ($N_{\rm GH} \times N_{\rm kin}$) minus the number of fitted parameters (6).  Columns 9 to 11 show the derived intrinsic shape parameters corresponding to the viewing angles. $\theta$ corresponds to the inclination of the galaxy. Column 12 is the dark matter scale radius, which can be calculated from the dark matter fractions. Columns 13 and 14 show quantities derived from our Schwarzschild models. Column 13 is the triaxiality at one effective radius defined as $T_{\rm Re}=(1-p_{\rm Re}^2)/(1-q_{\rm Re}^2)$ and column 14 is the derived stellar mass derived by adding the MGE components multiplied by the $M_*/L$ from column 3.
}
\end{sidewaystable}

\subsection{Triaxial Schwarzschild modelling setup}
\label{ss:schwarzschild}

We used triaxial Schwarzschild modelling to infer SMBH masses, the intrinsic shapes, and the dark matter fractions of our galaxies. This method is based on the numerical orbit-superposition method by \cite{Schwarzschild1979} and was described and applied in \cite{vandenBosch2008}. In this work, we used the DYnamics, Age, and Metallicity Indicators Tracing Evolution\footnote{\url{https://github.com/dynamics-of-stellar-systems/dynamite}} \citep[\texttt{DYNAMITE} v2.0;][]{Jethwa2020, Thater2022b} software to create our triaxial Schwarzschild models. The slow rotation and isophotal core profile of our galaxies indicate a complex triaxial shape, making \texttt{DYNAMITE} the best choice for our modelling. We followed the approach described in \cite{Thater2023} and provide here only a summary highlighting the differences between this and our previous study.

In the Schwarzschild modelling method, assuming a stationary gravitational potential for the galaxy, orbit libraries are created and then fitted to the observed kinematics to find a representative superposition of stellar orbits. This is repeated for many choices of the gravitational potential to find the models that best fit the observations. The gravitational potentials in our models are composed of the gravitational potential of the stellar mass distribution, a central black hole ($M_{\rm BH}$), and dark matter parametrised by a spherical halo following a Navarro-Frenk-White \cite[NFW; ][]{Navarro1996} radial profile\footnote{It has been shown that feedback can produce cored DM haloes as well. However, for SMBH mass measurements, the DM halo has an insignificant contribution as shown in \cite{Rusli2013} and \cite{Thater2022}. We therefore decided to use the commonly applied NFW parametrisation.}. The stellar potential is derived by deprojecting the two-dimensional surface brightness described by the MGE (Section \ref{ss:mass_models}) using the three viewing angles ($\theta, \phi, \psi$), and then multiplying by the stellar $M_*/L$ for each galaxy. The viewing angles are directly connected to the intrinsic shape parameters, which are p (intrinsic intermediate-to-major axis ratio), q (intrinsic minor-to-major axis ratio) and u (ratio between projected and intrinsic major axis). In fact, the flattest Gaussian component of the MGE (the component with the smallest projected axial ratio) dictates the allowed orientation of the deprojection. We therefore adopted ($p_{\rm min}$, $q_{\rm min}$, $u_{\rm min}$) which were varied in our dynamical models. We assumed a constant stellar $M_*/L$ in this work to constrain the hyperparameters of these galaxies without imposing additional complexity due to a varying $M_*/L$. Modelling a varying $M_*/L$ due to varying age, metallicity and initial mass function is discussed in our companion paper by \cite{Chaturvedi2025}. \\
The gravitational potential of the SMBH is included as a non-luminous spherical Plummer potential, which is parametrised by the black hole mass ($M_{\rm BH}$) and a very small softening length of $10^{-3}$ arcsec, that was introduced to avoid numerical issues when integrating orbits close to the SMBH and is not varied during the fit. The NFW profile that describes the dark matter halo is parametrized by the mass fraction at $R_{200}$, $f_{\rm DM}=M_{200}/M_{*}$, and the concentration parameter $c_{200} = R_{200}/R_{\rm s}$ where $R_{\rm s}$ describes the scale radius of the NFW profile. Due to the limited FOV of our kinematic observations, we applied the concentration mass relation of \cite{Dutton2014} to reduce the NFW parametrisation to one parameter. In summary, the gravitational potentials of our Schwarzschild models are described by six parameters ($p_{\rm min},q_{\rm min},u_{\rm min}, M_*/L, f_{\rm DM}, M_{\rm BH}$).

For each trial gravitational potential, we calculated two orbit libraries: one classical orbit library that includes mostly long- and short-axis tube orbits and a second one that includes box orbits. Similar to \cite{Thater2023}, we created our classical orbit library on 35 equipotential shells with logarithmically spaced radii covering from $0.01''$ to $3\sigma$ of the largest Gaussian of the MGE models (following Section 4.3 of \citealt{vandenBosch2008}). At each energy, we sampled 15 angular and 11 radial values. The initial set of orbits is doubled, accounting for prograde and retrograde orbital motion. The box orbits are sampled by $35 \times 15 \times 11$  combinations of binding energy and two spherical angles, respectively. Subsequently, each orbit was dithered by slightly changing its initial conditions 3$^3$ times and co-added to smooth the orbit libraries. 

The resulting 467 775 orbits were then integrated within the galaxy's potential. A \cite{Lawson1974} non-negative least squares (NNLS) method was used to assign weights to the orbits and fit the LOSVD (V, $\sigma$, h$_3$, h$_4$) derived from both SINFONI and large-scale observations (SINFONI or MUSE kinematics) while accounting for PSF effects and aperture binning. The 3-dimensional intrinsic and 2-dimensional aperture masses derived by integrating the MGE model were used as constraints in the NNLS to ensure that the light distribution of the galaxy can be recovered within 2\% (for a detailed mathematical explanation, see Section 5.1 in \citealt{vandenBosch2008}).

The high-resolution SINFONI observations were crucial in constraining the $M_{\rm BH}$. However, the uncertainties of the MUSE observations are on average three times lower than the SINFONI uncertainties but have a lower resolution. We therefore masked the MUSE kinematics within the central 2 arcsec by inflating the kinematic errors. 
We also masked bright stars in the MUSE observations. In total, we fitted between 1500 and 3500 kinematic bins depending on the galaxy.

\begin{figure*}[!htb]
  \centering
    \includegraphics[width=0.97\textwidth]{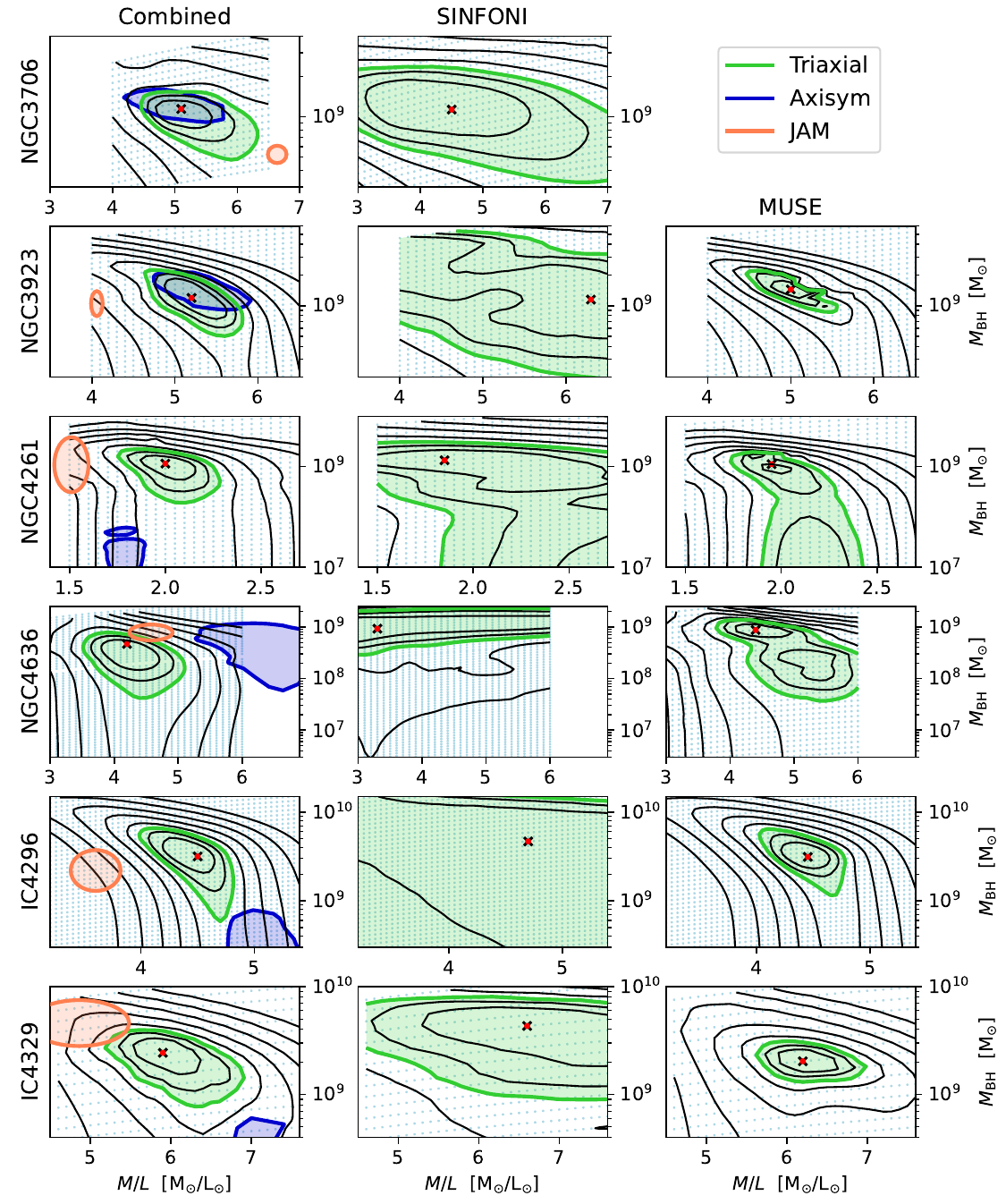}
      \caption{
      Grids of \texttt{DYNAMITE} Schwarzschild models (indicated by the lightblue dots) over different stellar mass-to-light ratios $M_*/L$ and black hole masses $M_{\rm BH}$. We show models constrained by the combined kinematics (left column), only SINFONI kinematics (middle column) and only MUSE kinematics (right column). Each row shows the model grids for one galaxy. Contours are the $\Delta \chi^2 = \chi^2 - \chi^2_{min}$ levels, where the thick green contour shows the $3\sigma$ level of the two-dimensional distribution. The contours were smoothed using an LOESS routine. The respective best-fitting model, derived as the minimum of $\chi^2$, is indicated by a red point.  In addition, we have added the 3-$\sigma$ limits of the best-fitting black hole masses of the JAM models (orange ellipses). We shifted the axisymmetric Schwarzschild result for NGC 4261 from 6.28 M$_{\odot}$/L$_{\odot}$ to 1.8 M$_{\odot}$/L$_{\odot}$ as the former measurements were done in a different filter. 
      }
      \label{ff:schwarzschild_grid3}
\end{figure*}

\begin{figure*}[!htb]
  \centering
    \includegraphics[width=0.90\textwidth]{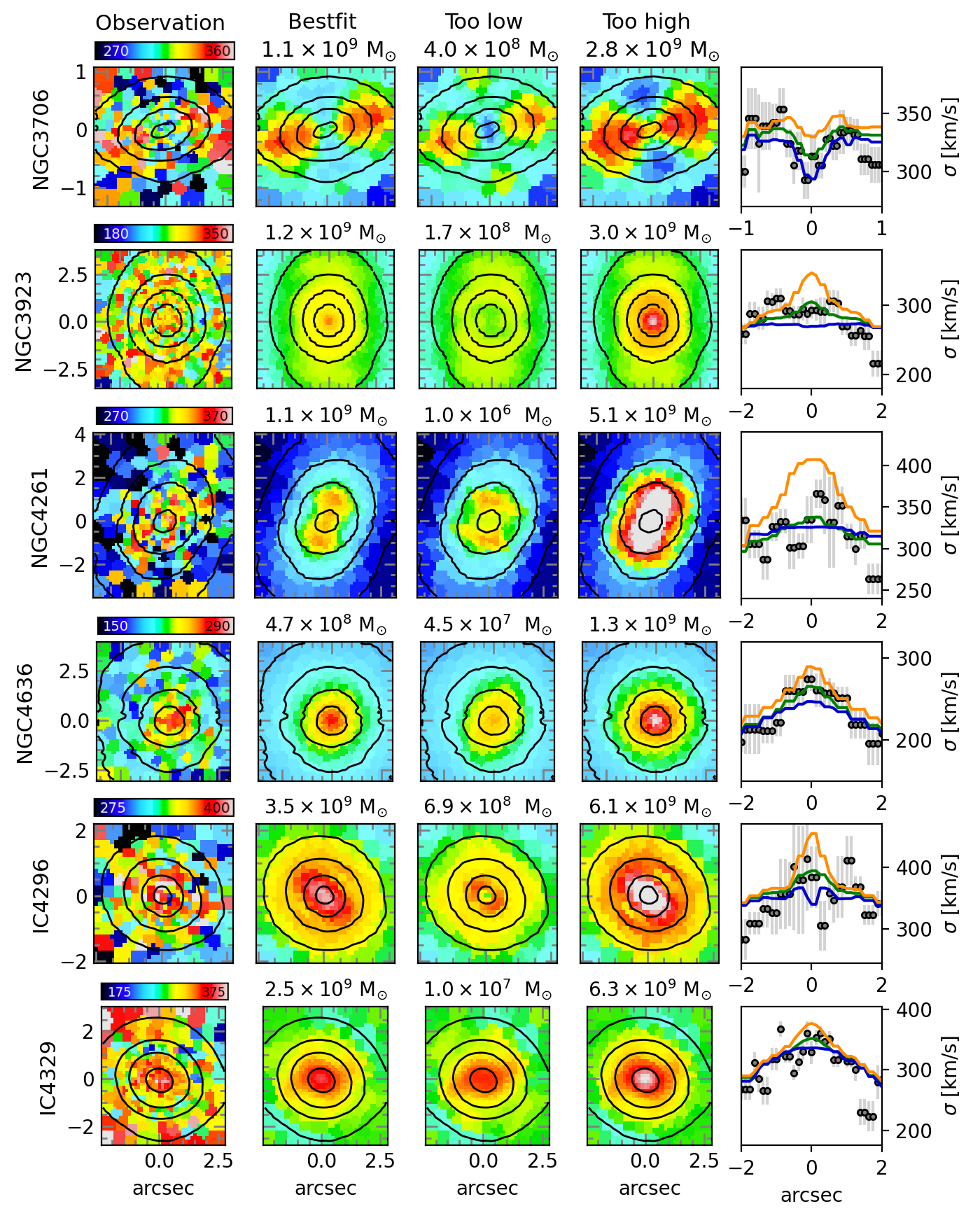}
      \caption{Comparison of the 
      velocity dispersion ($\sigma$) maps from the SINFONI observations and the Schwarzschild models. Each row shows the maps of one galaxy, respectively. From left to right, we present the observed $\sigma$ from the SINFONI data, and the $\sigma$ maps of the Schwarzschild models from the best fitting, a too low and a too high black hole mass. The black contours indicate the galaxy isophotes from the collapsed data cube and are chosen to be at an arbitrary level. The last column shows the $\sigma$ profiles along the x=0 axis. The too low (blue) and too high (orange) black hole masses are chosen to be just outside of the $3\sigma$ confidence intervals. All models are shown at the respective best-fitting $M_*/L$. The high- and low-mass models are clearly ruled out for most galaxies.}
      \label{ff:sigma_comparison}
\end{figure*}

\subsection{Modelling results: BH mass estimates}
We started our Schwarzschild models by first constraining the intrinsic shape parameters ($p_{\rm min}, q_{\rm min}$ and $u_{\rm min}$), the dark matter fraction and coarse values for $M_*/L$ and $M_{\rm BH}$ by simultaneously fitting the SINFONI and MUSE data. We therefore ran the "Legacy Grid Search" routine of \texttt{DYNAMITE}, which is an iterative parameter search. The best-fitting model for one iteration is used as a starting point to evaluate the set of models for the next iteration. We ran our models for 60-100 iterations, resulting in around 5000 models for each galaxy.  

We show triangle plots visualising the $\chi^2$ distributions in Figure \ref{ff:chi2}. The $\chi^2$ of each model was derived as the difference between the Schwarzschild model and observed kinematics divided by
the observational errors, summed over all kinematic bins and all Gauss-Hermite moments. 
In our previous work, we used the traditional $1\sigma$ confidence interval defined as $\Delta\chi^2 = \chi^2-\chi^2_{\rm min}$ < 2.3 for 2 degrees of freedom. However, the $1\sigma$ region is strongly underestimated when the number of observables (the number of Gauss-Hermite moments times the number of kinematic bins: $N_{\rm GH} \times N_{\rm kin}$) is large. We therefore used the approach of \cite{vandenBosch2008} and \cite{vandenBosch2009} and defined our $1\sigma$ confidence interval in this work as $\Delta\chi^2/(\sqrt{2\times N_{\rm GH} \times  N_{\rm kin}})$. Table 3 shows the best-fitting parameters of the "Legacy Grid Search" including the $1\sigma$ uncertainties. \cite{Lipka2021} found an inclination bias due to changing degrees of freedom in the case of axisymmetric Schwarzschild models. They account for this bias by including a penalty term in their parameter search.  \cite{Pilawa2024} apply the method of \cite{Lipka2021} for a triaxial galaxy and do not find significant changes due to the more complex orbit families in triaxial galaxy potentials. We follow the suggestion by \cite{Pilawa2024} and do not include a penalty term in our parameter search. 

A visual comparison between the observations and the data shows that our models can very well reproduce the complex kinematics for both SINFONI and MUSE observations (Figures S8 and S9 from the supplementary material). We also calculated for each galaxy the triaxiality parameter $T_{\rm Re}$ at one effective radius and the stellar mass derived from our Schwarzschild models. The triaxiality parameter $T_{\rm Re}$ is defined between 0 and 1, which corresponds to oblate and prolate shapes \citep{Franx1991}. Our derived triaxiality parameters confirm the clear triaxial shape of our six galaxies, as expected from the peculiar MUSE stellar kinematics. We will discuss the triaxiality and orbital distribution of our sample galaxies in detail in a subsequent paper.

In a second run, we created a regular fine grid for $M_{\rm BH}$ and $M_*/L$ centred on the global minimum of the iterative parameter search. The models were also constrained by simultaneously fitting the matched SINFONI and MUSE data. We noticed that for NGC 4261 and NGC 4636, the large-scale kinematics were strongly affecting the $\chi^2$ distribution, biasing the measurements to too low black hole masses (see Fig \ref{ff:chi2_rad} of Appendix \ref{ss:triangle}). After many careful checks, we decided to limit the calculation of the $\chi^2$ distribution in this second run to circular apertures in the kinematics with radius 3 arcsec for NGC 4261 and NGC 4636, and with radius 10 for IC 4329. For the remaining galaxies, we used the full kinematic FOV to calculate their $\chi^2$ distribution, as the $\chi^2$ distribution did not change significantly by applying the circular apertures. We emphasise that the dynamical modelling (orbit-integration, orbit weights) was performed on the full FOV of all kinematic datasets.
In Table~\ref{t:results} we summarise the results of the second run of the triaxial Schwarzschild modelling, including the results of the axisymmetric Schwarzschild and JAM modelling published in \cite{Thater2019b}.  

We present the $\chi^2$ distribution of our Schwarzschild model grids for the combined kinematics in the left column of Figure~\ref{ff:schwarzschild_grid3}. The topology of the $\chi^2$ contours was smoothed by applying the local regression smoothing algorithm LOESS \citep{Cleveland1979} as adapted for two dimensions\footnote{\url{https://pypi.org/project/loess/}} by \cite{Cappellari2013}. Our models of the combined kinematics can constrain the $ M_{\rm BH}$ for all six modelled galaxies. We also indicate the results of the axisymmetric Schwarzschild and Jeans models published in \cite{Thater2019b} (see Appendix~\ref{ss:jeans}) as blue contours and orange ellipses in these plots. A comparison of the different models shows mostly consistent $M_{\rm BH}$ measurements, while the $M_*/L$ are rather different.

To ensure that our models are not strongly biased towards the MUSE kinematics, which are of higher quality, we also ran a grid using only SINFONI kinematics and only MUSE kinematics (with normal errors in the central 2 arcsec). The model grids of these runs are shown in the middle and right columns of Figure~\ref{ff:schwarzschild_grid3}. The comparison of the three modelling runs shows that the set-ups give mostly consistent results. While the $M_*/L$ cannot be constrained for the SINFONI observations, the inferred $M_{\rm BH}$ aligns well with the mass indicated by the MUSE observations, strengthening our results.

\begin{sidewaystable}
\caption{Summary of dynamical modelling results}
\centering
\begin{tabular}{c|lcc|lcc|lcccc}
\hline\hline
&Triaxial & Schwarzschild   &  & Axisymmetric & Schwarzschild & &  JAM &  &  & &  \\
Galaxy  & $M_{\rm BH}$ & $M_*/L$  & $\chi^2/d.o.f.$ & $M_{\rm BH}$ & $M_*/L$  & $\chi^2/d.o.f.$ &    $M_{\rm BH}$ & $M_*/L$ & $\beta_{in}$ & $\beta_{out}$ & $\chi^2/d.o.f.$ \\
 &  $(\times 10^8~$M$_{\odot})$ &  (M$_{\odot}$/L$_{\odot})$ & 
   & $(\times 10^8~$M$_{\odot})$ 
 &  (M$_{\odot}$/L$_{\odot})$ & &  $(\times 10^8~$M$_{\odot})$ & (M$_{\odot}$/L$_{\odot})$ & &  \\
(1) & (2) & (3) & (4) & (5)  & (6) & (7) & (8) & (9) & (10) & (11) & (12) \\
\hline
 & &  &   & & & & \\
NGC 3706 & $11.4^{+4.1}_{-6.3}$ & $5.10^{+1.20}_{-0.60}$ & $1.23$  &  $12.3^{+3.5}_{-3.1}$ & $4.93^{+0.84}_{-0.71}$ &  $3.57$
&  $5.3^{+0.8}_{-0.8}$ & $6.64^{+0.11}_{-0.13}$  & $-0.06^{+0.04}_{-0.03}$ &  -- & $5.43$\\
 & &  &   & & & & \\
NGC 3923 &  $11.9^{+13.4}_{-8.0}$ & $5.20^{+0.60}_{-0.80}$  & $1.30$ & $17.5^{+3.2}_{-7.4}$  & $4.99^{+0.87}_{-0.22}$  & $2.54$ & $10.5^{+4.7}_{-1.8}$  & $4.06^{+0.06}_{-0.08}$ & $-0.08^{+0.08}_{-0.31}$ & $0.30^{+0.01}_{-0.02}$ & $5.95$\\
 & &  &   & & & & \\
NGC 4261 &  $11.4^{+10.8}_{-9.5}$ & $2.00^{+0.30}_{-0.25}$ & $1.43$ & $0.26^{+0.33}_{-0.18}$ & $6.28^{+0.30}_{-0.28}$   &  $2.28$ &   $10.9^{+24.3}_{-7.9}$ & $1.51^{+0.07}_{-0.11}$  &  $-0.16^{+0.16}_{-0.83}$ & $0.29^{+0.01}_{-0.09}$ & $1.96$\\
 & &  &   & & & & \\
NGC 4636 & $4.7^{+3.0}_{-4.3}$ & $4.20^{+0.90}_{-0.70}$ & $3.43$  & $6.06^{+5.45}_{-5.95}$  & $6.42^{+0.70}_{-1.02}$ & $1.53$ & $7.8^{+4.5}_{-1.7}$ & $4.57^{+0.43}_{-0.25}$ & $-0.08^{+0.08}_{-0.52}$ & $0.27^{+0.03}_{-0.14}$  & $1.89$\\
 & &  &   & & & & \\
IC 4296 & $35.1^{+33.7}_{-25.7}$ & $4.50^{+0.30}_{-00.50}$  & $1.58$  & $<10.5$ & $5.17^{+0.37}_{-0.51}$ & $0.72$  &  $22.0^{+20.2}_{-8.1}$ & $3.60^{+0.24}_{-0.20}$ & $-0.05^{+0.05}_{-0.34}$ & $0.28^{+0.02}_{-0.14}$ & $1.37$  \\
 & &  &   & & & & \\
IC 4329 & $24.3^{+15.3}_{-16.5}$ & $5.90^{+1.00}_{-0.70}$ & $1.20$ & $<4.99$ &  $6.53^{+0.32}_{-0.53}$ & $0.99$ &  $45.8^{+20.3}_{-21.0}$  & $4.86^{+0.36}_{-0.88}$ & $-0.11^{+0.10}_{-0.71}$ & $-0.62^{+0.89}_{-0.38}$ & $1.65$ \\

\hline

\end{tabular}
\\
\tablefoot{Column 1: galaxy name. Column 2-4: parameters of the triaxial Schwarzschild models (black hole mass, stellar mass-to-light ratio and the reduced $\chi^2$ of the best-fitting model). The other parameters were fixed in this run. Column 5-7: parameters of the axisymmetric Schwarzschild models (black hole mass, total mass-to-light ratio and the reduced $\chi^2$ of the best-fitting model). Column 8-12: parameters of the JAM models (black hole mass, total mass-to-light ratio, inner and outer anisotropy parameter and the reduced $\chi^2$ of the best-fitting model). The degrees of freedom (d.o.f.) are the sum of kinematic constraints ($N_{\rm GH} \times N_{\rm kin}$) minus the number of fitted parameters. We caution that the reduced $\chi^2$ between the different modelling methods should not be directly compared because different data sets were used in the different methods.}
\label{t:results}
\end{sidewaystable}

A visual comparison between the best-fit Schwarzschild model of the combined run and the SINFONI velocity dispersion ($\sigma$) maps is shown in Figure~\ref{ff:sigma_comparison}. We also show models of a too high and too low black hole mass for comparison. Although the overall kinematics (shape and height of the $\sigma$ peaks) can be reproduced by the best-fit models, a conclusion is difficult because of the large errors of the SINFONI kinematics.  A similar comparison of $\sigma$ maps, but for triaxial models fitting only the MUSE data, is shown in Figure ~\ref{ff:sigma_comparison_muse}.

NGC 3706 belongs to the best-recovered galaxies in our sample. The 2$\sigma$ peaks indicated by the observations can clearly be reproduced by the models. For a too low or too high black hole mass, the 2$\sigma$ peaks either fade or become too pronounced. In addition, a depression in the velocity dispersion appears between the peaks, or they start to connect for too low or too high $\rm M_{\rm BH}$ values, respectively. However, the kinematic uncertainties for NGC 3706 are large, resulting in large uncertainties for the black hole mass measurement. 

NGC 3923 has a very flat $\sigma$ peak at around 300 km s$^{-1}$ that suffers, however, from fluctuation due to strong telluric contamination. The best-fit Schwarzschild model is much smoother but can reproduce the flat $\sigma$ peak of NGC 3923. The $\sigma$ profiles show clearly that the too high and too low black hole masses cannot reproduce the central kinematics within the kinematic uncertainties. 

The $\sigma$ of NGC 4261 does not clearly peak in the centre but fluctuates around 340 km s$^{-1}$. The $\sigma$ of our model with formally best-fit $M_{\rm BH}$  of $1.14 \times 10^9$ M$_{\odot}$ traces the observations quite well. A few central $\sigma$ bins would indicate a black hole mass twice the formal best-fit value.
 Due to the large kinematic uncertainties, only a too high black hole mass can clearly be ruled out. While we can constrain the black hole mass using our $\chi^2$ distribution, this constraint is not very strong and is almost an upper limit.

NGC 4636 is the best modelled galaxy in our sample. The kinematics are derived using 4 CO bandheads in the near-infrared, and there is little telluric contamination. We also find a clear $\sigma$ peak in this galaxy, which is, however, not fully symmetric. From the $\chi^2$ distribution, we obtain a best-fit $M_{\rm BH}$ of $4.68 \times 10^8$ M$_{\odot}$ that reproduces the observed $\sigma$ peak well.  The $\chi^2$ distribution also predicts a second minimum at lower $M_{\rm BH}$. However, based on the visual inspection, this minimum is likely not caused by the central black hole. We also notice that the $\chi^2$/d.o.f. of this galaxy is exceptionally high, which is mostly caused by the overall positive $h_3$ Gauss-Hermite value. This is likely a spurious effect caused by the very small values for V.

IC 4296 shows a relatively clear $\sigma$ peak but also has large errors, which hardly constrain the data based on SINFONI alone. The best-fitting model can recover the central $\sigma$ but drops more slowly for larger distances from the centre than the observations. 

Finally, IC 4329 has again a clear but relatively flat $\sigma$ peak in its centre. The model of our formal best-fit $M_{\rm BH}$ of $2.43 \times 10^8$ M$_{\odot}$ gives a $\sigma$ that peaks slightly below the observations. Based on this visual inspection, the lower black hole mass in IC 4329 is barely constrained.

\begin{table*}
\caption{Grid results using MUSE only}
\centering
\begin{tabular}{lccc|cc}
\hline\hline
Galaxy & $M_{\rm BH}$ & $M_*/L$ & $\chi^2$/d.o.f. & PSF$_{\rm MUSE}$  & $R_{\rm SOI}/\sigma_{\rm PSF(MUSE)}$ \\
 & ($\times 10^8 \;$M$_{\odot}$) & (M$_{\odot}/$L$_{\odot}$) &  & (arcsec) &  \\
(1)  & (2) & (3) & (4) & (5) & (6)\\
\hline
\vspace{1mm}
NGC 3923 & $14.4^{+9.1}_{-6.3}$ & $5.00^{+0.60}_{-0.60}$ & $1.49$ & $0.98$ & $2.8$\\

\vspace{1mm}
NGC 4261 & $<19.1$ & $1.95^{+0.30}_{-0.20}$ & $1.50$  & $0.67$ & $3.0$ \\
\vspace{1mm}

NGC 4636 & $8.7^{+4.8}_{-8.3}$ & $4.40^{+1.60}_{-0.60}$ & $4.18$ & $1.06$ & $3.7$\\
\vspace{1mm}

IC 4296 & $31.3^{+32.5}_{-19.8}$ & $4.45^{+0.30}_{-0.40}$ & $1.63$  & $1.10$ & $1.2$\\

IC 4329 & $19.9^{+11.7}_{-7.6}$ & $6.10^{+0.70}_{-0.60}$ & $1.24$ & $1.02$ & $0.9$ \\

\hline
\\
\end{tabular}
\\
\tablefoot{This table summarises the results of the \texttt{DYNAMITE} run fitting only MUSE kinematics. Column 1: Galaxy name. Column 2: Best-fit black hole mass. Column 3: Best-fit mass-to-light ratio. Column 4: $\chi^2$ over degrees of freedom of the best-fitting model. Columns 5 and 6 give details about the detectability of the black hole mass with the MUSE observations. Column 5: Spatial resolution of the MUSE observations in units of full-width-half-maximum (FWHM) as derived in Section \ref{ss:MUSE PSF}. Column 6: Ratio between the radius of the sphere-of-influence and the MUSE spatial resolution expressed in units of $\sigma = FWHM/2.355$. For NGC 4261, we used the $M_{\rm BH}$ upper limits to derive this ratio.}
\label{tt:MBH_MUSE}
\end{table*}

\section{Discussion}

We give a summary of each galaxy, including a comparison with the literature in Section A of the Appendix.

\subsection{SMBH detectability with MUSE}

Our galaxies have SoI radii between 0.4\arcsec and 1\arcsec, as they have large velocity dispersions and are selected to be reasonably close. Resolving some of those SoI radii is, in theory, within the reach of good and stable ground-based seeing conditions. 
The right column of Figure~\ref{ff:schwarzschild_grid3} shows that $M_{\rm BH}$ for NGC 3923, NGC 4636, IC 4296 and IC 4329 can indeed be constrained using only the MUSE observations. The reason for this is in the quality of the MUSE data, specifically the dispersion of the Gaussian seeing ($\sigma_{\rm PSF(MUSE)}$), listed in Table~\ref{tt:MBH_MUSE}, and the obtained $M_{\rm BH}$ indicating massive SMBHs in these galaxies. 
The ratios between the actual radius of the SoI and the spatial resolution, $R_{\rm SOI}/\sigma_{\rm PSF(MUSE)}$, are also listed in Table~\ref{tt:MBH_MUSE}. They show that the MUSE observations resolve the SoI by a factor of 1 - 3 for NGC 3923, NGC 4261, NGC 4636 and IC 4296. 

In the case of IC 4329, $R_{\rm SOI}/\sigma_{\rm PSF(MUSE)}\approx 0.9$, meaning that the black hole SoI is not resolved, ($\sigma_{\rm PSF(MUSE)}=0.43$ arcsec). Nevertheless, MUSE data are of such quality that the dynamical models can distinguish between relatively small variations in $M_{\rm BH}$ and formally estimate the best $M_{\rm BH}$, which is, however, a factor of two smaller than the one estimated constraining the models with SINFONI-only data. The SINFONI-only $M_{\rm BH}\approx4\times10^9 M_\odot$, which would provide $R_{\rm SOI}/\sigma_{\rm PSF(MUSE)}\approx 1.4$ scaled down to the value driven by the large scale data.
The reason for this is in the relatively higher quality of the MUSE data compared with the SINFONI, both in the sense of lower uncertainties on the kinematics parameters and lower variations between the values in adjacent bins. Fig.~\ref{ff:sigma_comparison}, with the last row pertinent for IC 4329, demonstrates this issue for the models constrained using SINFONI and MUSE data. The $\sigma$ values (last panel) differ strongly along the pseudo-slit extracted along the major axis, and essentially all three models (with too-low, too-high and best fit $M_{\rm BH}$) agree with the data. Figure~\ref{ff:sigma_comparison_muse} is the equivalent of Fig.~\ref{ff:sigma_comparison}, but demonstrates the results of the models using only large-scale MUSE data. Here we see how the ``too-high'' $M_{\rm BH}$, which partially agrees with the SINFONI data, is excluded by the MUSE data.  Therefore, the models find a ``compromise'' between the lower-resolution, better-quality large-scale data and the higher-resolution, lower-quality small-scale data. The result is in a potentially under-estimated $M_{\rm BH}$, and a requirement for better high-spatial resolution data. 

Similar conclusions can be reached for other galaxies with MUSE observations, with the difference that now MUSE observations are actually able to resolve the SoI and exert less of a ``downward'' bias for $M_{\rm BH}$ (e.g. compare NGC3923, NGC 4636 and IC 4296 in Figs~\ref{ff:schwarzschild_grid3}, ~\ref{ff:sigma_comparison} and~\ref{ff:sigma_comparison_muse}).

\subsection{Long-axis rotation might bias $M_{\rm BH}$ measurements}
\label{ss:la-rot}

Contrary to other galaxies, for NGC 4636 and NGC 4261, we obtained very discrepant results when using the full FOV of our kinematics or a rather small aperture of 3 arcsec to calculate the $\chi^2$ distribution. In fact, our Schwarzschild models were only able to provide an upper limit for NGC 4261, but by limiting the considered kinematics, we were able to constrain the SMBH mass. For NGC 4636, both methods provided constrained SMBH masses, but the results differed by a factor of 10.

NGC 4261 and NGC 4636 have very complex kinematics with at least one change in the rotation axis. This suggests that the issue here might be the complexity of the stellar system. NGC4261 is distinguished by its unusual kinematics, even among other galaxies in Fig.~\ref{ff:kinematics_ls1}, specifically, by its fast long-axis rotation. However, looking at the major axis of the galaxy beyond 10\arcsec, one could also see that there is significant rotation around the short axis. The velocity map of NGC4261 can be visualised as a luminosity-weighted combination of stars moving on two orbital families: long-axis tubes and short-axis tubes, 
where each orbital family dominates at different regions of the galaxy. This is supported by the velocity dispersion map, which shows an unusual ``bow-tie'' feature of higher values. This feature indicates the regions where the two orbital families similarly contribute in terms of their stellar light (mass). Similar features were analysed using dynamical models in some other massive galaxies \citep[e.g.][]{vandenBosch2008, Krajnovic2015, Mulcahey2021, denBrok2021} and could be explained by such a combination of stellar orbital families. Similar features can be seen for NGC 4636.

A hypothesis, which we will further address in a future paper investigating the internal orbital structure of our galaxies, is that a particular combination of the long- and short-axis tubes, with potential inclusion of box orbits, might provide a highly degenerate environment for the Schwarzschild method, and limit its potential in robustly determining $M_{\rm BH}$. This would apply to triaxial galaxies with very complex orbital distributions. 

Further evidence can be provided by PGC 046832, also modelled by triaxial and axisymmetric Schwarzschild method \citep{denBrok2021}. This galaxy was mentioned earlier as having one of the most complex velocity maps with five spin reversals, of which two are prograde and retrograde long-axis rotation. Again, the triaxial Schwarzschild models were not able to constrain the black hole's mass, while the axisymmetric models (including JAM) were able to provide clear estimates, similar to the expectations from various $M_{\rm BH}$ scaling relations. Triaxial models were able to reproduce well the observed kinematics, and indicated that the mass fractions of short- and long-axis tubes are 42\% and 34\%, respectively. The reasons why axisymmetric Schwarzshild and JAM models obtained very different $M_{\rm BH}$ values are likely related to the fact that they did not ``see'' the long-axis tubes. The input kinematics for the axisymmetric Schwarzschild models was symmetrised around the minor axis, and the generally low level of rotation does not contribute significantly to the $V_{RMS}$ used to constrain JAM models. 

Important to note is that the long-axis rotation by itself is not necessarily an issue if there is no evidence for other types of rotation. SINFONI data for NGC 3923 clearly indicate long-axis rotation, but $M_{\rm BH}$ estimated using either or both data sets provide similar values. In this respect, NGC 3923 should perhaps be taken as a counter-example, as the velocity map of this galaxy seems to be dominated only by the long-axis tubes, with negligible rotation around the short-axis. More galaxies similar to NGC 3923 and NGC 4261 will need to be modelled to confirm or refute our hypothesis. 

\subsection{SMBH masses from different dynamical models}
\label{ss:comp}

The obtained $M_{\rm BH}$ are shown in Table~\ref{t:results} and visually in Fig.~\ref{ff:schwarzschild_grid3}, and the differences are clearly visible. Triaxial and axisymmetric Schwarzschild models agree remarkably well for NGC 3706 and NGC 3923, and disagree for others. Triaxial Schwarzschild and JAM models give consistent SMBH masses but disagree regarding the derived $M_*/L$. The axisymmetric Schwarzschild and JAM models typically do not agree with each other. While, as mentioned earlier, there are some global differences between the modelling approaches, such as not including dark matter or using different large-scale kinematics, the main difference is in the assumed symmetry and shape of modelled galaxies. Triaxial galaxies allow diverse orbital families, and the axisymmetric Schwarzschild models can not constrain them. When this is taken into account, together with the large-scale SAURON and VIMOS data used in \cite{Thater2019}, which were inferior in the spatial resolution and data quality compared to the MUSE data used here, it becomes clear why axisymmetric models were not able to constrain $M_{\rm BH}$ for half of the galaxies. 

JAM models, while axisymmetric in nature, do not suffer as much from the shape assumptions for two reasons. Firstly, they are constrained by SINFONI data only, within the central 1\arcsec. As the SoI radii show, this region is dominated by the black hole and therefore expected to have close to spherical symmetry. Secondly, the Jeans equations predict the second velocity moment, which is well approximated by the $V_{RMS} =\sqrt{V^2 + \sigma^2}$ from the IFU data \citep{Binney2005, Cappellari2007}. In our galaxies, $V_{RMS}$ is strongly dominated by the velocity dispersion, with little information on the underlying stellar streaming around different axes. In this respect, the disagreement between the JAM and triaxial models $M_{\rm BH}$ estimates stems primarily from the quality of SINFONI data.

\subsection{Scaling relations}

\begin{figure}
  \centering
    \includegraphics[width=0.5\textwidth]{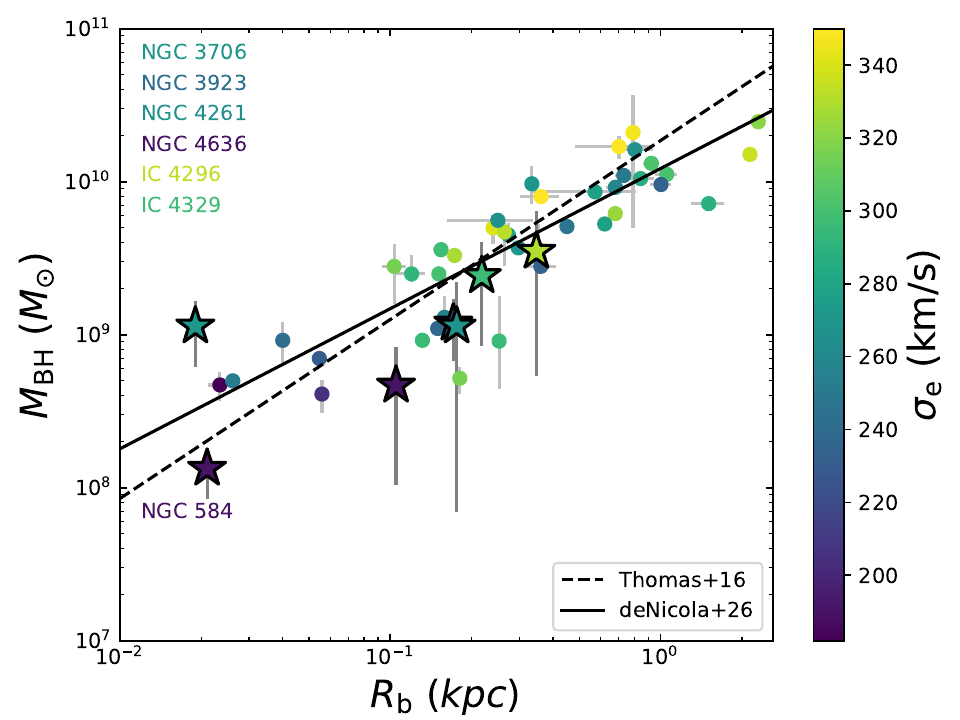}
    
      \caption{
      Relation between core-S\'ersic break radius $R_{\rm b}$ and black hole mass. We used the compilation of core galaxies from \cite{Rusli2013a}, and added several recent measurements (e.g., NGC 584 from \citealt{Thater2019}) and our newly analysed core galaxies (highlighted as stars). The colours indicate the effective velocity dispersion of the shown galaxies. The two overlapping stars are NGC 3923 and NGC 4261. The lines are the fits to the same sample by \cite{Thomas2016} and \cite{Nicola2025}.}
      \label{ff:breakr_blackhole_relation}
\end{figure}

\begin{figure}
  \centering
    \includegraphics[width=0.5\textwidth]{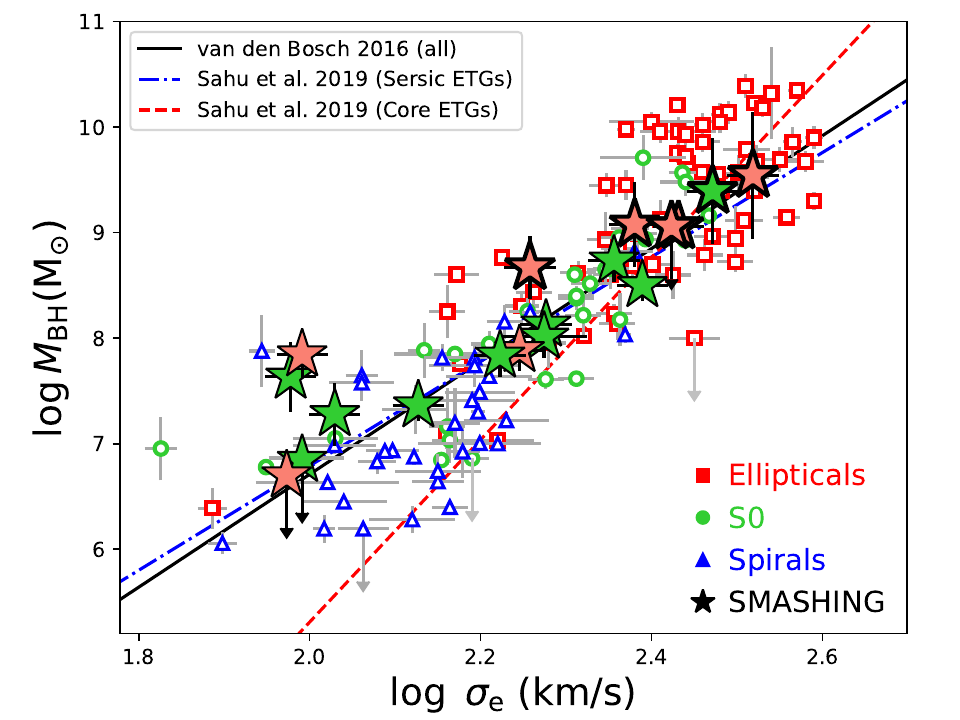}
    
      \caption{Supermassive black hole mass-effective velocity dispersion relation based on the compilation of \cite{Saglia2016}. The colour scheme indicates the morphological type of the galaxies: elliptical (red), lenticular (green) and spiral (blue). In order to visualise the general trend, we have added the global scaling relation by \cite{Bosch2016} for all galaxy types (black solid line) as well as the relations by \cite{Sahu2019b} for S\'ersic (dashed-dotted blue line) and core galaxies (dashed-dashed red line). Our new measurements (highlighted as stars with bold edges) lie very well on the scaling relations. The figure includes the other SMASHING galaxies from \cite{Krajnovic2018} and \cite{Thater2019}, indicated with the star symbol.}
      \label{ff:mbh_sigma}
\end{figure}

Our results of this work, reported in Table \ref{t:results}, provide six black hole mass measurements for the scaling relations. Five measurements are revisions using a more complex modelling method, and for IC 4329, we have measured the black hole mass for the first time. For a detailed discussion on the revision, we refer to Section A of the Appendix. Moreover, our sample populates the high-mass region of the scaling relation, which shows an increased intrinsic scatter possibly caused by different galaxy evolutionary tracks or limitations in the modelling methods. Many of the very massive galaxies have triaxial shapes, while most of the stellar-based black hole mass measurements were derived using axisymmetric Schwarzschild modelling. Specifically, galaxies which have complex kinematics, such as those presented here, might require a revision of their SMBH mass determinations.

We show our results from the triaxial Schwarzschild models in Figs.~\ref{ff:breakr_blackhole_relation} and~\ref{ff:mbh_sigma}, and compare them with a compilation of literature values. Figure~\ref{ff:breakr_blackhole_relation} shows the relation between core-S\'ersic break radius and black hole mass: $M_{\rm BH} - R_b$ \citep{Rusli2013a, Dullo2014, Thomas2016,Rantala2018}. We used the compilation of core galaxies from \cite{Rusli2013a} and added NGC 584 from \cite{Thater2019}, NGC 708 from \cite{Nicola2024}, NGC 1272 from \cite{Saglia2024} and 16 preliminary $M_{\rm BH}$ measurements from \cite{Nicola2026}.  Our new measurements are well aligned with previously measured black hole masses. NGC 3706 is over-massive compared to the scaling relation, and NGC 4636 and IC 4329 lie below the line. For NGC 4636, the break radius might be incorrect and might require a revision \citep[see, e.g.][for discussion on NGC 4636]{Graham2003}. For IC 4329, a more massive black hole mass (we find a second $\chi^2$ minimum a factor of two higher) would bring this galaxy closer to the scaling relations.

We also show our results in the context of the $M_{\rm BH} - \sigma_{\rm e}$ relation in Fig.~\ref{ff:mbh_sigma}. All of our measurements align well with literature measurements and the shown scaling relations. We do not find any significant over-massive black hole in the high-mass region as was discussed in previous work. The most over-massive case, NGC 4636, actually falls in the intermediate $\sigma$ range, which is still a region where ellipticals generally seem to be offset from S0 galaxies, possibly due to different evolutionary scenarios. 
Recently, \cite{Dullo2021} and \cite{Nicola2025} showed that massive core galaxies follow the $M_{\rm BH} - M_{\rm *}$ relation better than the $M_{\rm BH} - \sigma_{\rm e}$ relation, because effective velocity dispersions seem to stagnate around 400 km s$^{-1}$. We compared our measurements to the $M_{\rm BH} - M_{\rm *}$ relation by \cite{Sahu2019} and find that our new measurements follow both relations equally well. The reason why our measurements follow the $M_{\rm BH} - \sigma_{\rm e}$ relation well is likely that our galaxies do not have very large depleted cores \citep{Dullo2021}.

\section{Conclusions}

In this study, we present black hole mass estimates for six massive galaxies (NGC 3706, NGC 3923, NGC 4261, NGC 4636, IC 4296, and IC 4329). These galaxies were selected based on their expected resolvable spheres of influence (SoI) with the SINFONI IFU. We combine adaptive-optics-assisted near-infrared SINFONI observations with optical MUSE IFU data to construct triaxial Schwarzschild orbit-superposition models using the \texttt{DYNAMITE} software. Stellar kinematics were extracted from both the SINFONI and MUSE data using the Python-based \texttt{pPXF} software (Section~\ref{s:kin}). For the dynamical modelling, we first built stellar mass models using the MGE method, carefully accounting for dust features present in the galaxy centres, and then used these to construct triaxial dynamical models. The triaxial modelling allows us to constrain the intrinsic shape, dark matter content, and black hole mass for each galaxy.

Our main findings and conclusions are as follows:

\begin{itemize}
    
\item The sample galaxies exhibit interesting and complex kinematic features, visible in both the SINFONI kinematic maps (Figure~\ref{ff:sinfoni_kinematics}) and large-scale MUSE data (Figure~\ref{ff:kinematics_ls1}). Notably, NGC 3706, NGC 4636, and IC 4296 show clear kinematically decoupled cores (KDCs) in their nuclei. The measured rotational velocities are typically low, with three galaxies—NGC 3923, NGC 4636, and IC 4329—exhibiting long-axis rotation.

\item We derive black hole masses for all six galaxies: $M_{\rm BH}$ (NGC 3706) = $\rm (1.14^{+0.41}_{-0.63})  \times 10^9\,M_{\odot}$, $M_{\rm BH}$ (NGC 3923) = $(1.19^{+1.34}_{-0.80}) \times 10^9\,M_{\odot}$, $M_{\rm BH}$ (NGC 4261) = $(1.14^{+1.08}_{-0.95}) \times 10^9$ M$_{\odot}$, $M_{\rm BH}$ (NGC 4636) = $(4.68^{+2.99}_{-4.26}) \times 10^8\,M_{\odot}$, $\rm M_{\rm BH}$ (IC 4296) = $(3.51^{+3.37}_{-2.57}) 10^9\,M_{\odot}$, and $\rm M_{\rm BH}$ (IC 4329) = $(2.43^{+1.53}_{-1.65}) \times 10^9\,M_{\odot}$ (see Table~\ref{t:results}). These values are broadly consistent with previous measurements obtained using axisymmetric mass modelling, although the inclusion of triaxial geometry and extended kinematic coverage improves the robustness of the estimates.

\item Using our triaxial models, we compare the black hole masses with previous estimates from JAM and axisymmetric Schwarzschild models \citep{Thater2019b}. While the overall trends across methods are generally consistent, significant differences are found for individual galaxies. This underscores the importance of accounting for galaxy geometry and having high-resolution kinematic data when estimating black hole masses.

\item For NGC 4261 and NGC 4636, we find a strong discrepancy when using the full FOV or a small central aperture when deriving the SMBHS mass. These two galaxies show the most complex kinematics of our studied sample, and we presume that the long-axis rotation in these galaxies might bias the mass measurement. 
\end{itemize}

Our findings contribute to refining the black hole census in massive galaxies and offer insight into improving SMBH–host galaxy scaling relations at the high-mass end, where increased scatter and structural complexity challenge conventional modelling approaches. In future papers, we will investigate the effects of spatially varying mass-to-light ratios and stellar population gradients on black hole mass measurements. In addition, we will explore the internal orbital structure of these systems to better understand their dynamical states and evolutionary histories. 

\section*{Data availability}
The Supplementary material is accessible under \url{https://doi.org/10.5281/zenodo.20158190}. The reduced SINFONI and MUSE data cubes (FITS files) are available at the CDS via anonymous ftp to cdsarc.u-strasbg.fr (130.79.128.5) or via http://cdsweb.u-strasbg.fr/cgi-bin/qcat?J/A+A/.. 

\begin{acknowledgements} 
We thank the members of the Vienna Dynamics Team for their support and valuable discussions. We also thank Aaron Barth and Ben Boizelle for providing reduced and calibrated images for some of our sample galaxies. SK acknowledges funding via STFC Small Grant ST/Y001133/1. 
T.N. acknowledges the support of the Deutsche Forschungsgemeinschaft (DFG, German Research Foundation) under Germany's Excellence Strategy - EXC-2094 - 390783311 of the DFG Cluster of Excellence "ORIGINS". The computational results have been achieved using the Austrian Scientific Computing (ASC) infrastructure. This work is based on observations collected at the European Organisation for Astronomical Research in the Southern Hemisphere and also based on observations made with the NASA/ESA Hubble Space Telescope, obtained from the Hubble Legacy Archive, which is a collaboration between the Space Telescope Science Institute (STScI/NASA), the Space Telescope European Coordinating Facility (STECF/ESA) and the Canadian Astronomy Data Centre (CADC/NRC/CSA). This research has made use of the NASA/IPAC Extragalactic Database (NED), which is operated by the Jet Propulsion Laboratory, California Institute of Technology, under contract with the National Aeronautics and Space Administration. This research is partially based on data from the MILES project. 
\end{acknowledgements}

\bibliography{papers}

@Article{Alam2015,
  Title                    = {{The Eleventh and Twelfth Data Releases of the Sloan Digital Sky Survey: Final Data from SDSS-III}},
  Author                   = {{Alam}, S. and {Albareti}, F.~D. and {Allende Prieto}, C. and {Anders}, F. and {Anderson}, S.~F. and {Anderton}, T. and {Andrews}, B.~H. and {Armengaud}, E. and {Aubourg}, {\'E}. and {Bailey}, S. and et al.},
  Journal                  = {\apjs},
  Year                     = {2015},

  Month                    = jul,
  Pages                    = {12},
  Volume                   = {219},
  Adsurl                   = {http://adsabs.harvard.edu/abs/2015ApJS..219...12A},
  Archiveprefix            = {arXiv},
  Doi                      = {10.1088/0067-0049/219/1/12},
  Eid                      = {12},
  Eprint                   = {1501.00963},
  Primaryclass             = {astro-ph.IM}
}

@Article{Bilek2016,
  author        = {{B{\'{\i}}lek}, M. and {Cuillandre}, J.-C. and {Gwyn}, S. and {Ebrov{\'a}}, I. and {Barto{\v s}kov{\'a}}, K. and {Jungwiert}, B. and {J{\'{\i}}lkov{\'a}}, L.},
  title         = {{Deep imaging of the shell elliptical galaxy NGC 3923 with MegaCam}},
  journal       = {\aap},
  year          = {2016},
  volume        = {588},
  pages         = {A77},
  month         = apr,
  adsurl        = {http://adsabs.harvard.edu/abs/2016A%26A...588A..77B},
  archiveprefix = {arXiv},
  doi           = {10.1051/0004-6361/201526608},
  eid           = {A77},
  eprint        = {1505.07146},
}

@InProceedings{Bacon2010,
  author    = {{Bacon}, R. and {Accardo}, M. and {Adjali}, L. and {Anwand}, H. and {Bauer}, S. and {Biswas}, I. and {Blaizot}, J. and {Boudon}, D. and {Brau-Nogue}, S. and {Brinchmann}, J. and {Caillier}, P. and {Capoani}, L. and {Carollo}, C.~M. and {Contini}, T. and {Couderc}, P. and {Daguis{\'e}}, E. and {Deiries}, S. and {Delabre}, B. and {Dreizler}, S. and {Dubois}, J. and {Dupieux}, M. and {Dupuy}, C. and {Emsellem}, E. and {Fechner}, T. and {Fleischmann}, A. and {Fran{\c c}ois}, M. and {Gallou}, G. and {Gharsa}, T. and {Glindemann}, A. and {Gojak}, D. and {Guiderdoni}, B. and {Hansali}, G. and {Hahn}, T. and {Jarno}, A. and {Kelz}, A. and {Koehler}, C. and {Kosmalski}, J. and {Laurent}, F. and {Le Floch}, M. and {Lilly}, S.~J. and {Lizon}, J.-L. and {Loupias}, M. and {Manescau}, A. and {Monstein}, C. and {Nicklas}, H. and {Olaya}, J.-C. and {Pares}, L. and {Pasquini}, L. and {P{\'e}contal-Rousset}, A. and {Pell{\'o}}, R. and {Petit}, C. and {Popow}, E. and {Reiss}, R. and {Remillieux}, A. and {Renault}, E. and {Roth}, M. and {Rupprecht}, G. and {Serre}, D. and {Schaye}, J. and {Soucail}, G. and {Steinmetz}, M. and {Streicher}, O. and {Stuik}, R. and {Valentin},, H. and {Vernet}, J. and {Weilbacher}, P. and {Wisotzki}, L. and {Yerle}, N.},
  title     = {{The MUSE second-generation VLT instrument}},
  booktitle = {Ground-based and Airborne Instrumentation for Astronomy III},
  year      = {2010},
  volume    = {7735},
  series    = {\procspie},
  pages     = {773508},
  month     = jul,
  adsurl    = {http://adsabs.harvard.edu/abs/2010SPIE.7735E..08B},
  doi       = {10.1117/12.856027},
  eid       = {773508},
}

@Article{Bacon2001,
  author    = {{Bacon}, R. and {Copin}, Y. and {Monnet}, G. and {Miller}, B.~W. and {Allington-Smith}, J.~R. and {Bureau}, M. and {Carollo}, C.~M. and {Davies}, R.~L. and {Emsellem}, E. and {Kuntschner}, H. and {Peletier}, R.~F. and {Verolme}, E.~K. and {de Zeeuw}, P.~T.},
  title     = {{The SAURON project - I. The panoramic integral-field spectrograph}},
  journal   = {\mnras},
  year      = {2001},
  volume    = {326},
  pages     = {23-35},
  month     = sep,
  adsurl    = {http://ads.nao.ac.jp/abs/2001MNRAS.326...23B},
  doi       = {10.1046/j.1365-8711.2001.04612.x},
  eprint    = {astro-ph/0103451},
}

@Article{Beifiori2012,
  author        = {{Beifiori}, A. and {Courteau}, S. and {Corsini}, E.~M. and {Zhu}, Y.},
  journal       = {\mnras},
  title         = {{On the correlations between galaxy properties and supermassive black hole mass}},
  year          = {2012},
  month         = jan,
  pages         = {2497-2528},
  volume        = {419},
  adsurl        = {https://ui.adsabs.harvard.edu/abs/2012MNRAS.419.2497B},
  archiveprefix = {arXiv},
  doi           = {10.1111/j.1365-2966.2011.19903.x},
  eprint        = {1109.6265},
}

@Article{Bonfini2018,
  Title                    = {{Connecting traces of galaxy evolution: the missing core mass-morphological fine structure relation}},
  Author                   = {{Bonfini}, P. and {Bitsakis}, T. and {Zezas}, A. and {Duc}, P.-A. and {Iodice}, E. and {Gonz{\'a}lez-Mart{\'{\i}}n}, O. and {Bruzual}, G. and {Gonz{\'a}lez Sanoja}, A.~J.},
  Journal                  = {\mnras},
  Year                     = {2018},

  Month                    = jan,
  Pages                    = {L94-L100},
  Volume                   = {473},
  Adsurl                   = {http://adsabs.harvard.edu/abs/2018MNRAS.473L..94B},
  Archiveprefix            = {arXiv},
  Doi                      = {10.1093/mnrasl/slx169},
  Eprint                   = {1710.05025}
}

@Article{Bonnet2004,
  author  = {{Bonnet}, H. and {Abuter}, R. and {Baker}, A. and {Bornemann}, W. and {Brown}, A. and {Castillo}, R. and {Conzelmann}, R. and {Damster}, R. and {Davies}, R. and {Delabre}, B. and {Donaldson}, R. and {Dumas}, C. and {Eisenhauer}, F. and {Elswijk}, E. and {Fedrigo}, E. and {Finger}, G. and {Gemperlein}, H. and {Genzel}, R. and {Gilbert}, A. and {Gillet}, G. and {Goldbrunner}, A. and {Horrobin}, M. and {Ter Horst}, R. and {Huber}, S. and {Hubin}, N. and {Iserlohe}, C. and {Kaufer}, A. and {Kissler-Patig}, M. and {Kragt}, J. and {Kroes}, G. and {Lehnert}, M. and {Lieb}, W. and {Liske}, J. and {Lizon}, J.-L. and {Lutz}, D. and {Modigliani}, A. and {Monnet}, G. and {Nesvadba}, N. and {Patig}, J. and {Pragt}, J. and {Reunanen}, J. and {R{\"o}hrle}, C. and {Rossi}, S. and {Schmutzer}, R. and {Schoenmaker}, T. and {Schreiber}, J. and {Stroebele}, S. and {Szeifert}, T. and {Tacconi}, L. and {Tecza}, M. and {Thatte}, N. and {Tordo}, S. and {van der Werf}, P. and {Weisz}, H.},
  title   = {{First light of SINFONI at the VLT}},
  journal = {The Messenger},
  year    = {2004},
  volume  = {117},
  pages   = {17-24},
  month   = sep,
  adsurl  = {http://adsabs.harvard.edu/abs/2004Msngr.117...17B},
}

@Article{Bosch2016,
  Title                    = {UNIFICATION OF THE FUNDAMENTAL PLANE {AND} SUPER MASSIVE BLACK HOLE MASSES},
  Author                   = {van den Bosch, Remco C. E.},
  Journal                  = {The Astrophysical Journal},
  Year                     = {2016},

  Month                    = {Nov},
  Number                   = {2},
  Pages                    = {134},
  Volume                   = {831},

  Doi                      = {10.3847/0004-637x/831/2/134},
  ISSN                     = {1538-4357},
  Publisher                = {American Astronomical Society},
  Url                      = {http://dx.doi.org/10.3847/0004-637X/831/2/134}
}

@Article{Capetti2005,
  author    = {{Capetti}, A. and {Marconi}, A. and {Macchetto}, D. and {Axon}, D.},
  journal   = {\aap},
  title     = {{The supermassive black hole in the Seyfert 2 galaxy NGC 5252}},
  year      = {2005},
  month     = feb,
  pages     = {465-475},
  volume    = {431},
  adsurl    = {http://adsabs.harvard.edu/abs/2005A%26A...431..465C},
  doi       = {10.1051/0004-6361:20041701},
  eprint    = {astro-ph/0411081},
}

@Article{Cappellari2017,
  author        = {{Cappellari}, M.},
  title         = {{Improving the full spectrum fitting method: accurate convolution with Gauss-Hermite functions}},
  journal       = {\mnras},
  year          = {2017},
  volume        = {466},
  pages         = {798-811},
  month         = apr,
  adsurl        = {http://adsabs.harvard.edu/abs/2017MNRAS.466..798C},
  archiveprefix = {arXiv},
  doi           = {10.1093/mnras/stw3020},
  eprint        = {1607.08538},
}

@Article{Cappellari2016,
  author        = {{Cappellari}, Michele},
  journal       = {\araa},
  title         = {{Structure and Kinematics of Early-Type Galaxies from Integral Field Spectroscopy}},
  year          = {2016},
  month         = sep,
  pages         = {597-665},
  volume        = {54},
  adsurl        = {https://ui.adsabs.harvard.edu/abs/2016ARA&A..54..597C},
  archiveprefix = {arXiv},
  doi           = {10.1146/annurev-astro-082214-122432},
  eprint        = {1602.04267},
  primaryclass  = {astro-ph.GA},
}

@Article{Cappellari2008,
  author        = {{Cappellari}, M.},
  journal       = {\mnras},
  title         = {{Measuring the inclination and mass-to-light ratio of axisymmetric galaxies via anisotropic Jeans models of stellar kinematics}},
  year          = {2008},
  month         = {Oct},
  pages         = {71-86},
  volume        = {390},
  adsurl        = {http://adsabs.harvard.edu/abs/2008MNRAS.390...71C},
  archiveprefix = {arXiv},
  comment       = {JAM},
  doi           = {10.1111/j.1365-2966.2008.13754.x},
  eprint        = {0806.0042},
}

@Article{Cappellari2002,
  author        = {{Cappellari}, Michele},
  journal       = {\mnras},
  title         = {{Efficient multi-Gaussian expansion of galaxies}},
  year          = {2002},
  month         = jun,
  number        = {2},
  pages         = {400-410},
  volume        = {333},
  adsurl        = {https://ui.adsabs.harvard.edu/abs/2002MNRAS.333..400C},
  archiveprefix = {arXiv},
  doi           = {10.1046/j.1365-8711.2002.05412.x},
  eprint        = {astro-ph/0201430},
  primaryclass  = {astro-ph},
}

@Article{Cappellari2006,
  author    = {{Cappellari}, M. and {Bacon}, R. and {Bureau}, M. and {Damen}, M.~C. and {Davies}, R.~L. and {de Zeeuw}, P.~T. and {Emsellem}, E. and {Falc{\'o}n-Barroso}, J. and {Krajnovi{\'c}}, D. and {Kuntschner}, H. and {McDermid}, R.~M. and {Peletier}, R.~F. and {Sarzi}, M. and {van den Bosch}, R.~C.~E. and {van de Ven}, G.},
  title     = {{The SAURON project - IV. The mass-to-light ratio, the virial mass estimator and the Fundamental Plane of elliptical and lenticular galaxies}},
  journal   = {\mnras},
  year      = {2006},
  volume    = {366},
  pages     = {1126-1150},
  month     = mar,
  adsurl    = {http://adsabs.harvard.edu/abs/2006MNRAS.366.1126C},
  comment   = {sigma_e interpolation},
  doi       = {10.1111/j.1365-2966.2005.09981.x},
  eprint    = {astro-ph/0505042},
}

@Article{Cappellari2003,
  author    = {{Cappellari}, M. and {Copin}, Y.},
  title     = {{Adaptive spatial binning of integral-field spectroscopic data using Voronoi tessellations}},
  journal   = {\mnras},
  year      = {2003},
  volume    = {342},
  pages     = {345-354},
  month     = jun,
  adsurl    = {http://adsabs.harvard.edu/abs/2003MNRAS.342..345C},
  comment   = {Voronoi},
  doi       = {10.1046/j.1365-8711.2003.06541.x},
  eprint    = {astro-ph/0302262},
}

@Article{Cappellari2004,
  author    = {{Cappellari}, M. and {Emsellem}, E.},
  title     = {{Parametric Recovery of Line-of-Sight Velocity Distributions from Absorption-Line Spectra of Galaxies via Penalized Likelihood}},
  journal   = {\pasp},
  year      = {2004},
  volume    = {116},
  pages     = {138-147},
  month     = feb,
  adsurl    = {http://adsabs.harvard.edu/abs/2004PASP..116..138C},
  comment   = {PPXF},
  doi       = {10.1086/381875},
  eprint    = {astro-ph/0312201},
}

@Article{Cappellari2007,
  Title                    = {{The SAURON project - X. The orbital anisotropy of elliptical and lenticular galaxies: revisiting the (V/{$\sigma$}, {$\epsilon$}) diagram with integral-field stellar kinematics}},
  Author                   = {{Cappellari}, M. and {Emsellem}, E. and {Bacon}, R. and {Bureau}, M. and {Davies}, R.~L. and {de Zeeuw}, P.~T. and {Falc{\'o}n-Barroso}, J. and {Krajnovi{\'c}}, D. and {Kuntschner}, H. and {McDermid}, R.~M. and {Peletier}, R.~F. and {Sarzi}, M. and {van den Bosch}, R.~C.~E. and {van de Ven}, G.},
  Journal                  = {\mnras},
  Year                     = {2007},

  Month                    = aug,
  Pages                    = {418-444},
  Volume                   = {379},
  Adsurl                   = {http://adsabs.harvard.edu/abs/2007MNRAS.379..418C},
  Comment                  = {Schwarzschild},
  Doi                      = {10.1111/j.1365-2966.2007.11963.x},
  Eprint                   = {astro-ph/0703533}
}

@Article{Cappellari2011,
  Title                    = {{The ATLAS$^{3D}$ project - I. A volume-limited sample of 260 nearby early-type galaxies: science goals and selection criteria}},
  Author                   = {{Cappellari}, M. and {Emsellem}, E. and {Krajnovi{\'c}}, D. and {McDermid}, R.~M. and {Scott}, N. and {Verdoes Kleijn}, G.~A. and {Young}, L.~M. and {Alatalo}, K. and {Bacon}, R. and {Blitz}, L. and {Bois}, M. and {Bournaud}, F. and {Bureau}, M. and {Davies}, R.~L. and {Davis}, T.~A. and {de Zeeuw}, P.~T. and {Duc}, P.-A. and {Khochfar}, S. and {Kuntschner}, H. and {Lablanche}, P.-Y. and {Morganti}, R. and {Naab}, T. and {Oosterloo}, T. and {Sarzi}, M. and {Serra}, P. and {Weijmans}, A.-M.},
  Journal                  = {\mnras},
  Year                     = {2011},

  Month                    = may,
  Pages                    = {813-836},
  Volume                   = {413},
  Adsurl                   = {http://adsabs.harvard.edu/abs/2011MNRAS.413..813C},
  Archiveprefix            = {arXiv},
  Doi                      = {10.1111/j.1365-2966.2010.18174.x},
  Eprint                   = {1012.1551}
}

@Article{Cappellari2013b,
  author        = {{Cappellari}, M. and {McDermid}, R.~M. and {Alatalo}, K. and {Blitz}, L. and {Bois}, M. and {Bournaud}, F. and {Bureau}, M. and {Crocker}, A.~F. and {Davies}, R.~L. and {Davis}, T.~A. and {de Zeeuw}, P.~T. and {Duc}, P.-A. and {Emsellem}, E. and {Khochfar}, S. and {Krajnovi{\'c}}, D. and {Kuntschner}, H. and {Morganti}, R. and {Naab}, T. and {Oosterloo}, T. and {Sarzi}, M. and {Scott}, N. and {Serra}, P. and {Weijmans}, A.-M. and {Young}, L.~M.},
  title         = {{The ATLAS$^{3D}$ project - XX. Mass-size and mass-{$\sigma$} distributions of early-type galaxies: bulge fraction drives kinematics, mass-to-light ratio, molecular gas fraction and stellar initial mass function}},
  journal       = {\mnras},
  year          = {2013},
  volume        = {432},
  pages         = {1862-1893},
  month         = jul,
  adsurl        = {http://adsabs.harvard.edu/abs/2013MNRAS.432.1862C},
  archiveprefix = {arXiv},
  comment       = {LOESS},
  doi           = {10.1093/mnras/stt644},
  eprint        = {1208.3523},
  primaryclass  = {astro-ph.CO},
}

@Article{Cappellari2015,
  author        = {{Cappellari}, M. and {Romanowsky}, A.~J. and {Brodie}, J.~P. and {Forbes}, D.~A. and {Strader}, J. and {Foster}, C. and {Kartha}, S.~S. and {Pastorello}, N. and {Pota}, V. and {Spitler}, L.~R. and {Usher}, C. and {Arnold}, J.~A.},
  title         = {{Small Scatter and Nearly Isothermal Mass Profiles to Four Half-light Radii from Two-dimensional Stellar Dynamics of Early-type Galaxies}},
  journal       = {\apjl},
  year          = {2015},
  volume        = {804},
  pages         = {L21},
  month         = may,
  adsurl        = {http://adsabs.harvard.edu/abs/2015ApJ...804L..21C},
  archiveprefix = {arXiv},
  comment       = {JAM Navarro profile},
  doi           = {10.1088/2041-8205/804/1/L21},
  eid           = {L21},
  eprint        = {1504.00075},
}

@Article{Cappellari2013,
  Title                    = {{The ATLAS$^{3D}$ project - XV. Benchmark for early-type galaxies scaling relations from 260 dynamical models: mass-to-light ratio, dark matter, Fundamental Plane and Mass Plane}},
  Author                   = {{Cappellari}, M. and {Scott}, N. and {Alatalo}, K. and {Blitz}, L. and {Bois}, M. and {Bournaud}, F. and {Bureau}, M. and {Crocker}, A.~F. and {Davies}, R.~L. and {Davis}, T.~A. and {de Zeeuw}, P.~T. and {Duc}, P.-A. and {Emsellem}, E. and {Khochfar}, S. and {Krajnovi{\'c}}, D. and {Kuntschner}, H. and {McDermid}, R.~M. and {Morganti}, R. and {Naab}, T. and {Oosterloo}, T. and {Sarzi}, M. and {Serra}, P. and {Weijmans}, A.-M. and {Young}, L.~M.},
  Journal                  = {\mnras},
  Year                     = {2013},

  Month                    = jul,
  Pages                    = {1709-1741},
  Volume                   = {432},
  Adsurl                   = {http://adsabs.harvard.edu/abs/2013MNRAS.432.1709C},
  Archiveprefix            = {arXiv},
  Comment                  = {ETG, beta, DM},
  Doi                      = {10.1093/mnras/stt562},
  Eprint                   = {1208.3522},
  Primaryclass             = {astro-ph.CO}
}

@Article{Carlsten2017,
  author        = {{Carlsten}, S.~G. and {Hau}, G.~K.~T. and {Zenteno}, A.},
  title         = {{Stellar populations of shell galaxies}},
  journal       = {\mnras},
  year          = {2017},
  volume        = {472},
  pages         = {2889-2905},
  month         = dec,
  adsurl        = {http://adsabs.harvard.edu/abs/2017MNRAS.472.2889C},
  archiveprefix = {arXiv},
  doi           = {10.1093/mnras/stx2182},
  eprint        = {1611.05437},
}

@Article{Carter1998,
  Title                    = {{Minor axis rotation and the intrinsic shape of the shell elliptical NGC 3923}},
  Author                   = {{Carter}, D. and {Thomson}, R.~C. and {Hau}, G.~K.~T.},
  Journal                  = {\mnras},
  Year                     = {1998},

  Month                    = feb,
  Pages                    = {182},
  Volume                   = {294},
  Adsurl                   = {https://ui.adsabs.harvard.edu/abs/1998MNRAS.294..182C},
  Doi                      = {10.1046/j.1365-8711.1998.01287.x}
}

@Article{Chae2019,
  Title                    = {{Modeling Nearly Spherical Pure-bulge Galaxies with a Stellar Mass-to-light Ratio Gradient under the {$\Lambda$}CDM and MOND Paradigms. II. The Orbital Anisotropy of Slow Rotators within the Effective Radius}},
  Author                   = {{Chae}, K.-H. and {Bernardi}, M. and {Sheth}, R.~K.},
  Journal                  = {\apj},
  Year                     = {2019},

  Month                    = mar,
  Pages                    = {41},
  Volume                   = {874},
  Adsurl                   = {https://ui.adsabs.harvard.edu/abs/2019ApJ...874...41C},
  Archiveprefix            = {arXiv},
  Doi                      = {10.3847/1538-4357/ab09fd},
  Eid                      = {41},
  Eprint                   = {1902.09350}
}

@Article{Cleveland1979,
  Title                    = {Robust locally weighted regression and smoothing scatterplots},
  Author                   = {Cleveland, William S.},
  Journal                  = {Journal of the American Statistical Association},
  Year                     = {1979},
  Pages                    = {829--836},
  Volume                   = {74}
}

@Article{DallaBonta2009,
  Title                    = {{The High-Mass End of the Black Hole Mass Function: Mass Estimates in Brightest Cluster Galaxies}},
  Author                   = {{Dalla Bont{\`a}}, E. and {Ferrarese}, L. and {Corsini}, E.~M. and {Miralda-Escud{\'e}}, J. and {Coccato}, L. and {Sarzi}, M. and {Pizzella}, A. and {Beifiori}, A.},
  Journal                  = {\apj},
  Year                     = {2009},

  Month                    = jan,
  Pages                    = {537-559},
  Volume                   = {690},
  Adsurl                   = {http://adsabs.harvard.edu/abs/2009ApJ...690..537D},
  Archiveprefix            = {arXiv},
  Doi                      = {10.1088/0004-637X/690/1/537},
  Eprint                   = {0809.0766}
}

@Article{Davies1986,
  author    = {{Davies}, R.~L. and {Birkinshaw}, M.},
  journal   = {\apjl},
  title     = {{NGC 4261 - A prolate elliptical galaxy}},
  year      = {1986},
  month     = apr,
  pages     = {L45-L49},
  volume    = {303},
  adsurl    = {https://ui.adsabs.harvard.edu/abs/1986ApJ...303L..45D},
  doi       = {10.1086/184650},
}

@Book{deVaucouleurs1991rc3,
  Title                    = {{Third Reference Catalogue of Bright Galaxies. Volume I: Explanations and references. Volume II: Data for galaxies between 0$^{h}$ and 12$^{h}$. Volume III: Data for galaxies between 12$^{h}$ and 24$^{h}$.}},
  Author                   = {{de Vaucouleurs}, G. and {de Vaucouleurs}, A. and {Corwin}, Jr., H.~G. and {Buta}, R.~J. and {Paturel}, G. and {Fouqu{\'e}}, P.},
  Year                     = {1991},
  Adsurl                   = {http://adsabs.harvard.edu/abs/1991rc3..book.....D},
  Booktitle                = {Third Reference Catalogue of Bright Galaxies.~Volume I: Explanations and references.~ Volume II: Data for galaxies between 0$^{h}$ and 12$^{h}$.~ Volume III: Data for galaxies between 12$^{h}$ and 24$^{h}$., by de Vaucouleurs, G.; de Vaucouleurs, A.; Corwin, H.~G., Jr.; Buta, R.~J.; Paturel, G.; Fouqu{\'e}, P..~Springer, New York, NY (USA), 1991, 2091 p., ISBN 0-387-97552-7, Price US$ 198.00. ISBN 3-540-97552-7, Price DM 448.00. ISBN 0-387-97549-7 (Vol. I), ISBN 0-387-97550-0 (Vol. II), ISBN 0-387-97551-9 (Vol. III).},
  Comment                  = {ngc 4414}
}

@Article{Dirsch2005,
  Title                    = {{A wide-field photometric study of the globular cluster system of NGC 4636}},
  Author                   = {{Dirsch}, B. and {Schuberth}, Y. and {Richtler}, T.},
  Journal                  = {\aap},
  Year                     = {2005},

  Month                    = apr,
  Pages                    = {43-56},
  Volume                   = {433},
  Adsurl                   = {https://ui.adsabs.harvard.edu/abs/2005A%26A...433...43D},
  Doi                      = {10.1051/0004-6361:20035737},
  Eprint                   = {astro-ph/0503283}
}

@Article{Dolphin2009,
  Title                    = {{A Revised Characterization of the WFPC2 CTE Loss}},
  Author                   = {{Dolphin}, A.~E.},
  Journal                  = {\pasp},
  Year                     = {2009},

  Month                    = jun,
  Pages                    = {655},
  Volume                   = {121},
  Adsurl                   = {http://ads.nao.ac.jp/abs/2009PASP..121..655D},
  Archiveprefix            = {arXiv},
  Comment                  = {correction of HST2 magnitudes},
  Doi                      = {10.1086/600028},
  Eprint                   = {0906.3557},
  Primaryclass             = {astro-ph.IM}
}

@Article{Drehmer2015,
  author        = {{Drehmer}, D.~A. and {Storchi-Bergmann}, T. and {Ferrari}, F. and {Cappellari}, M. and {Riffel}, R.~A.},
  title         = {{The benchmark black hole in NGC 4258: dynamical models from high-resolution two-dimensional stellar kinematics}},
  journal       = {\mnras},
  year          = {2015},
  volume        = {450},
  pages         = {128-144},
  month         = jun,
  adsurl        = {http://adsabs.harvard.edu/abs/2015MNRAS.450..128D},
  archiveprefix = {arXiv},
  comment       = {Davor Empfehlung},
  doi           = {10.1093/mnras/stv536},
  eprint        = {1503.04540},
}

@Article{Dullo2019,
  author        = {{Dullo}, B.~T. and {Chamorro-Cazorla}, M. and {Gil de Paz}, A. and {Castillo-Morales}, {\'A}. and {Gallego}, J. and {Carrasco}, E. and {Iglesias-P{\'a}ramo}, J. and {Cedazo}, R. and {Garc{\'{\i}}a-Vargas}, M.~L. and {Pascual}, S. and {Cardiel}, N. and {P{\'e}rez-Calpena}, A. and {G{\'o}mez-Alvarez}, P. and {Mart{\'{\i}}nez-Delgado}, I. and {Catal{\'a}n-Torrecilla}, C.},
  journal       = {\apj},
  title         = {{High-resolution MEGARA Integral-field Unit Spectroscopy and Structural Analysis of a Fast-rotating, Disky Bulge in NGC 7025}},
  year          = {2019},
  month         = jan,
  pages         = {9},
  volume        = {871},
  adsurl        = {http://adsabs.harvard.edu/abs/2019ApJ...871....9D},
  archiveprefix = {arXiv},
  doi           = {10.3847/1538-4357/aaf424},
  eid           = {9},
  eprint        = {1811.10440},
}

@Article{Dullo2014,
  author        = {{Dullo}, B.~T. and {Graham}, A.~W.},
  title         = {{Depleted cores, multicomponent fits, and structural parameter relations for luminous early-type galaxies}},
  journal       = {\mnras},
  year          = {2014},
  volume        = {444},
  pages         = {2700-2722},
  month         = nov,
  adsurl        = {http://ads.nao.ac.jp/abs/2014MNRAS.444.2700D},
  archiveprefix = {arXiv},
  doi           = {10.1093/mnras/stu1590},
  eprint        = {1310.5867},
}

@Article{Dullo2013,
  Title                    = {{Central Stellar Mass Deficits in the Bulges of Local Lenticular Galaxies, and the Connection with Compact z \~{} 1.5 Galaxies}},
  Author                   = {{Dullo}, B.~T. and {Graham}, A.~W.},
  Journal                  = {\apj},
  Year                     = {2013},

  Month                    = may,
  Pages                    = {36},
  Volume                   = {768},
  Adsurl                   = {https://ui.adsabs.harvard.edu/abs/2013ApJ...768...36D},
  Archiveprefix            = {arXiv},
  Doi                      = {10.1088/0004-637X/768/1/36},
  Eid                      = {36},
  Eprint                   = {1303.1273}
}

@Article{Ebrova2017,
  Title                    = {{Galaxies with Prolate Rotation in Illustris}},
  Author                   = {{Ebrov{\'a}}, I. and {{\L}okas}, E.~L.},
  Journal                  = {\apj},
  Year                     = {2017},

  Month                    = dec,
  Pages                    = {144},
  Volume                   = {850},
  Adsurl                   = {https://ui.adsabs.harvard.edu/abs/2017ApJ...850..144E},
  Archiveprefix            = {arXiv},
  Doi                      = {10.3847/1538-4357/aa96ff},
  Eid                      = {144},
  Eprint                   = {1708.03311}
}

@Article{Ebrova2015,
  Title                    = {{The Origin of Prolate Rotation in Dwarf Spheroidal Galaxies Formed by Mergers of Disky Dwarfs}},
  Author                   = {{Ebrov{\'a}}, I. and {{\L}okas}, E.~L.},
  Journal                  = {\apj},
  Year                     = {2015},

  Month                    = nov,
  Pages                    = {10},
  Volume                   = {813},
  Adsurl                   = {https://ui.adsabs.harvard.edu/abs/2015ApJ...813...10E},
  Archiveprefix            = {arXiv},
  Doi                      = {10.1088/0004-637X/813/1/10},
  Eid                      = {10},
  Eprint                   = {1505.05412}
}

@InProceedings{Eisenhauer2003,
  author    = {{Eisenhauer}, F. and {Abuter}, R. and {Bickert}, K. and {Biancat-Marchet}, F. and {Bonnet}, H. and {Brynnel}, J. and {Conzelmann}, R.~D. and {Delabre}, B. and {Donaldson}, R. and {Farinato}, J. and {Fedrigo}, E. and {Genzel}, R. and {Hubin}, N.~N. and {Iserlohe}, C. and {Kasper}, M.~E. and {Kissler-Patig}, M. and {Monnet}, G.~J. and {Roehrle}, C. and {Schreiber}, J. and {Stroebele}, S. and {Tecza}, M. and {Thatte}, N.~A. and {Weisz}, H.},
  title     = {{SINFONI - Integral field spectroscopy at 50 milli-arcsecond resolution with the ESO VLT}},
  booktitle = {Instrument Design and Performance for Optical/Infrared Ground-based Telescopes},
  year      = {2003},
  editor    = {{Iye}, M. and {Moorwood}, A.~F.~M.},
  volume    = {4841},
  series    = {\procspie},
  pages     = {1548-1561},
  month     = mar,
  adsurl    = {http://adsabs.harvard.edu/abs/2003SPIE.4841.1548E},
  doi       = {10.1117/12.459468},
  eprint    = {astro-ph/0306191},
}

@Article{Emsellem2011,
  Title                    = {{The ATLAS$^{3D}$ project - III. A census of the stellar angular momentum within the effective radius of early-type galaxies: unveiling the distribution of fast and slow rotators}},
  Author                   = {{Emsellem}, E. and {Cappellari}, M. and {Krajnovi{\'c}}, D. and {Alatalo}, K. and {Blitz}, L. and {Bois}, M. and {Bournaud}, F. and {Bureau}, M. and {Davies}, R.~L. and {Davis}, T.~A. and {de Zeeuw}, P.~T. and {Khochfar}, S. and {Kuntschner}, H. and {Lablanche}, P.-Y. and {McDermid}, R.~M. and {Morganti}, R. and {Naab}, T. and {Oosterloo}, T. and {Sarzi}, M. and {Scott}, N. and {Serra}, P. and {van de Ven}, G. and {Weijmans}, A.-M. and {Young}, L.~M.},
  Journal                  = {\mnras},
  Year                     = {2011},

  Month                    = jun,
  Pages                    = {888-912},
  Volume                   = {414},
  Adsurl                   = {https://ui.adsabs.harvard.edu/abs/2011MNRAS.414..888E},
  Archiveprefix            = {arXiv},
  Doi                      = {10.1111/j.1365-2966.2011.18496.x},
  Eprint                   = {1102.4444}
}

@Article{Emsellem1994,
  author    = {{Emsellem}, E. and {Monnet}, G. and {Bacon}, R.},
  title     = {{The multi-gaussian expansion method: a tool for building realistic photometric and kinematical models of stellar systems I. The formalism}},
  journal   = {\aap},
  year      = {1994},
  volume    = {285},
  month     = may,
  adsurl    = {http://adsabs.harvard.edu/abs/1994A%26A...285..723E},
}

@Article{Ene2019,
  author        = {{Ene}, Irina and {Ma}, Chung-Pei and {McConnell}, Nicholas J. and {Walsh}, Jonelle L. and {Kempski}, Philipp and {Greene}, Jenny E. and {Thomas}, Jens and {Blakeslee}, John P.},
  journal       = {\apj},
  title         = {{The MASSIVE Survey XIII. Spatially Resolved Stellar Kinematics in the Central 1 kpc of 20 Massive Elliptical Galaxies with the GMOS-North Integral Field Spectrograph}},
  year          = {2019},
  month         = jun,
  number        = {1},
  pages         = {57},
  volume        = {878},
  adsurl        = {https://ui.adsabs.harvard.edu/abs/2019ApJ...878...57E},
  archiveprefix = {arXiv},
  doi           = {10.3847/1538-4357/ab1f04},
  eid           = {57},
  eprint        = {1904.08929},
  primaryclass  = {astro-ph.GA},
}

@Article{Ene2018,
  author        = {{Ene}, Irina and {Ma}, Chung-Pei and {Veale}, Melanie and {Greene}, Jenny E. and {Thomas}, Jens and {Blakeslee}, John P. and {Foster}, Caroline and {Walsh}, Jonelle L. and {Ito}, Jennifer and {Goulding}, Andy D.},
  journal       = {\mnras},
  title         = {{The MASSIVE Survey - X. Misalignment between kinematic and photometric axes and intrinsic shapes of massive early-type galaxies}},
  year          = {2018},
  month         = sep,
  number        = {2},
  pages         = {2810-2826},
  volume        = {479},
  adsurl        = {https://ui.adsabs.harvard.edu/abs/2018MNRAS.479.2810E},
  archiveprefix = {arXiv},
  doi           = {10.1093/mnras/sty1649},
  eprint        = {1802.00014},
  primaryclass  = {astro-ph.GA},
}

@Article{Faber1997,
  Title                    = {{The Centers of Early-Type Galaxies with HST. IV. Central Parameter Relations.}},
  Author                   = {{Faber}, S.~M. and {Tremaine}, S. and {Ajhar}, E.~A. and {Byun}, Y.-I. and {Dressler}, A. and {Gebhardt}, K. and {Grillmair}, C. and {Kormendy}, J. and {Lauer}, T.~R. and {Richstone}, D.},
  Journal                  = {\aj},
  Year                     = {1997},

  Month                    = nov,
  Pages                    = {1771},
  Volume                   = {114},
  Adsurl                   = {https://ui.adsabs.harvard.edu/abs/1997AJ....114.1771F},
  Doi                      = {10.1086/118606},
  Eprint                   = {astro-ph/9610055}
}

@Article{Ferrarese1996,
  author    = {{Ferrarese}, L. and {Ford}, H.~C. and {Jaffe}, W.},
  journal   = {\apj},
  title     = {{Evidence for a Massive Black Hole in the Active Galaxy NGC 4261 from Hubble Space Telescope Images and Spectra}},
  year      = {1996},
  month     = oct,
  pages     = {444},
  volume    = {470},
  adsurl    = {http://adsabs.harvard.edu/abs/1996ApJ...470..444F},
  doi       = {10.1086/177876},
}

@Article{Ferrarese1994,
  Title                    = {{Hubble Space Telescope photometry of the central regions of Virgo cluster elliptical galaxies. 3: Brightness profiles}},
  Author                   = {{Ferrarese}, L. and {van den Bosch}, F.~C. and {Ford}, H.~C. and {Jaffe}, W. and {O'Connell}, R.~W.},
  Journal                  = {\aj},
  Year                     = {1994},

  Month                    = nov,
  Pages                    = {1598-1609},
  Volume                   = {108},
  Adsurl                   = {https://ui.adsabs.harvard.edu/abs/1994AJ....108.1598F},
  Doi                      = {10.1086/117180}
}

@InProceedings{Ford1998,
  Title                    = {{Advanced camera for the Hubble Space Telescope}},
  Author                   = {{Ford}, H.~C. and {Bartko}, F. and {Bely}, P.~Y. and {Broadhurst}, T. and {Burrows}, C.~J. and {Cheng}, E.~S. and {Clampin}, M. and {Crocker}, J.~H. and {Feldman}, P.~D. and {Golimowski}, D.~A. and {Hartig}, G.~F. and {Illingworth}, G. and {Kimble}, R.~A. and {Lesser}, M.~P. and {Miley}, G. and {Neff}, S.~G. and {Postman}, M. and {Sparks}, W.~B. and {Tsvetanov}, Z. and {White}, R.~L. and {Sullivan}, P. and {Krebs}, C.~A. and {Leviton}, D.~B. and {La Jeunesse}, T. and {Burmester}, W. and {Fike}, S. and {Johnson}, R. and {Slusher}, R.~B. and {Volmer}, P. and {Woodruff}, R.~A.},
  Booktitle                = {Space Telescopes and Instruments V},
  Year                     = {1998},
  Editor                   = {{Bely}, P.~Y. and {Breckinridge}, J.~B.},
  Month                    = aug,
  Pages                    = {234-248},
  Series                   = {\procspie},
  Volume                   = {3356},
  Adsurl                   = {http://adsabs.harvard.edu/abs/1998SPIE.3356..234F},
  Doi                      = {10.1117/12.324464}
}

@Article{Franx1989,
  Title                    = {{Major and minor axis kinematics of 22 ellipticals}},
  Author                   = {{Franx}, M. and {Illingworth}, G. and {Heckman}, T.},
  Journal                  = {\apj},
  Year                     = {1989},

  Month                    = sep,
  Pages                    = {613-636},
  Volume                   = {344},
  Adsurl                   = {https://ui.adsabs.harvard.edu/abs/1989ApJ...344..613F},
  Doi                      = {10.1086/167830}
}

@Article{Gueltekin2014,
  author        = {{G{\"u}ltekin}, K. and {Gebhardt}, K. and {Kormendy}, J. and {Lauer}, T.~R. and {Bender}, R. and {Tremaine}, S. and {Richstone}, D.~O.},
  title         = {{The Black Hole Mass and the Stellar Ring in NGC 3706}},
  journal       = {\apj},
  year          = {2014},
  volume        = {781},
  pages         = {112},
  month         = feb,
  adsurl        = {http://adsabs.harvard.edu/abs/2014ApJ...781..112G},
  archiveprefix = {arXiv},
  doi           = {10.1088/0004-637X/781/2/112},
  eid           = {112},
  eprint        = {1312.4799},
}

@Article{Gerhard1993,
  author    = {{Gerhard}, O.~E.},
  title     = {{Line-of-sight velocity profiles in spherical galaxies: breaking the degeneracy between anisotropy and mass.}},
  journal   = {\mnras},
  year      = {1993},
  volume    = {265},
  pages     = {213},
  month     = nov,
  adsurl    = {http://adsabs.harvard.edu/abs/1993MNRAS.265..213G},
  comment   = {Hermite},
  doi       = {10.1093/mnras/265.1.213},
}

@Article{Graham2015,
  Title                    = {{How well can we identify pseudobulges?}},
  Author                   = {{Graham}, A.},
  Journal                  = {Highlights of Astronomy},
  Year                     = {2015},

  Month                    = mar,
  Pages                    = {360-360},
  Volume                   = {16},
  Adsurl                   = {https://ui.adsabs.harvard.edu/abs/2015HiA....16..360G},
  Doi                      = {10.1017/S1743921314011326}
}

@Article{Graham2016,
  author        = {{Graham}, A.~W.},
  title         = {{Galaxy Bulges and Their Massive Black Holes: A Review}},
  journal       = {Galactic Bulges},
  year          = {2016},
  volume        = {418},
  pages         = {263},
  adsurl        = {http://adsabs.harvard.edu/abs/2016ASSL..418..263G},
  archiveprefix = {arXiv},
  doi           = {10.1007/978-3-319-19378-6_11},
  eprint        = {1501.02937},
}

@Article{Graham2003,
  author    = {{Graham}, A.~W. and {Erwin}, P. and {Trujillo}, I. and {Asensio Ramos}, A.},
  journal   = {\aj},
  title     = {{A New Empirical Model for the Structural Analysis of Early-Type Galaxies, and A Critical Review of the Nuker Model}},
  year      = {2003},
  month     = jun,
  pages     = {2951-2963},
  volume    = {125},
  adsurl    = {https://ui.adsabs.harvard.edu/abs/2003AJ....125.2951G},
  doi       = {10.1086/375320},
  eprint    = {astro-ph/0306023},
}

@Article{Ho1997,
  Title                    = {{A Search for ``Dwarf'' Seyfert Nuclei. III. Spectroscopic Parameters and Properties of the Host Galaxies}},
  Author                   = {{Ho}, L.~C. and {Filippenko}, A.~V. and {Sargent}, W.~L.~W.},
  Journal                  = {\apjs},
  Year                     = {1997},

  Month                    = oct,
  Pages                    = {315-390},
  Volume                   = {112},
  Adsurl                   = {http://adsabs.harvard.edu/abs/1997ApJS..112..315H},
  Doi                      = {10.1086/313041},
  Eprint                   = {astro-ph/9704107}
}

@Article{Ho2011,
  author    = {Ho, Luis C. and Li, Zhao-Yu and Barth, Aaron J. and Seigar, Marc S. and Peng, Chien Y.scale},
  title     = {THE CARNEGIE-{IR}VINE GALAXY SURVEY. I. OVERVIEW {AND} ATLAS OF OPTICAL IMAGES},
  journal   = {The Astrophysical Journal Supplement Series},
  year      = {2011},
  volume    = {197},
  number    = {2},
  pages     = {21},
  month     = {Nov},
  issn      = {1538-4365},
  doi       = {10.1088/0067-0049/197/2/21},
  publisher = {IOP Publishing},
  url       = {http://dx.doi.org/10.1088/0067-0049/197/2/21},
}

@Article{Holtzman1995,
  author   = {{Holtzman}, J.~A. and {Burrows}, C.~J. and {Casertano}, S. and {Hester}, J.~J. and {Trauger}, J.~T. and {Watson}, A.~M. and {Worthey}, G.},
  journal  = {\pasp},
  title    = {{The Photometric Performance and Calibration of WFPC2}},
  year     = {1995},
  month    = nov,
  pages    = {1065},
  volume   = {107},
  adsurl   = {http://adsabs.harvard.edu/abs/1995PASP..107.1065H},
  doi      = {10.1086/133664},
}

@Article{Huang2013,
  Title                    = {THE CARNEGIE-{IR}VINE GALAXY SURVEY. III. THE THREE-COMPONENT STRUCTURE OF NEARBY ELLIPTICAL GALAXIES},
  Author                   = {Huang, Song and Ho, Luis C. and Peng, Chien Y. and Li, Zhao-Yu and Barth, Aaron J.},
  Journal                  = {The Astrophysical Journal},
  Year                     = {2013},

  Month                    = {Mar},
  Number                   = {1},
  Pages                    = {47},
  Volume                   = {766},

  Doi                      = {10.1088/0004-637x/766/1/47},
  ISSN                     = {1538-4357},
  Publisher                = {IOP Publishing},
  Url                      = {http://dx.doi.org/10.1088/0004-637X/766/1/47}
}

@Article{Jaffe1996,
  author    = {{Jaffe}, W. and {Ford}, H. and {Ferrarese}, L. and {van den Bosch}, F. and {O'Connell}, R.~W.},
  journal   = {\apj},
  title     = {{The Nuclear Disk of NGC 4261: Hubble Space Telescope Images and Ground-based Spectra}},
  year      = {1996},
  month     = mar,
  pages     = {214},
  volume    = {460},
  adsurl    = {https://ui.adsabs.harvard.edu/abs/1996ApJ...460..214J},
  doi       = {10.1086/176963},
}

@Article{Jarrett2000,
  Title                    = {{2MASS Extended Source Catalog: Overview and Algorithms}},
  Author                   = {{Jarrett}, T.~H. and {Chester}, T. and {Cutri}, R. and {Schneider}, S. and {Skrutskie}, M. and {Huchra}, J.~P.},
  Journal                  = {\aj},
  Year                     = {2000},

  Month                    = may,
  Pages                    = {2498-2531},
  Volume                   = {119},
  Adsurl                   = {http://adsabs.harvard.edu/abs/2000AJ....119.2498J},
  Doi                      = {10.1086/301330},
  Eprint                   = {astro-ph/0004318}
}

@Article{Kormendy2013,
  author    = {Kormendy, John and Ho, Luis C.},
  title     = {Coevolution (Or Not) of Supermassive Black Holes and Host Galaxies},
  journal   = {Annu. Rev. Astro. Astrophys.},
  year      = {2013},
  volume    = {51},
  number    = {1},
  pages     = {511–653},
  month     = {Aug},
  issn      = {1545-4282},
  comment   = {Review},
  doi       = {10.1146/annurev-astro-082708-101811},
  publisher = {Annual Reviews},
  url       = {http://dx.doi.org/10.1146/annurev-astro-082708-101811},
}

@Article{Krajnovic2005,
  author    = {{Krajnovi{\'c}}, D. and {Cappellari}, M. and {Emsellem}, E. and {McDermid}, R.~M. and {de Zeeuw}, P.~T.},
  title     = {{Dynamical modelling of stars and gas in NGC 2974: determination of mass-to-light ratio, inclination and orbital structure using the Schwarzschild method}},
  journal   = {\mnras},
  year      = {2005},
  volume    = {357},
  pages     = {1113-1133},
  month     = mar,
  adsurl    = {http://adsabs.harvard.edu/abs/2005MNRAS.357.1113K},
  comment   = {2d velocity, deprojection},
  doi       = {10.1111/j.1365-2966.2005.08715.x},
  eprint    = {astro-ph/0412186},
}

@Article{Krajnovic2018b,
  author        = {{Krajnovi{\'c}}, D. and {Cappellari}, M. and {McDermid}, R.~M.},
  title         = {{Two channels of supermassive black hole growth as seen on the galaxies mass-size plane}},
  journal       = {\mnras},
  year          = {2018},
  volume        = {473},
  pages         = {5237-5247},
  month         = feb,
  adsurl        = {http://adsabs.harvard.edu/abs/2018MNRAS.473.5237K},
  archiveprefix = {arXiv},
  doi           = {10.1093/mnras/stx2704},
  eprint        = {1707.04274},
}

@Article{Krajnovic2018,
  author        = {{Krajnovi{\'c}}, D. and {Cappellari}, M. and {McDermid}, R.~M. and {Thater}, S. and {Nyland}, K. and {de Zeeuw}, P.~T. and {Falc{\'o}n-Barroso}, J. and {Khochfar}, S. and {Kuntschner}, H. and {Sarzi}, M. and {Young}, L.~M.},
  title         = {{A quartet of black holes and a missing duo: probing the low end of the M$_{BH}$-{$\sigma$} relation with the adaptive optics assisted integral-field spectroscopy}},
  journal       = {\mnras},
  year          = {2018},
  volume        = {477},
  pages         = {3030-3064},
  month         = jul,
  adsurl        = {http://adsabs.harvard.edu/abs/2018MNRAS.477.3030K},
  archiveprefix = {arXiv},
  doi           = {10.1093/mnras/sty778},
  eprint        = {1803.08055},
}

@Article{Krajnovic2011,
  author        = {{Krajnovi{\'c}}, D. and {Emsellem}, E. and {Cappellari}, M. and {Alatalo}, K. and {Blitz}, L. and {Bois}, M. and {Bournaud}, F. and {Bureau}, M. and {Davies}, R.~L. and {Davis}, T.~A. and {de Zeeuw}, P.~T. and {Khochfar}, S. and {Kuntschner}, H. and {Lablanche}, P.-Y. and {McDermid}, R.~M. and {Morganti}, R. and {Naab}, T. and {Oosterloo}, T. and {Sarzi}, M. and {Scott}, N. and {Serra}, P. and {Weijmans}, A.-M. and {Young}, L.~M.},
  title         = {{The ATLAS$^{3D}$ project - II. Morphologies, kinemetric features and alignment between photometric and kinematic axes of early-type galaxies}},
  journal       = {\mnras},
  year          = {2011},
  volume        = {414},
  pages         = {2923-2949},
  month         = jul,
  adsurl        = {http://adsabs.harvard.edu/abs/2011MNRAS.414.2923K},
  archiveprefix = {arXiv},
  doi           = {10.1111/j.1365-2966.2011.18560.x},
  eprint        = {1102.3801},
}

@Article{Krajnovic2018a,
  author        = {{Krajnovi{\'c}}, D. and {Emsellem}, E. and {den Brok}, M. and {Marino}, R.~A. and {Schmidt}, K.~B. and {Steinmetz}, M. and {Weilbacher}, P.~M.},
  title         = {{Climbing to the top of the galactic mass ladder: evidence for frequent prolate-like rotation among the most massive galaxies}},
  journal       = {\mnras},
  year          = {2018},
  volume        = {477},
  pages         = {5327-5337},
  month         = jul,
  adsurl        = {http://adsabs.harvard.edu/abs/2018MNRAS.477.5327K},
  archiveprefix = {arXiv},
  doi           = {10.1093/mnras/sty1031},
  eprint        = {1802.02591},
}

@Article{Krajnovic2009,
  author        = {{Krajnovi{\'c}}, D. and {McDermid}, R.~M. and {Cappellari}, M. and {Davies}, R.~L.},
  title         = {{Determination of masses of the central black holes in NGC 524 and 2549 using laser guide star adaptive optics}},
  journal       = {\mnras},
  year          = {2009},
  volume        = {399},
  pages         = {1839-1857},
  month         = nov,
  adsurl        = {http://adsabs.harvard.edu/abs/2009MNRAS.399.1839K},
  archiveprefix = {arXiv},
  doi           = {10.1111/j.1365-2966.2009.15415.x},
  eprint        = {0907.3748},
}

@Article{Krist2001,
  Title                    = {{The Tiny Tim User's Manual, version 6.3}},
  Author                   = {{Krist}, J. and {Hook}, R.},
  Journal                  = {{The Tiny Tim User's Manual, version 6.3}},
  Year                     = {2001}
}

@Article{Kulier2015,
  Title                    = {{Understanding Black Hole Mass Assembly via Accretion and Mergers at Late Times in Cosmological Simulations}},
  Author                   = {{Kulier}, A. and {Ostriker}, J.~P. and {Natarajan}, P. and {Lackner}, C.~N. and {Cen}, R.},
  Journal                  = {\apj},
  Year                     = {2015},

  Month                    = feb,
  Pages                    = {178},
  Volume                   = {799},
  Adsurl                   = {http://adsabs.harvard.edu/abs/2015ApJ...799..178K},
  Archiveprefix            = {arXiv},
  Doi                      = {10.1088/0004-637X/799/2/178},
  Eid                      = {178},
  Eprint                   = {1307.3684}
}

@Article{Laine2003,
  Title                    = {{Hubble Space Telescope Imaging of Brightest Cluster Galaxies}},
  Author                   = {{Laine}, S. and {van der Marel}, R.~P. and {Lauer}, T.~R. and {Postman}, M. and {O'Dea}, C.~P. and {Owen}, F.~N.},
  Journal                  = {\aj},
  Year                     = {2003},

  Month                    = feb,
  Pages                    = {478-505},
  Volume                   = {125},
  Adsurl                   = {https://ui.adsabs.harvard.edu/abs/2003AJ....125..478L},
  Doi                      = {10.1086/345823},
  Eprint                   = {astro-ph/0211074}
}

@Article{Lauer1985,
  Title                    = {{The cores of elliptical galaxies}},
  Author                   = {{Lauer}, T.~R.},
  Journal                  = {\apj},
  Year                     = {1985},

  Month                    = may,
  Pages                    = {104-121},
  Volume                   = {292},
  Adsurl                   = {https://ui.adsabs.harvard.edu/abs/1985ApJ...292..104L},
  Doi                      = {10.1086/163136}
}

@Article{Lauer1995,
  Title                    = {{The Centers of Early-Type Galaxies with HST.I.An Observational Survey}},
  Author                   = {{Lauer}, T.~R. and {Ajhar}, E.~A. and {Byun}, Y.-I. and {Dressler}, A. and {Faber}, S.~M. and {Grillmair}, C. and {Kormendy}, J. and {Richstone}, D. and {Tremaine}, S.},
  Journal                  = {\aj},
  Year                     = {1995},

  Month                    = dec,
  Pages                    = {2622},
  Volume                   = {110},
  Adsurl                   = {https://ui.adsabs.harvard.edu/abs/1995AJ....110.2622L},
  Doi                      = {10.1086/117719}
}

@Article{Lauer2005,
  Title                    = {{The Centers of Early-Type Galaxies with Hubble Space Telescope. V. New WFPC2 Photometry}},
  Author                   = {{Lauer}, T.~R. and {Faber}, S.~M. and {Gebhardt}, K. and {Richstone}, D. and {Tremaine}, S. and {Ajhar}, E.~A. and {Aller}, M.~C. and {Bender}, R. and {Dressler}, A. and {Filippenko}, A.~V. and {Green}, R. and {Grillmair}, C.~J. and {Ho}, L.~C. and {Kormendy}, J. and {Magorrian}, J. and {Pinkney}, J. and {Siopis}, C.},
  Journal                  = {\aj},
  Year                     = {2005},

  Month                    = may,
  Pages                    = {2138-2185},
  Volume                   = {129},
  Adsurl                   = {https://ui.adsabs.harvard.edu/abs/2005AJ....129.2138L},
  Doi                      = {10.1086/429565},
  Eprint                   = {astro-ph/0412040}
}

@Book{Lawson1974,
  Title                    = {{Solving least squares problems}},
  Author                   = {{Lawson}, C.~L. and {Hanson}, R.~J.},
  Year                     = {1974},
  Adsurl                   = {http://adsabs.harvard.edu/abs/1974slsp.book.....L},
  Booktitle                = {Prentice-Hall Series in Automatic Computation, Englewood Cliffs: Prentice-Hall, 1974},
  Comment                  = {Schwarzschild}
}

@Article{Lee2010,
  Title                    = {{The Globular Cluster System of the Virgo Giant Elliptical Galaxy NGC 4636. II. Kinematics of the Globular Cluster System}},
  Author                   = {{Lee}, M.~G. and {Park}, H.~S. and {Hwang}, H.~S. and {Arimoto}, N. and {Tamura}, N. and {Onodera}, M.},
  Journal                  = {\apj},
  Year                     = {2010},

  Month                    = feb,
  Pages                    = {1083-1099},
  Volume                   = {709},
  Adsurl                   = {https://ui.adsabs.harvard.edu/abs/2010ApJ...709.1083L},
  Archiveprefix            = {arXiv},
  Doi                      = {10.1088/0004-637X/709/2/1083},
  Eprint                   = {0912.1728}
}

@Article{Li2016,
  author        = {{Li}, H. and {Li}, R. and {Mao}, S. and {Xu}, D. and {Long}, R.~J. and {Emsellem}, E.},
  title         = {{Assessing the Jeans Anisotropic Multi-Gaussian Expansion method with the Illustris simulation}},
  journal       = {\mnras},
  year          = {2016},
  volume        = {455},
  pages         = {3680-3692},
  month         = feb,
  adsurl        = {http://adsabs.harvard.edu/abs/2016MNRAS.455.3680L},
  archiveprefix = {arXiv},
  doi           = {10.1093/mnras/stv2565},
  eprint        = {1511.00789},
}

@Article{Li2011,
  author    = {Li, Zhao-Yu and Ho, Luis C. and Barth, Aaron J. and Peng, Chien Y.},
  title     = {THE CARNEGIE-{IR}VINE GALAXY SURVEY. II. {ISO}PHOTAL ANALYSIS},
  journal   = {The Astrophysical Journal Supplement Series},
  year      = {2011},
  volume    = {197},
  number    = {2},
  pages     = {22},
  month     = {Nov},
  issn      = {1538-4365},
  doi       = {10.1088/0067-0049/197/2/22},
  publisher = {IOP Publishing},
  url       = {http://dx.doi.org/10.1088/0067-0049/197/2/22},
}

@Article{Loewenstein2003,
  Title                    = {{The nature of dark matter in elliptical galaxies: Chandra observations of NGC 4636}},
  Author                   = {{Loewenstein}, M. and {Mushotzky}, R.},
  Journal                  = {Nuclear Physics B Proceedings Supplements},
  Year                     = {2003},

  Month                    = jul,
  Pages                    = {91-94},
  Volume                   = {124},
  Adsurl                   = {https://ui.adsabs.harvard.edu/abs/2003NuPhS.124...91L},
  Doi                      = {10.1016/S0920-5632(03)02085-1},
  Eprint                   = {astro-ph/0205359}
}

@Article{Malin1983,
  Title                    = {{A catalog of elliptical galaxies with shells}},
  Author                   = {{Malin}, D.~F. and {Carter}, D.},
  Journal                  = {\apj},
  Year                     = {1983},

  Month                    = nov,
  Pages                    = {534-540},
  Volume                   = {274},
  Adsurl                   = {https://ui.adsabs.harvard.edu/abs/1983ApJ...274..534M},
  Doi                      = {10.1086/161467}
}

@Article{Mazzalay2016,
  author        = {{Mazzalay}, X. and {Thomas}, J. and {Saglia}, R.~P. and {Wegner}, G.~A. and {Bender}, R. and {Erwin}, P. and {Fabricius}, M.~H. and {Rusli}, S.~P.},
  journal       = {\mnras},
  title         = {{The supermassive black hole and double nucleus of the core elliptical NGC 5419}},
  year          = {2016},
  month         = nov,
  pages         = {2847-2860},
  volume        = {462},
  adsurl        = {https://ui.adsabs.harvard.edu/abs/2016MNRAS.462.2847M},
  archiveprefix = {arXiv},
  doi           = {10.1093/mnras/stw1802},
  eprint        = {1607.06466},
}

@Article{Mezcua2017,
  Title                    = {{Observational evidence for intermediate-mass black holes}},
  Author                   = {{Mezcua}, M.},
  Journal                  = {International Journal of Modern Physics D},
  Year                     = {2017},
  Pages                    = {1730021},
  Volume                   = {26},
  Adsurl                   = {http://adsabs.harvard.edu/abs/2017IJMPD..2630021M},
  Archiveprefix            = {arXiv},
  Doi                      = {10.1142/S021827181730021X},
  Eid                      = {1730021},
  Eprint                   = {1705.09667}
}

@Article{Milosavljevic2002,
  Title                    = {{Galaxy cores as relics of black hole mergers}},
  Author                   = {{Milosavljevi{\'c}}, M. and {Merritt}, D. and {Rest}, A. and {van den Bosch}, F.~C.},
  Journal                  = {\mnras},
  Year                     = {2002},

  Month                    = apr,
  Pages                    = {L51-L55},
  Volume                   = {331},
  Adsurl                   = {https://ui.adsabs.harvard.edu/abs/2002MNRAS.331L..51M},
  Doi                      = {10.1046/j.1365-8711.2002.05436.x},
  Eprint                   = {astro-ph/0110185}
}

@Article{Modigliani2007,
  author        = {{Modigliani}, Andrea and {Hummel}, Wolfgang and {Abuter}, Roberto and {Amico}, Paola and {Ballester}, Pascal and {Davies}, Richard and {Dumas}, Christophe and {Horrobin}, Mattew and {Neeser}, Mark and {Kissler-Patig}, Markus and {Peron}, Michele and {Rehunanen}, Juha and {Schreiber}, Juergen and {Szeifert}, Thomas},
  title         = {{The SINFONI pipeline}},
  journal       = {arXiv e-prints},
  year          = {2007},
  pages         = {astro-ph/0701297},
  month         = {Jan},
  adsurl        = {https://ui.adsabs.harvard.edu/\#abs/2007astro.ph..1297M},
  archiveprefix = {arXiv},
  eid           = {astro-ph/0701297},
  eprint        = {astro-ph/0701297},
  primaryclass  = {astro-ph},
}

@Article{Navarro1996,
  Title                    = {{The Structure of Cold Dark Matter Halos}},
  Author                   = {{Navarro}, J.~F. and {Frenk}, C.~S. and {White}, S.~D.~M.},
  Journal                  = {\apj},
  Year                     = {1996},

  Month                    = may,
  Pages                    = {563},
  Volume                   = {462},
  Adsurl                   = {http://adsabs.harvard.edu/abs/1996ApJ...462..563N},
  Doi                      = {10.1086/177173},
  Eprint                   = {astro-ph/9508025}
}

@Article{Nguyen2017,
  author        = {{Nguyen}, D.~D. and {Seth}, A.~C. and {den Brok}, M. and {Neumayer}, N. and {Cappellari}, M. and {Barth}, A.~J. and {Caldwell}, N. and {Williams}, B.~F. and {Binder}, B.},
  journal       = {\apj},
  title         = {{Improved Dynamical Constraints on the Mass of the Central Black Hole in NGC 404}},
  year          = {2017},
  month         = feb,
  pages         = {237},
  volume        = {836},
  adsurl        = {http://adsabs.harvard.edu/abs/2017ApJ...836..237N},
  archiveprefix = {arXiv},
  doi           = {10.3847/1538-4357/aa5cb4},
  eid           = {237},
  eprint        = {1610.09385},
}

@Article{Nguyen2018,
  author        = {{Nguyen}, D.~D. and {Seth}, A.~C. and {Neumayer}, N. and {Kamann}, S. and {Voggel}, K.~T. and {Cappellari}, M. and {Picotti}, A. and {Nguyen}, P.~M. and {B{\"o}ker}, T. and {Debattista}, V. and {Caldwell}, N. and {McDermid}, R. and {Bastian}, N. and {Ahn}, C.~C. and {Pechetti}, R.},
  journal       = {\apj},
  title         = {{Nearby Early-type Galactic Nuclei at High Resolution: Dynamical Black Hole and Nuclear Star Cluster Mass Measurements}},
  year          = {2018},
  month         = may,
  pages         = {118},
  volume        = {858},
  adsurl        = {http://adsabs.harvard.edu/abs/2018ApJ...858..118N},
  archiveprefix = {arXiv},
  doi           = {10.3847/1538-4357/aabe28},
  eid           = {118},
  eprint        = {1711.04314},
}

@Article{Poci2016,
  author         = {A. Poci and M. Cappellari and R. M. McDermid},
  title          = {Systematic trends in total-mass profiles from dynamical models of early-type galaxies},
  journal        = {MNRAS},
  year           = {2016},
  volume         = {467:},
  pages          = {1397,2017},
  month          = dec,
  comments       = {17 pages, 11 figures, 3 tables. Published in MNRAS},
  eprint         = {1612.05805},
  oai2identifier = {1612.05805},
}

@Article{Prieur1988,
  author    = {{Prieur}, J.-L.},
  title     = {{The shell system around NGC 3923 and its implications for the potential of the galaxy}},
  journal   = {\apj},
  year      = {1988},
  volume    = {326},
  pages     = {596-615},
  month     = mar,
  adsurl    = {http://adsabs.harvard.edu/abs/1988ApJ...326..596P},
  doi       = {10.1086/166120},
}

@Article{Rantala2019,
  author        = {{Rantala}, A. and {Johansson}, P.~H. and {Naab}, T. and {Thomas}, J. and {Frigo}, M.},
  journal       = {\apjl},
  title         = {{The Simultaneous Formation of Cored, Tangentially Biased, and Kinematically Decoupled Centers in Massive Early-type Galaxies}},
  year          = {2019},
  month         = feb,
  pages         = {L17},
  volume        = {872},
  adsurl        = {http://adsabs.harvard.edu/abs/2019ApJ...872L..17R},
  archiveprefix = {arXiv},
  doi           = {10.3847/2041-8213/ab04b1},
  eid           = {L17},
  eprint        = {1812.02732},
}

@Article{Rantala2018,
  author        = {{Rantala}, A. and {Johansson}, P.~H. and {Naab}, T. and {Thomas}, J. and {Frigo}, M.},
  journal       = {\apj},
  title         = {{The Formation of Extremely Diffuse Galaxy Cores by Merging Supermassive Black Holes}},
  year          = {2018},
  month         = sep,
  pages         = {113},
  volume        = {864},
  adsurl        = {http://adsabs.harvard.edu/abs/2018ApJ...864..113R},
  archiveprefix = {arXiv},
  doi           = {10.3847/1538-4357/aada47},
  eid           = {113},
  eprint        = {1805.10295},
}

@Article{Rest2001,
  author    = {{Rest}, A. and {van den Bosch}, F.~C. and {Jaffe}, W. and {Tran}, H. and {Tsvetanov}, Z. and {Ford}, H.~C. and {Davies}, J. and {Schafer}, J.},
  journal   = {\aj},
  title     = {{WFPC2 Images of the Central Regions of Early-Type Galaxies. I. The Data}},
  year      = {2001},
  month     = may,
  pages     = {2431-2482},
  volume    = {121},
  adsurl    = {https://ui.adsabs.harvard.edu/abs/2001AJ....121.2431R},
  doi       = {10.1086/320370},
  eprint    = {astro-ph/0102286},
}

@Article{Rusli2013a,
  Title                    = {{Depleted Galaxy Cores and Dynamical Black Hole Masses}},
  Author                   = {{Rusli}, S.~P. and {Erwin}, P. and {Saglia}, R.~P. and {Thomas}, J. and {Fabricius}, M. and {Bender}, R. and {Nowak}, N.},
  Journal                  = {\aj},
  Year                     = {2013},

  Month                    = dec,
  Pages                    = {160},
  Volume                   = {146},
  Adsurl                   = {https://ui.adsabs.harvard.edu/abs/2013AJ....146..160R},
  Archiveprefix            = {arXiv},
  Doi                      = {10.1088/0004-6256/146/6/160},
  Eid                      = {160},
  Eprint                   = {1310.5310}
}

@Article{Rusli2013,
  author        = {{Rusli}, S.~P. and {Thomas}, J. and {Saglia}, R.~P. and {Fabricius}, M. and {Erwin}, P. and {Bender}, R. and {Nowak}, N. and {Lee}, C.~H. and {Riffeser}, A. and {Sharp}, R.},
  title         = {{The Influence of Dark Matter Halos on Dynamical Estimates of Black Hole Mass: 10 New Measurements for High-{$\sigma$} Early-type Galaxies}},
  journal       = {\aj},
  year          = {2013},
  volume        = {146},
  pages         = {45},
  month         = sep,
  adsurl        = {http://adsabs.harvard.edu/abs/2013AJ....146...45R},
  archiveprefix = {arXiv},
  comment       = {BH talk 2014},
  doi           = {10.1088/0004-6256/146/3/45},
  eid           = {45},
  eprint        = {1306.1124},
}

@Article{Sanchez-Blazquez2006,
  author    = {{S{\'a}nchez-Bl{\'a}zquez}, P. and {Peletier}, R.~F. and {Jim{\'e}nez-Vicente}, J. and {Cardiel}, N. and {Cenarro}, A.~J. and {Falc{\'o}n-Barroso}, J. and {Gorgas}, J. and {Selam}, S. and {Vazdekis}, A.},
  title     = {{Medium-resolution Isaac Newton Telescope library of empirical spectra}},
  journal   = {\mnras},
  year      = {2006},
  volume    = {371},
  pages     = {703-718},
  month     = sep,
  adsurl    = {http://adsabs.harvard.edu/abs/2006MNRAS.371..703S},
  comment   = {MILES},
  doi       = {10.1111/j.1365-2966.2006.10699.x},
  eprint    = {astro-ph/0607009},
}

@Article{Saglia1993,
  Title                    = {{Stellar dynamical evidence for dark halos in elliptical galaxies - The case of NGC 4472, IC 4296, and NGC 7144}},
  Author                   = {{Saglia}, R.~P. and {Bertin}, G. and {Bertola}, F. and {Danziger}, J. and {Dejonghe}, H. and {Sadler}, E.~M. and {Stiavelli}, M. and {de Zeeuw}, P.~T. and {Zeilinger}, W.~W.},
  Journal                  = {\apj},
  Year                     = {1993},

  Month                    = feb,
  Pages                    = {567-572},
  Volume                   = {403},
  Adsurl                   = {https://ui.adsabs.harvard.edu/abs/1993ApJ...403..567S},
  Doi                      = {10.1086/172226}
}

@Article{Saglia2016,
  author        = {{Saglia}, R.~P. and {Opitsch}, M. and {Erwin}, P. and {Thomas}, J. and {Beifiori}, A. and {Fabricius}, M. and {Mazzalay}, X. and {Nowak}, N. and {Rusli}, S.~P. and {Bender}, R.},
  title         = {{The SINFONI Black Hole Survey: The Black Hole Fundamental Plane Revisited and the Paths of (Co)evolution of Supermassive Black Holes and Bulges}},
  journal       = {\apj},
  year          = {2016},
  volume        = {818},
  pages         = {47},
  month         = feb,
  adsurl        = {http://adsabs.harvard.edu/abs/2016ApJ...818...47S},
  archiveprefix = {arXiv},
  doi           = {10.3847/0004-637X/818/1/47},
  eid           = {47},
  eprint        = {1601.00974},
}

@Article{Sahu2019,
  author        = {{Sahu}, Nandini and {Graham}, Alister W. and {Davis}, Benjamin L.},
  journal       = {\apj},
  title         = {{Black Hole Mass Scaling Relations for Early-type Galaxies. I. M $_{BH}$-M $_{*,}$ $_{sph}$ and M $_{BH}$-M $_{*,gal}$}},
  year          = {2019},
  month         = may,
  number        = {2},
  pages         = {155},
  volume        = {876},
  adsurl        = {https://ui.adsabs.harvard.edu/abs/2019ApJ...876..155S},
  archiveprefix = {arXiv},
  doi           = {10.3847/1538-4357/ab0f32},
  eid           = {155},
  eprint        = {1903.04738},
  primaryclass  = {astro-ph.GA},
}

@Article{Schuberth2006,
  Title                    = {{Dynamics of the NGC 4636 globular cluster system. An extremely dark matter dominated galaxy?}},
  Author                   = {{Schuberth}, Y. and {Richtler}, T. and {Dirsch}, B. and {Hilker}, M. and {Larsen}, S.~S. and {Kissler-Patig}, M. and {Mebold}, U.},
  Journal                  = {\aap},
  Year                     = {2006},

  Month                    = nov,
  Pages                    = {391-406},
  Volume                   = {459},
  Adsurl                   = {https://ui.adsabs.harvard.edu/abs/2006A%26A...459..391S},
  Doi                      = {10.1051/0004-6361:20053134},
  Eprint                   = {astro-ph/0604309}
}

@Article{Schwarzschild1979,
  author    = {{Schwarzschild}, M.},
  title     = {{A numerical model for a triaxial stellar system in dynamical equilibrium}},
  journal   = {\apj},
  year      = {1979},
  volume    = {232},
  pages     = {236-247},
  month     = aug,
  adsurl    = {http://adsabs.harvard.edu/abs/1979ApJ...232..236S},
  doi       = {10.1086/157282},
}

@Article{Shen2010,
  Title                    = {{The Supermassive Black Hole and Dark Matter Halo of NGC 4649 (M60)}},
  Author                   = {{Shen}, J. and {Gebhardt}, K.},
  Journal                  = {\apj},
  Year                     = {2010},

  Month                    = mar,
  Pages                    = {484-494},
  Volume                   = {711},
  Adsurl                   = {https://ui.adsabs.harvard.edu/abs/2010ApJ...711..484S},
  Archiveprefix            = {arXiv},
  Doi                      = {10.1088/0004-637X/711/1/484},
  Eprint                   = {0910.4168}
}

@Article{Temi2003,
  author    = {{Temi}, P. and {Mathews}, W.~G. and {Brighenti}, F. and {Bregman}, J.~D.},
  journal   = {\apjl},
  title     = {{Dust in Hot Gas: Far-Infrared Emission from Three Local Elliptical Galaxies}},
  year      = {2003},
  month     = mar,
  pages     = {L121-L124},
  volume    = {585},
  adsurl    = {https://ui.adsabs.harvard.edu/abs/2003ApJ...585L.121T},
  doi       = {10.1086/374326},
  eprint    = {astro-ph/0301452},
}

@Article{Thater2019,
  author        = {{Thater}, Sabine and {Krajnovi{\'c}}, Davor and {Cappellari}, Michele and {Davis}, Timothy A. and {de Zeeuw}, P. Tim and {McDermid}, Richard M. and {Sarzi}, Marc},
  journal       = {\aap},
  title         = {{Six new supermassive black hole mass determinations from adaptive-optics assisted SINFONI observations}},
  year          = {2019},
  month         = may,
  pages         = {A62},
  volume        = {625},
  adsurl        = {https://ui.adsabs.harvard.edu/abs/2019A&A...625A..62T},
  archiveprefix = {arXiv},
  doi           = {10.1051/0004-6361/201834808},
  eid           = {A62},
  eprint        = {1902.10175},
  primaryclass  = {astro-ph.GA},
}

@Article{Thomas2016,
  Title                    = {{A 17-billion-solar-mass black hole in a group galaxy with a diffuse core}},
  Author                   = {{Thomas}, J. and {Ma}, C.-P. and {McConnell}, N.~J. and {Greene}, J.~E. and {Blakeslee}, J.~P. and {Janish}, R.},
  Journal                  = {\nat},
  Year                     = {2016},

  Month                    = apr,
  Pages                    = {340-342},
  Volume                   = {532},
  Adsurl                   = {http://adsabs.harvard.edu/abs/2016Natur.532..340T},
  Archiveprefix            = {arXiv},
  Doi                      = {10.1038/nature17197},
  Eprint                   = {1604.01400}
}

@Article{Thomas2014,
  Title                    = {{The Dynamical Fingerprint of Core Scouring in Massive Elliptical Galaxies}},
  Author                   = {{Thomas}, J. and {Saglia}, R.~P. and {Bender}, R. and {Erwin}, P. and {Fabricius}, M.},
  Journal                  = {\apj},
  Year                     = {2014},

  Month                    = feb,
  Pages                    = {39},
  Volume                   = {782},
  Adsurl                   = {https://ui.adsabs.harvard.edu/abs/2014ApJ...782...39T},
  Archiveprefix            = {arXiv},
  Doi                      = {10.1088/0004-637X/782/1/39},
  Eid                      = {39},
  Eprint                   = {1311.3783}
}

@Article{Tremaine2002,
  author    = {{Tremaine}, S. and {Gebhardt}, K. and {Bender}, R. and {Bower}, G. and {Dressler}, A. and {Faber}, S.~M. and {Filippenko}, A.~V. and {Green}, R. and {Grillmair}, C. and {Ho}, L.~C. and {Kormendy}, J. and {Lauer}, T.~R. and {Magorrian}, J. and {Pinkney}, J. and {Richstone}, D.},
  journal   = {\apj},
  title     = {{The Slope of the Black Hole Mass versus Velocity Dispersion Correlation}},
  year      = {2002},
  month     = aug,
  pages     = {740-753},
  volume    = {574},
  adsurl    = {http://adsabs.harvard.edu/abs/2002ApJ...574..740T},
  doi       = {10.1086/341002},
  eprint    = {astro-ph/0203468},
}

@Article{Tsatsi2017,
  author        = {{Tsatsi}, A. and {Lyubenova}, M. and {van de Ven}, G. and {Chang}, J. and {Aguerri}, J.~A.~L. and {Falc{\'o}n-Barroso}, J. and {Macci{\`o}}, A.~V.},
  journal       = {\aap},
  title         = {{CALIFA reveals prolate rotation in massive early-type galaxies: A polar galaxy merger origin?}},
  year          = {2017},
  month         = oct,
  pages         = {A62},
  volume        = {606},
  adsurl        = {https://ui.adsabs.harvard.edu/abs/2017A%26A...606A..62T},
  archiveprefix = {arXiv},
  doi           = {10.1051/0004-6361/201630218},
  eid           = {A62},
  eprint        = {1707.05130},
}

@Article{vandenBosch2010,
  author        = {{van den Bosch}, R.~C.~E. and {de Zeeuw}, P.~T.},
  title         = {{Estimating black hole masses in triaxial galaxies}},
  journal       = {\mnras},
  year          = {2010},
  volume        = {401},
  pages         = {1770-1780},
  month         = jan,
  adsurl        = {http://adsabs.harvard.edu/abs/2010MNRAS.401.1770V},
  archiveprefix = {arXiv},
  doi           = {10.1111/j.1365-2966.2009.15832.x},
  eprint        = {0910.0844},
  primaryclass  = {astro-ph.CO},
}

@Article{vandenBosch2008,
  author        = {{van den Bosch}, R.~C.~E. and {van de Ven}, G. and {Verolme}, E.~K. and {Cappellari}, M. and {de Zeeuw}, P.~T.},
  journal       = {\mnras},
  title         = {{Triaxial orbit based galaxy models with an application to the (apparent) decoupled core galaxy NGC 4365}},
  year          = {2008},
  month         = apr,
  pages         = {647-666},
  volume        = {385},
  adsurl        = {http://adsabs.harvard.edu/abs/2008MNRAS.385..647V},
  archiveprefix = {arXiv},
  doi           = {10.1111/j.1365-2966.2008.12874.x},
  eprint        = {0712.0113},
}

@Article{vanderMarel1993,
  author    = {{van der Marel}, R.~P. and {Franx}, M.},
  title     = {{A new method for the identification of non-Gaussian line profiles in elliptical galaxies}},
  journal   = {\apj},
  year      = {1993},
  volume    = {407},
  pages     = {525-539},
  month     = apr,
  adsurl    = {http://adsabs.harvard.edu/abs/1993ApJ...407..525V},
  doi       = {10.1086/172534},
}

@Article{vanDokkum2001,
  Title                    = {{Cosmic-Ray Rejection by Laplacian Edge Detection}},
  Author                   = {{van Dokkum}, P.~G.},
  Journal                  = {\pasp},
  Year                     = {2001},

  Month                    = nov,
  Pages                    = {1420-1427},
  Volume                   = {113},
  Adsurl                   = {http://adsabs.harvard.edu/abs/2001PASP..113.1420V},
  Doi                      = {10.1086/323894},
  Eprint                   = {astro-ph/0108003}
}

@Article{Volonteri2010,
  Title                    = {Formation of supermassive black holes},
  Author                   = {Volonteri, Marta},
  Journal                  = {The Astronomy and Astrophysics Review},
  Year                     = {2010},

  Month                    = {Apr},
  Number                   = {3},
  Pages                    = {279–315},
  Volume                   = {18},

  Doi                      = {10.1007/s00159-010-0029-x},
  ISSN                     = {1432-0754},
  Publisher                = {Springer Nature},
  Url                      = {http://dx.doi.org/10.1007/s00159-010-0029-x}
}

@Article{Wagner1988,
  Title                    = {{Minor-axis rotation and triaxiality in elliptical galaxies}},
  Author                   = {{Wagner}, S.~J. and {Bender}, R. and {Moellenhoff}, C.},
  Journal                  = {\aap},
  Year                     = {1988},

  Month                    = apr,
  Pages                    = {L5-L8},
  Volume                   = {195},
  Adsurl                   = {https://ui.adsabs.harvard.edu/abs/1988A%26A...195L...5W}
}

@Article{Weilbacher2015,
  author         = {Peter M. Weilbacher and Ole Streicher and Tanya Urrutia and Arlette Pécontal-Rousset and Aurélien Jarno and Roland Bacon},
  title          = {The MUSE Data Reduction Pipeline: Status after Preliminary Acceptance Europe},
  journal        = {485, p. 451},
  year           = {2015},
  volume         = {2014},
  pages          = {ADASSXXIII.MansetandForshay(eds.),ASPconf.ser.,vol.},
  month          = jul,
  comments       = {Proceedings paper on the MUSE pipeline; the software has become public in the meantime, and available from ESO at http://www.eso.org/sci/software/pipelines/muse/muse-pipe-recipes.html},
  eprint         = {1507.00034},
  oai2identifier = {1507.00034},
}

@Article{Willmer2018,
  author        = {{Willmer}, C.~N.~A.},
  journal       = {\apjs},
  title         = {{c}},
  year          = {2018},
  month         = jun,
  pages         = {47},
  volume        = {236},
  adsurl        = {http://adsabs.harvard.edu/abs/2018ApJS..236...47W},
  archiveprefix = {arXiv},
  doi           = {10.3847/1538-4365/aabfdf},
  eid           = {47},
  eprint        = {1804.07788},
  primaryclass  = {astro-ph.SR},
}

@Article{Winge2009,
  author        = {{Winge}, C. and {Riffel}, R.~A. and {Storchi-Bergmann}, T.},
  title         = {{The Gemini Spectral Library of Near-IR Late-Type Stellar Templates and Its Application for Velocity Dispersion Measurements}},
  journal       = {\apjs},
  year          = {2009},
  volume        = {185},
  pages         = {186-197},
  month         = nov,
  adsurl        = {http://adsabs.harvard.edu/abs/2009ApJS..185..186W},
  archiveprefix = {arXiv},
  doi           = {10.1088/0067-0049/185/1/186},
  eprint        = {0910.2619},
}

@Article{Yoo2007,
  Title                    = {{The Most Massive Black Holes in the Universe: Effects of Mergers in Massive Galaxy Clusters}},
  Author                   = {{Yoo}, J. and {Miralda-Escud{\'e}}, J. and {Weinberg}, D.~H. and {Zheng}, Z. and {Morgan}, C.~W.},
  Journal                  = {\apj},
  Year                     = {2007},

  Month                    = oct,
  Pages                    = {813-825},
  Volume                   = {667},
  Adsurl                   = {http://adsabs.harvard.edu/abs/2007ApJ...667..813Y},
  Doi                      = {10.1086/521015},
  Eprint                   = {astro-ph/0702199}
}

@Article{Mehrgan2019,
  author        = {{Mehrgan}, Kianusch and {Thomas}, Jens and {Saglia}, Roberto and {Mazzalay}, Ximena and {Erwin}, Peter and {Bender}, Ralf and {Kluge}, Matthias and {Fabricius}, Maximilian},
  title         = {{A 40 Billion Solar-mass Black Hole in the Extreme Core of Holm 15A, the Central Galaxy of Abell 85}},
  journal       = {\apj},
  year          = {2019},
  volume        = {887},
  number        = {2},
  pages         = {195},
  month         = dec,
  adsurl        = {https://ui.adsabs.harvard.edu/abs/2019ApJ...887..195M},
  archiveprefix = {arXiv},
  doi           = {10.3847/1538-4357/ab5856},
  eid           = {195},
  eprint        = {1907.10608},
  primaryclass  = {astro-ph.GA},
}

@Article{vandenBosch2009,
  author        = {{van den Bosch}, Remco C.~E. and {van de Ven}, Glenn},
  title         = {{Recovering the intrinsic shape of early-type galaxies}},
  journal       = {\mnras},
  year          = {2009},
  volume        = {398},
  number        = {3},
  pages         = {1117-1128},
  month         = sep,
  adsurl        = {https://ui.adsabs.harvard.edu/abs/2009MNRAS.398.1117V},
  archiveprefix = {arXiv},
  doi           = {10.1111/j.1365-2966.2009.15177.x},
  eprint        = {0811.3474},
  primaryclass  = {astro-ph},
}

@Article{Boizelle2021,
  author        = {{Boizelle}, Benjamin D. and {Walsh}, Jonelle L. and {Barth}, Aaron J. and {Buote}, David A. and {Baker}, Andrew J. and {Darling}, Jeremy and {Ho}, Luis C. and {Cohn}, Jonathan and {Kabasares}, Kyle M.},
  journal       = {\apj},
  title         = {{Black Hole Mass Measurements of Radio Galaxies NGC 315 and NGC 4261 Using ALMA CO Observations}},
  year          = {2021},
  month         = feb,
  number        = {1},
  pages         = {19},
  volume        = {908},
  adsurl        = {https://ui.adsabs.harvard.edu/abs/2021ApJ...908...19B},
  archiveprefix = {arXiv},
  doi           = {10.3847/1538-4357/abd24d},
  eid           = {19},
  eprint        = {2012.04669},
  primaryclass  = {astro-ph.GA},
}

@Article{Cappellari2020,
  author        = {{Cappellari}, Michele},
  journal       = {\mnras},
  title         = {{Efficient solution of the anisotropic spherically aligned axisymmetric Jeans equations of stellar hydrodynamics for galactic dynamics}},
  year          = {2020},
  month         = jun,
  number        = {4},
  pages         = {4819-4837},
  volume        = {494},
  adsurl        = {https://ui.adsabs.harvard.edu/abs/2020MNRAS.494.4819C},
  archiveprefix = {arXiv},
  doi           = {10.1093/mnras/staa959},
  eprint        = {1907.09894},
  primaryclass  = {astro-ph.GA},
}

@Article{Sahu2020,
  author        = {{Sahu}, Nandini and {Graham}, Alister W. and {Davis}, Benjamin L.},
  journal       = {\apj},
  title         = {{Defining the (Black Hole)-Spheroid Connection with the Discovery of Morphology-dependent Substructure in the M$_{BH}$-n$_{sph}$ and M$_{BH}$-R$_{e,sph}$ Diagrams: New Tests for Advanced Theories and Realistic Simulations}},
  year          = {2020},
  month         = nov,
  number        = {2},
  pages         = {97},
  volume        = {903},
  adsurl        = {https://ui.adsabs.harvard.edu/abs/2020ApJ...903...97S},
  archiveprefix = {arXiv},
  doi           = {10.3847/1538-4357/abb675},
  eid           = {97},
  eprint        = {2101.04895},
  primaryclass  = {astro-ph.GA},
}

@Article{Sahu2019b,
  author        = {{Sahu}, Nandini and {Graham}, Alister W. and {Davis}, Benjamin L.},
  journal       = {\apj},
  title         = {{Revealing Hidden Substructures in the M $_{BH}$-{\ensuremath{\sigma}} Diagram, and Refining the Bend in the L-{\ensuremath{\sigma}} Relation}},
  year          = {2019},
  month         = dec,
  number        = {1},
  pages         = {10},
  volume        = {887},
  adsurl        = {https://ui.adsabs.harvard.edu/abs/2019ApJ...887...10S},
  archiveprefix = {arXiv},
  doi           = {10.3847/1538-4357/ab50b7},
  eid           = {10},
  eprint        = {1908.06838},
  primaryclass  = {astro-ph.GA},
}

@Article{Lipka2021,
  author        = {{Lipka}, Mathias and {Thomas}, Jens},
  journal       = {\mnras},
  title         = {{A novel approach to optimize the regularization and evaluation of dynamical models using a model selection framework}},
  year          = {2021},
  month         = jul,
  number        = {3},
  pages         = {4599-4625},
  volume        = {504},
  adsurl        = {https://ui.adsabs.harvard.edu/abs/2021MNRAS.504.4599L},
  archiveprefix = {arXiv},
  doi           = {10.1093/mnras/stab1092},
  eprint        = {2104.10168},
  primaryclass  = {astro-ph.GA},
}

@Article{Dutton2014,
  author        = {{Dutton}, Aaron A. and {Macci{\`o}}, Andrea V.},
  journal       = {\mnras},
  title         = {{Cold dark matter haloes in the Planck era: evolution of structural parameters for Einasto and NFW profiles}},
  year          = {2014},
  month         = jul,
  number        = {4},
  pages         = {3359-3374},
  volume        = {441},
  adsurl        = {https://ui.adsabs.harvard.edu/abs/2014MNRAS.441.3359D},
  archiveprefix = {arXiv},
  doi           = {10.1093/mnras/stu742},
  eprint        = {1402.7073},
  primaryclass  = {astro-ph.CO},
}

@Article{denBrok2021,
  author        = {{den Brok}, Mark and {Krajnovi{\'c}}, Davor and {Emsellem}, Eric and {Brinchmann}, Jarle and {Maseda}, Michael},
  journal       = {\mnras},
  title         = {{Dynamical modelling of the twisted galaxy PGC 046832}},
  year          = {2021},
  month         = dec,
  number        = {4},
  pages         = {4786-4805},
  volume        = {508},
  adsurl        = {https://ui.adsabs.harvard.edu/abs/2021MNRAS.508.4786D},
  archiveprefix = {arXiv},
  doi           = {10.1093/mnras/stab2852},
  eprint        = {2109.14640},
  primaryclass  = {astro-ph.GA},
}

@Article{Jethwa2020,
  author        = {{Jethwa}, Prashin and {Thater}, Sabine and {Maindl}, Thomas and {Van de Ven}, Glenn},
  journal       = {Astrophysical Source Code Library},
  title         = {{DYNAMITE: DYnamics, Age and Metallicity Indicators Tracing Evolution}},
  year          = {2020},
  month         = nov,
  pages         = {record ascl:2011.007},
  adsurl        = {https://ui.adsabs.harvard.edu/abs/2020ascl.soft11007J},
  archiveprefix = {ascl},
  eid           = {ascl:2011.007},
  eprint        = {2011.007},
}

@Article{Thater2022,
  author        = {{Thater}, Sabine and {Krajnovi{\'c}}, Davor and {Weilbacher}, Peter M. and {Nguyen}, Dieu D. and {Bureau}, Martin and {Cappellari}, Michele and {Davis}, Timothy A. and {Iguchi}, Satoru and {McDermid}, Richard and {Onishi}, Kyoko and {Sarzi}, Marc and {van de Ven}, Glenn},
  journal       = {\mnras},
  title         = {{Cross-checking SMBH mass estimates in NGC 6958 - I. Stellar dynamics from adaptive optics-assisted MUSE observations}},
  year          = {2022},
  month         = feb,
  number        = {4},
  pages         = {5416-5436},
  volume        = {509},
  adsurl        = {https://ui.adsabs.harvard.edu/abs/2022MNRAS.509.5416T},
  archiveprefix = {arXiv},
  doi           = {10.1093/mnras/stab3210},
  eprint        = {2111.01620},
  primaryclass  = {astro-ph.GA},
}

@Article{Thater2022b,
  author        = {{Thater}, Sabine and {Jethwa}, Prashin and {Tahmasebzadeh}, Behzad and {Zhu}, Ling and {den Brok}, Mark and {Santucci}, Giulia and {Ding}, Yuchen and {Poci}, Adriano and {Lilley}, Edward and {Tim de Zeeuw}, P. and {Zocchi}, Alice and {Maindl}, Thomas I. and {Rigamonti}, Fabio and {Yang}, Meng and {Fahrion}, Katja and {van de Ven}, Glenn},
  journal       = {\aap},
  title         = {{Testing the robustness of DYNAMITE triaxial Schwarzschild modelling: The effects of correcting the orbit mirroring}},
  year          = {2022},
  month         = nov,
  pages         = {A51},
  volume        = {667},
  adsurl        = {https://ui.adsabs.harvard.edu/abs/2022A&A...667A..51T},
  archiveprefix = {arXiv},
  doi           = {10.1051/0004-6361/202243926},
  eid           = {A51},
  eprint        = {2205.04165},
  primaryclass  = {astro-ph.GA},
}

@Article{Thater2023,
  author        = {Sabine Thater and Mariya Lyubenova and Katja Fahrion and Ignacio Martín-Navarro and Prashin Jethwa and Dieu D. Nguyen and Glenn van de Ven and {Thater}, Sabine and {Lyubenova}, Mariya and {Fahrion}, Katja and {Mart{\'\i}n-Navarro}, Ignacio and {Jethwa}, Prashin and {Nguyen}, Dieu D. and {van de Ven}, Glenn},
  journal       = {\aap},
  title         = {{Effect of the initial mass function on the dynamical SMBH mass estimate in the nucleated early-type galaxy FCC 47}},
  year          = {2023},
  month         = jul,
  pages         = {A18},
  volume        = {675},
  adsurl        = {https://ui.adsabs.harvard.edu/abs/2023A&A...675A..18T},
  archiveprefix = {arXiv},
  doi           = {10.1051/0004-6361/202245362},
  eid           = {A18},
  eprint        = {2304.13310},
  primaryclass  = {astro-ph.GA},
}

@Article{Cappellari2023,
  author        = {{Cappellari}, Michele},
  journal       = {\mnras},
  title         = {{Full spectrum fitting with photometry in PPXF: stellar population versus dynamical masses, non-parametric star formation history and metallicity for 3200 LEGA-C galaxies at redshift z {\ensuremath{\approx}} 0.8}},
  year          = {2023},
  month         = dec,
  number        = {3},
  pages         = {3273-3300},
  volume        = {526},
  adsurl        = {https://ui.adsabs.harvard.edu/abs/2023MNRAS.526.3273C},
  archiveprefix = {arXiv},
  doi           = {10.1093/mnras/stad2597},
  eprint        = {2208.14974},
  primaryclass  = {astro-ph.GA},
}

@Article{Ruffa2023,
  author        = {{Ruffa}, Ilaria and {Davis}, Timothy A. and {Cappellari}, Michele and {Bureau}, Martin and {Elford}, Jacob and {Iguchi}, Satoru and {Lelli}, Federico and {Liang}, Fu-Heng and {Liu}, Lijie and {Lu}, Anan and {Sarzi}, Marc and {Williams}, Thomas G.},
  journal       = {\mnras},
  title         = {{WISDOM project - XIV. SMBH mass in the early-type galaxies NGC 0612, NGC 1574, and NGC 4261 from CO dynamical modelling}},
  year          = {2023},
  month         = jul,
  number        = {4},
  pages         = {6170-6195},
  volume        = {522},
  adsurl        = {https://ui.adsabs.harvard.edu/abs/2023MNRAS.522.6170R},
  archiveprefix = {arXiv},
  doi           = {10.1093/mnras/stad1119},
  eprint        = {2304.06117},
  primaryclass  = {astro-ph.GA},
}

@InProceedings{Markwardt2009,
  author        = {{Markwardt}, C.~B.},
  booktitle     = {Astronomical Data Analysis Software and Systems XVIII},
  title         = {{Non-linear Least-squares Fitting in IDL with MPFIT}},
  year          = {2009},
  editor        = {{Bohlender}, D.~A. and {Durand}, D. and {Dowler}, P.},
  month         = sep,
  pages         = {251},
  series        = {Astronomical Society of the Pacific Conference Series},
  volume        = {411},
  adsurl        = {https://ui.adsabs.harvard.edu/abs/2009ASPC..411..251M},
  archiveprefix = {arXiv},
  doi           = {10.48550/arXiv.0902.2850},
  eprint        = {0902.2850},
  primaryclass  = {astro-ph.IM},
}

@Article{Agarwal2013,
  author        = {{Agarwal}, Bhaskar and {Davis}, Andrew J. and {Khochfar}, Sadegh and {Natarajan}, Priyamvada and {Dunlop}, James S.},
  journal       = {\mnras},
  title         = {{Unravelling obese black holes in the first galaxies}},
  year          = {2013},
  month         = jul,
  number        = {4},
  pages         = {3438-3444},
  volume        = {432},
  adsurl        = {https://ui.adsabs.harvard.edu/abs/2013MNRAS.432.3438A},
  archiveprefix = {arXiv},
  doi           = {10.1093/mnras/stt696},
  eprint        = {1302.6996},
  primaryclass  = {astro-ph.CO},
}

@Article{Latif2022,
  author        = {{Latif}, M.~A. and {Whalen}, D.~J. and {Khochfar}, S. and {Herrington}, N.~P. and {Woods}, T.~E.},
  journal       = {\nat},
  title         = {{Turbulent cold flows gave birth to the first quasars}},
  year          = {2022},
  month         = jul,
  number        = {7917},
  pages         = {48-51},
  volume        = {607},
  adsurl        = {https://ui.adsabs.harvard.edu/abs/2022Natur.607...48L},
  archiveprefix = {arXiv},
  doi           = {10.1038/s41586-022-04813-y},
  eprint        = {2207.05093},
  primaryclass  = {astro-ph.GA},
}

@Article{Hirschmann2010,
  author        = {{Hirschmann}, Michaela and {Khochfar}, Sadegh and {Burkert}, Andreas and {Naab}, Thorsten and {Genel}, Shy and {Somerville}, Rachel S.},
  journal       = {\mnras},
  title         = {{On the evolution of the intrinsic scatter in black hole versus galaxy mass relations}},
  year          = {2010},
  month         = sep,
  number        = {2},
  pages         = {1016-1032},
  volume        = {407},
  adsurl        = {https://ui.adsabs.harvard.edu/abs/2010MNRAS.407.1016H},
  archiveprefix = {arXiv},
  doi           = {10.1111/j.1365-2966.2010.17006.x},
  eprint        = {1005.2100},
  primaryclass  = {astro-ph.GA},
}

@Article{Graham2022,
  author        = {{Graham}, Alister W. and {Sahu}, Nandini},
  journal       = {\mnras},
  title         = {{Reading the tea leaves in the M$_{bh}$-M$_{*,sph}$ and M$_{bh}$-R$_{e,sph}$ diagrams: dry and gaseous mergers with remnant angular momentum}},
  year          = {2023},
  month         = apr,
  number        = {2},
  pages         = {1975-1996},
  volume        = {520},
  adsurl        = {https://ui.adsabs.harvard.edu/abs/2023MNRAS.520.1975G},
  archiveprefix = {arXiv},
  doi           = {10.1093/mnras/stad087},
  eprint        = {2210.09557},
  primaryclass  = {astro-ph.GA},
}

@Article{Khochfar2004,
  author        = {{Khochfar}, S. and {Burkert}, A.},
  journal       = {\mnras},
  title         = {{On the origin of isophotal shapes in elliptical galaxies}},
  year          = {2005},
  month         = jun,
  number        = {4},
  pages         = {1379-1385},
  volume        = {359},
  adsurl        = {https://ui.adsabs.harvard.edu/abs/2005MNRAS.359.1379K},
  archiveprefix = {arXiv},
  doi           = {10.1111/j.1365-2966.2005.08988.x},
  eprint        = {astro-ph/0409705},
  primaryclass  = {astro-ph},
}

@Article{Schawinski2006,
  author        = {{Schawinski}, Kevin and {Khochfar}, Sadegh and {Kaviraj}, Sugata and {Yi}, Sukyoung K. and {Boselli}, Alessandro and {Barlow}, Tom and {Conrow}, Tim and {Forster}, Karl and {Friedman}, Peter G. and {Martin}, D. Chris and {Morrissey}, Patrick and {Neff}, Susan and {Schiminovich}, David and {Seibert}, Mark and {Small}, Todd and {Wyder}, Ted K. and {Bianchi}, Luciana and {Donas}, Jose and {Heckman}, Tim and {Lee}, Young-Wook and {Madore}, Barry and {Milliard}, Bruno and {Rich}, R. Michael and {Szalay}, Alex},
  journal       = {\nat},
  title         = {{Suppression of star formation in early-type galaxies by feedback from supermassive black holes}},
  year          = {2006},
  month         = aug,
  number        = {7105},
  pages         = {888-891},
  volume        = {442},
  adsurl        = {https://ui.adsabs.harvard.edu/abs/2006Natur.442..888S},
  archiveprefix = {arXiv},
  doi           = {10.1038/nature04934},
  eprint        = {astro-ph/0608517},
  primaryclass  = {astro-ph},
}

@Article{Nicola2024,
  author        = {{de Nicola}, Stefano and {Thomas}, Jens and {Saglia}, Roberto P. and {Snigula}, Jan and {Kluge}, Matthias and {Bender}, Ralf},
  journal       = {\mnras},
  title         = {{Triaxial Schwarzschild models of NGC 708: a 10-billion solar mass black hole in a low-dispersion galaxy with a Kroupa IMF}},
  year          = {2024},
  month         = may,
  number        = {1},
  pages         = {1035-1053},
  volume        = {530},
  adsurl        = {https://ui.adsabs.harvard.edu/abs/2024MNRAS.530.1035D},
  archiveprefix = {arXiv},
  doi           = {10.1093/mnras/stae806},
  eprint        = {2403.12144},
  primaryclass  = {astro-ph.GA},
}

@Article{Binney1978a,
  author   = {{Binney}, J.},
  journal  = {Comments on Astrophysics},
  title    = {{Elliptical galaxies: prolate, oblate or triaxial?}},
  year     = {1978},
  month    = jan,
  number   = {2},
  pages    = {27-36},
  volume   = {8},
  adsurl   = {https://ui.adsabs.harvard.edu/abs/1978ComAp...8...27B},
}

@Article{Binney1978b,
  author   = {{Binney}, J.},
  journal  = {\mnras},
  title    = {{On the rotation of elliptical galaxies.}},
  year     = {1978},
  month    = may,
  pages    = {501-514},
  volume   = {183},
  adsurl   = {https://ui.adsabs.harvard.edu/abs/1978MNRAS.183..501B},
  doi      = {10.1093/mnras/183.3.501},
}

@Article{Liepold2025,
  author        = {{Liepold}, Emily R. and {Ma}, Chung-Pei and {Walsh}, Jonelle L.},
  journal       = {\apj},
  title         = {{A 22 Billion M$_{{\ensuremath{\odot}}}$ Black Hole in Holmberg 15A with Keck KCWI Spectroscopy and Triaxial Orbit Modeling}},
  year          = {2025},
  month         = feb,
  number        = {1},
  pages         = {58},
  volume        = {980},
  adsurl        = {https://ui.adsabs.harvard.edu/abs/2025ApJ...980...58L},
  archiveprefix = {arXiv},
  doi           = {10.3847/1538-4357/ada4b0},
  eid           = {58},
  eprint        = {2501.01493},
  primaryclass  = {astro-ph.GA},
}

@Article{Dullo2019b,
  author        = {{Dullo}, Bililign T.},
  journal       = {\apj},
  title         = {{The Most Massive Galaxies with Large Depleted Cores: Structural Parameter Relations and Black Hole Masses}},
  year          = {2019},
  month         = dec,
  number        = {2},
  pages         = {80},
  volume        = {886},
  adsurl        = {https://ui.adsabs.harvard.edu/abs/2019ApJ...886...80D},
  archiveprefix = {arXiv},
  doi           = {10.3847/1538-4357/ab4d4f},
  eid           = {80},
  eprint        = {1910.10240},
  primaryclass  = {astro-ph.GA},
}

@Article{Franx1991,
  author   = {{Franx}, Marijn and {Illingworth}, Garth and {de Zeeuw}, Tim},
  journal  = {\apj},
  title    = {{The Ordered Nature of Elliptical Galaxies: Implications for Their Intrinsic Angular Momenta and Shapes}},
  year     = {1991},
  month    = dec,
  pages    = {112},
  volume   = {383},
  adsurl   = {https://ui.adsabs.harvard.edu/abs/1991ApJ...383..112F},
  doi      = {10.1086/170769},
}

@Article{Krajnovic2015,
  author        = {{Krajnovi{\'c}}, Davor and {Weilbacher}, Peter M. and {Urrutia}, Tanya and {Emsellem}, Eric and {Carollo}, C. Marcella and {Shirazi}, Maryam and {Bacon}, Roland and {Contini}, Thierry and {Epinat}, Beno{\^\i}t and {Kamann}, Sebastian and {Martinsson}, Thomas and {Steinmetz}, Matthias},
  journal       = {\mnras},
  title         = {{Unveiling the counter-rotating nature of the kinematically distinct core in NGC 5813 with MUSE}},
  year          = {2015},
  month         = sep,
  number        = {1},
  pages         = {2-18},
  volume        = {452},
  adsurl        = {https://ui.adsabs.harvard.edu/abs/2015MNRAS.452....2K},
  archiveprefix = {arXiv},
  doi           = {10.1093/mnras/stv958},
  eprint        = {1505.06226},
  primaryclass  = {astro-ph.GA},
}

@Article{Mulcahey2021,
  author        = {{Mulcahey}, C.~R. and {Prichard}, L.~J. and {Krajnovi{\'c}}, D. and {Jorgenson}, R.~A.},
  journal       = {\mnras},
  title         = {{Deciphering the origin of ionized gas in IC 1459 with VLT/MUSE}},
  year          = {2021},
  month         = jul,
  number        = {4},
  pages         = {5087-5097},
  volume        = {504},
  adsurl        = {https://ui.adsabs.harvard.edu/abs/2021MNRAS.504.5087M},
  archiveprefix = {arXiv},
  doi           = {10.1093/mnras/stab1137},
  eprint        = {2104.09600},
  primaryclass  = {astro-ph.GA},
}

@Article{Binney2005,
  author        = {{Binney}, James},
  journal       = {\mnras},
  title         = {{Rotation and anisotropy of galaxies revisited}},
  year          = {2005},
  month         = nov,
  number        = {3},
  pages         = {937-942},
  volume        = {363},
  adsurl        = {https://ui.adsabs.harvard.edu/abs/2005MNRAS.363..937B},
  archiveprefix = {arXiv},
  doi           = {10.1111/j.1365-2966.2005.09495.x},
  eprint        = {astro-ph/0504387},
  primaryclass  = {astro-ph},
}

@PhdThesis{Thater2019b,
  author  = {{Thater}, Sabine},
  school  = {University of Potsdam, Germany},
  title   = {{The interplay between supermassive black holes and their host galaxies}},
  year    = {2019},
  month   = jan,
  adsurl  = {https://ui.adsabs.harvard.edu/abs/2020PhDT.........9T},
}

@Article{Rantala2024,
  author        = {{Rantala}, Antti and {Rawlings}, Alexander and {Naab}, Thorsten and {Thomas}, Jens and {Johansson}, Peter H.},
  journal       = {\mnras},
  title         = {{The supermassive black hole merger-driven evolution of high-redshift red nuggets into present-day cored early-type galaxies}},
  year          = {2024},
  month         = nov,
  number        = {1},
  pages         = {1202-1227},
  volume        = {535},
  adsurl        = {https://ui.adsabs.harvard.edu/abs/2024MNRAS.535.1202R},
  archiveprefix = {arXiv},
  doi           = {10.1093/mnras/stae2424},
  eprint        = {2407.18303},
  primaryclass  = {astro-ph.GA},
}

@Article{Neureiter2022,
  author        = {{Neureiter}, B. and {de Nicola}, S. and {Thomas}, J. and {Saglia}, R. and {Bender}, R. and {Rantala}, A.},
  journal       = {\mnras},
  title         = {{Accuracy and precision of triaxial orbit models I: SMBH mass, stellar mass, and dark-matter halo}},
  year          = {2023},
  month         = feb,
  number        = {2},
  pages         = {2004-2016},
  volume        = {519},
  adsurl        = {https://ui.adsabs.harvard.edu/abs/2023MNRAS.519.2004N},
  archiveprefix = {arXiv},
  doi           = {10.1093/mnras/stac3652},
  eprint        = {2212.06173},
  primaryclass  = {astro-ph.GA},
}

@Article{Frigo2021,
  author        = {{Frigo}, M. and {Naab}, T. and {Rantala}, A. and {Johansson}, P.~H. and {Neureiter}, B. and {Thomas}, J. and {Rizzuto}, F.},
  journal       = {\mnras},
  title         = {{The two phases of core formation - orbital evolution in the centres of ellipticals with supermassive black hole binaries}},
  year          = {2021},
  month         = dec,
  number        = {3},
  pages         = {4610-4624},
  volume        = {508},
  adsurl        = {https://ui.adsabs.harvard.edu/abs/2021MNRAS.508.4610F},
  archiveprefix = {arXiv},
  doi           = {10.1093/mnras/stab2754},
  eprint        = {2109.09996},
  primaryclass  = {astro-ph.GA},
}

@Article{Rawlings2024,
  author        = {{Rawlings}, Alexander and {Keitaanranta}, Atte and {Mattero}, Max and {Soininen}, Sonja and {Wright}, Ruby J. and {Kallioinen}, Noa and {Liao}, Shihong and {Rantala}, Antti and {Johansson}, Peter H. and {Naab}, Thorsten and {Irodotou}, Dimitrios},
  journal       = {\mnras},
  title         = {{Identifying supermassive black hole recoil in elliptical galaxies}},
  year          = {2025},
  month         = mar,
  number        = {4},
  pages         = {3421-3447},
  volume        = {537},
  adsurl        = {https://ui.adsabs.harvard.edu/abs/2025MNRAS.537.3421R},
  archiveprefix = {arXiv},
  doi           = {10.1093/mnras/staf238},
  eprint        = {2410.13942},
  primaryclass  = {astro-ph.GA},
}

@Article{Mannerkoski2021,
  author        = {{Mannerkoski}, Matias and {Johansson}, Peter H. and {Rantala}, Antti and {Naab}, Thorsten and {Liao}, Shihong and {Rawlings}, Alexander},
  journal       = {\apj},
  title         = {{Signatures of the Many Supermassive Black Hole Mergers in a Cosmologically Forming Massive Early-type Galaxy}},
  year          = {2022},
  month         = apr,
  number        = {2},
  pages         = {167},
  volume        = {929},
  adsurl        = {https://ui.adsabs.harvard.edu/abs/2022ApJ...929..167M},
  archiveprefix = {arXiv},
  doi           = {10.3847/1538-4357/ac5f0b},
  eid           = {167},
  eprint        = {2112.03576},
  primaryclass  = {astro-ph.GA},
}

@Article{Simon2023,
  author        = {{Simon}, David A. and {Cappellari}, Michele and {Hartke}, Johanna},
  journal       = {\mnras},
  title         = {{Supermassive black hole mass in the massive elliptical galaxy M87 from integral-field stellar dynamics using OASIS and MUSE with adaptive optics: assessing systematic uncertainties}},
  year          = {2024},
  month         = jan,
  number        = {2},
  pages         = {2341-2361},
  volume        = {527},
  adsurl        = {https://ui.adsabs.harvard.edu/abs/2024MNRAS.527.2341S},
  archiveprefix = {arXiv},
  doi           = {10.1093/mnras/stad3309},
  eprint        = {2303.18229},
  primaryclass  = {astro-ph.GA},
}

@InProceedings{Cappellari2025,
  author        = {{Cappellari}, Michele},
  booktitle     = {Encyclopedia of Astrophysics},
  title         = {{Early-type galaxies: Elliptical and S0 galaxies, or fast and slow rotators}},
  year          = {2026},
  month         = jan,
  pages         = {122-152},
  volume        = {4},
  adsurl        = {https://ui.adsabs.harvard.edu/abs/2026enap....4..122C},
  archiveprefix = {arXiv},
  doi           = {10.1016/B978-0-443-21439-4.00109-7},
  eprint        = {2503.02746},
  primaryclass  = {astro-ph.GA},
}

@Article{Pilawa2024,
  author        = {{Pilawa}, Jacob and {Liepold}, Emily R. and {Ma}, Chung-Pei},
  journal       = {\apj},
  title         = {{TriOS Schwarzschild Orbit Modeling: Robustness of Parameter Inference for Masses and Shapes of Triaxial Galaxies with Supermassive Black Holes}},
  year          = {2024},
  month         = may,
  number        = {2},
  pages         = {205},
  volume        = {966},
  adsurl        = {https://ui.adsabs.harvard.edu/abs/2024ApJ...966..205P},
  archiveprefix = {arXiv},
  doi           = {10.3847/1538-4357/ad3935},
  eid           = {205},
  eprint        = {2403.07996},
  primaryclass  = {astro-ph.GA},
}

@Article{Saglia2024,
  author        = {{Saglia}, R. and {Mehrgan}, K. and {de Nicola}, S. and {Thomas}, J. and {Kluge}, M. and {Bender}, R. and {Delley}, D. and {Erwin}, P. and {Fabricius}, M. and {Neureiter}, B. and {Andreon}, S. and {Baccigalupi}, C. and {Baldi}, M. and {Bardelli}, S. and {Bonino}, D. and {Branchini}, E. and {Brescia}, M. and {Brinchmann}, J. and {Caillat}, A. and {Camera}, S. and {Capobianco}, V. and {Carbone}, C. and {Carretero}, J. and {Casas}, S. and {Castellano}, M. and {Castignani}, G. and {Cavuoti}, S. and {Cimatti}, A. and {Colodro-Conde}, C. and {Congedo}, G. and {Conselice}, C.~J. and {Conversi}, L. and {Copin}, Y. and {Courbin}, F. and {Courtois}, H.~M. and {Degaudenzi}, H. and {De Lucia}, G. and {Dinis}, J. and {Dupac}, X. and {Dusini}, S. and {Farina}, M. and {Farrens}, S. and {Faustini}, F. and {Ferriol}, S. and {Fourmanoit}, N. and {Frailis}, M. and {Franceschi}, E. and {Fumana}, M. and {Galeotta}, S. and {George}, K. and {Gillis}, B. and {Giocoli}, C. and {Grazian}, A. and {Grupp}, F. and {Guzzo}, L. and {Haugan}, S.~V.~H. and {Hoar}, J. and {Holmes}, W. and {Hormuth}, F. and {Hornstrup}, A. and {Jahnke}, K. and {Jhabvala}, M. and {Keih{\"a}nen}, E. and {Kermiche}, S. and {Kiessling}, A. and {Kilbinger}, M. and {Kubik}, B. and {K{\"u}mmel}, M. and {Kunz}, M. and {Kurki-Suonio}, H. and {Le Mignant}, D. and {Ligori}, S. and {Lilje}, P.~B. and {Lindholm}, V. and {Lloro}, I. and {Mainetti}, G. and {Maiorano}, E. and {Mansutti}, O. and {Marggraf}, O. and {Markovic}, K. and {Martinelli}, M. and {Martinet}, N. and {Marulli}, F. and {Massey}, R. and {Medinaceli}, E. and {Melchior}, M. and {Mellier}, Y. and {Meneghetti}, M. and {Merlin}, E. and {Meylan}, G. and {Moresco}, M. and {Moscardini}, L. and {Munari}, E. and {Nakajima}, R. and {Neissner}, C. and {Nichol}, R.~C. and {Niemi}, S.-M. and {Nightingale}, J.~W. and {Padilla}, C. and {Paltani}, S. and {Pasian}, F. and {Pedersen}, K. and {Percival}, W.~J. and {Pettorino}, V. and {Pires}, S. and {Polenta}, G. and {Poncet}, M. and {Popa}, L.~A. and {Pozzetti}, L. and {Raison}, F. and {Rebolo}, R. and {Renzi}, A. and {Rhodes}, J. and {Riccio}, G. and {Romelli}, E. and {Roncarelli}, M. and {Rossetti}, E. and {Sakr}, Z. and {S{\'a}nchez}, A.~G. and {Sapone}, D. and {Sartoris}, B. and {Schirmer}, M. and {Schneider}, P. and {Schrabback}, T. and {Secroun}, A. and {Seiffert}, M. and {Serrano}, S. and {Sirignano}, C. and {Sirri}, G. and {Skottfelt}, J. and {Stanco}, L. and {Steinwagner}, J. and {Tallada-Cresp{\'\i}}, P. and {Tavagnacco}, D. and {Taylor}, A.~N. and {Tereno}, I. and {Toledo-Moreo}, R. and {Torradeflot}, F. and {Tutusaus}, I. and {Valenziano}, L. and {Vassallo}, T. and {Verdoes Kleijn}, G. and {Wang}, Y. and {Weller}, J. and {Zamorani}, G. and {Zucca}, E. and {Burigana}, C. and {Scottez}, V. and {Ferrarese}, L. and {Lusso}, E. and {Scott}, D.},
  journal       = {\aap},
  title         = {{Euclid: The r$_{b}$‑M$_{*}$ relation as a function of redshift: I. The 5 {\texttimes} {}10$^{9}$ M$_{{\ensuremath{\odot}}}$ black hole in NGC 1272}},
  year          = {2024},
  month         = dec,
  pages         = {A124},
  volume        = {692},
  adsurl        = {https://ui.adsabs.harvard.edu/abs/2024A&A...692A.124S},
  archiveprefix = {arXiv},
  doi           = {10.1051/0004-6361/202452548},
  eid           = {A124},
  eprint        = {2411.01927},
  primaryclass  = {astro-ph.GA},
}

@Article{Nicola2025,
  author        = {{de Nicola}, Stefano and {Thomas}, Jens and {Saglia}, Roberto P. and {Kluge}, Matthias and {Snigula}, Jan and {Bender}, Ralf},
  journal       = {arXiv e-prints},
  title         = {{Eight New Ultramassive Black Hole Masses confirm Best Correlation with Galaxy Core Sizes}},
  year          = {2025},
  month         = dec,
  pages         = {arXiv:2512.04178},
  adsurl        = {https://ui.adsabs.harvard.edu/abs/2025arXiv251204178D},
  archiveprefix = {arXiv},
  doi           = {10.48550/arXiv.2512.04178},
  eid           = {arXiv:2512.04178},
  eprint        = {2512.04178},
  primaryclass  = {astro-ph.GA},
}

@Article{Krajnovic2026,
  author        = {{Krajnovi{\'c}}, Davor and {Emsellem}, Eric and {Bacon}, Roland and {Boogaard}, Leindert A. and {Weilbacher}, Peter M. and {Wisotzki}, Lutz},
  journal       = {\aap},
  title         = {{Multi-spin stellar velocity maps of the most massive galaxies}},
  year          = {2026},
  month         = mar,
  pages         = {A15},
  volume        = {708},
  adsurl        = {https://ui.adsabs.harvard.edu/abs/2026A&A...708A..15K},
  archiveprefix = {arXiv},
  doi           = {10.1051/0004-6361/202558624},
  eid           = {A15},
  eprint        = {2602.04968},
  primaryclass  = {astro-ph.GA},
}

@Article{Dullo2021,
  author        = {{Dullo}, Bililign T. and {Gil de Paz}, Armando and {Knapen}, Johan H.},
  journal       = {\apj},
  title         = {{Ultramassive Black Holes in the Most Massive Galaxies: M$_{BH}$-{\ensuremath{\sigma}} versus M$_{BH}$-R$_{b}$}},
  year          = {2021},
  month         = feb,
  number        = {2},
  pages         = {134},
  volume        = {908},
  adsurl        = {https://ui.adsabs.harvard.edu/abs/2021ApJ...908..134D},
  archiveprefix = {arXiv},
  doi           = {10.3847/1538-4357/abceae},
  eid           = {134},
  eprint        = {2012.04471},
  primaryclass  = {astro-ph.GA},
}

@Article{Nicola2026,
  author        = {{de Nicola}, Stefano and {Saglia}, Roberto P. and {Thomas}, Jens and {Snigula}, Jan and {Kluge}, Matthias and {Bender}, Ralf},
  journal       = {arXiv e-prints},
  title         = {{Triaxial Schwarzschild Models of Brightest Cluster Galaxies with Long-Slit LBT Data}},
  year          = {2026},
  month         = mar,
  pages         = {arXiv:2603.01827},
  adsurl        = {https://ui.adsabs.harvard.edu/abs/2026arXiv260301827D},
  archiveprefix = {arXiv},
  doi           = {10.48550/arXiv.2603.01827},
  eid           = {arXiv:2603.01827},
  eprint        = {2603.01827},
  primaryclass  = {astro-ph.GA},
}

@Article{Loubser2022,
  author        = {{Loubser}, S.~I. and {Lagos}, P. and {Babul}, A. and {O'Sullivan}, E. and {Jung}, S.~L. and {Olivares}, V. and {Kolokythas}, K.},
  journal       = {\mnras},
  title         = {{Merger histories of brightest group galaxies from MUSE stellar kinematics}},
  year          = {2022},
  month         = sep,
  number        = {1},
  pages         = {1104-1121},
  volume        = {515},
  adsurl        = {https://ui.adsabs.harvard.edu/abs/2022MNRAS.515.1104L},
  archiveprefix = {arXiv},
  doi           = {10.1093/mnras/stac1781},
  eprint        = {2206.13215},
  primaryclass  = {astro-ph.GA},
}

@Misc{Chaturvedi2025,
  author = {{Chaturvedi}, Avinash and {Thater}, Sabine and {Krajnovi{\'c}}, Davor},
  month  = aug,
  note   = {Submitted for publication to \emph{Astronomy \& Astrophysics}},
  title  = {Influence of variable initial mass function on the dynamical black hole mass estimates in massive early-type galaxies},
  year   = {2025},
}
\bibliographystyle{aa}

\newpage
\begin{appendix}

\section{Notes on the individual objects}
We give a summary of each galaxy, including a comparison with the literature, in this section.

\subsection{NGC 3706} 
NGC 3706 is a core galaxy with a central ring component, which is tilted with respect to the outer disk, and is clearly visible in the HST images. In our kinematic maps, we report the counter-rotation ($V_{\rm max}\approx 70$ km s$^{-1}$) of this disk, which is kinematically decoupled from the rotation of the galaxy and produces two $\sigma$ peaks in the velocity dispersion which reach up to $340$ km/s. Our derived stellar kinematics are in agreement with the slit-spectroscopic measurements from \cite{Gueltekin2014}. The formation of the about 20 pc sized stellar ring is likely the consequence of large amounts of gas being delivered to the galaxy centre by a recent satellite-galaxy merger \citep[see][]{Bonfini2018}, and thus triggering star formation. The randomly distributed circular arcs of NGC 3706's shell system confirm such a recent merger event \citep{Malin1983}. \cite{Carlsten2017} investigate the stellar populations of NGC 3706 and find an overall old galaxy with a 4 Gyr old metal-rich sub-component, which might result from a recent starburst. 

\cite{Gueltekin2014} derive a black hole mass\footnote{They assume a distance of 46 Mpc and report $M_{\rm BH} = (6.0 ^{+6.0}_{-4.6}) \times 10^8 M_{\odot}$ with $3\sigma$ confidence. The value we report here is downscaled to a distance of 43.1 as assumed in this paper.} of $M_{\rm BH} = (5.6^{+5.6}_{4.2}) \times 10^8 M_{\odot}$ and a $M/L_{\rm V} \approx 6\,M_{\odot}/L_{\odot}$ from axisymmetric Schwarzschild models using HST STIS long-slit kinematics. Our triaxial Schwarzschild models of the combined high-resolution and larger-scale SINFONI kinematics predict $M_{\rm BH} = (1.14^{+0.41}_{-0.63}) \times 10^9$ M$_{\odot}$, which is a factor of 2 more massive than the previous result but consistent within the $3\sigma$ uncertainties. \cite{Gueltekin2014} argue that their result is undermassive compared to the black hole scaling relations, while our mass measurement puts NGC 3706 back in agreement with the scaling relations. We also note that our SMBH mass derived from axisymmetric Schwarzschild modelling \citep{Thater2019b} is very close to our measurement from the triaxial models. On the other hand, the estimated black hole mass from Jeans modelling \citep{Thater2019b} $\rm M_{ BH} = (5.3 \pm 0.8) \times 10^8 M_{\odot}$ based on the central $1''$ of the SINFONI data is closer to the value by \cite{Gueltekin2014}. We caution here, however, that the axisymmetric models cannot reproduce the co- and counterrotating kinematical features of the galaxy independently.

\subsection{NGC 3923} 
NGC 3923 is likely in the late phase of a merger between an elliptical galaxy and one or more dwarf galaxies. It is a famous example of a shell galaxy having 42 stellar shells, which extend to large radii and are likely remnants of the tidal interaction between NGC 3923 and its accreted galaxies \citep{Prieur1988, Bilek2016}. The galaxy shows evidence for slow long-axis rotation with a $V_{\rm max}\approx 30$ km s$^{-1}$ \citep{Carter1998} which has recently also been observed in several other massive galaxies in IFU galaxy surveys \citep[e.g.,][]{Tsatsi2017, Ene2018, Krajnovic2018a}. Our MUSE data reveal that the long-axis rotation quickly falls off to only 10 km s$^{-1}$ outside of 10 arcsec from the centre. Our derived velocity dispersion does not show a pronounced peak, but a slight decrease along the minor axis. Comparing our kinematic data with long-slit data from \cite{Carter1998}, all moments match the previously derived kinematics except for the h$_4$ moment. Our values are unusually high, but similar values are also found for the MUSE observations. 
\cite{Saglia2016} used SINFONI data to dynamically derive the black hole mass with axisymmetric Schwarzschild models without taking dark matter into account and obtained $M_{\rm BH} = (2.81 \pm 0.85) \times 10^9 M_{\odot}$ and a $M/L \approx 4.22\,M_{\odot}/L_{\odot}$ in the z-band. Our derived black hole mass from triaxial Schwarzschild modelling, including dark matter $M_{\rm BH} = (1.19^{+1.34}_{-0.80}) \times 10^9$ M$_{\odot}$ is a factor of 2 lower, but still consistent with the measurement by \cite{Saglia2016}. We also measure a similar value using our axisymmetric Schwarzschild measurements, while our Jeans models cannot well constrain the black hole mass.

\subsection{NGC 4261} 
NGC 4261 is an elliptical galaxy with prominent radio jets emanating from the nucleus, which can be classified as type 2 LINER \citep{Ho1997}. Perpendicular to the radio jet, a $2''$ long, thick nuclear dust disk is clearly visible \citep{Jaffe1996}. The galaxy was classified as a core galaxy, but \cite{Bonfini2018} shows that nuclear dust can masquerade the galaxy to look like a core galaxy. In order to mitigate the dust in our modelling, we used the stellar light model by \cite{Boizelle2021}, which is based on near-infrared H-band data. 
Both our SINFONI and MUSE stellar kinematics reveal almost no rotation in the centre, while long-axis rotation dominates the larger scales, which was already noted in earlier studies \citep{Davies1986, Wagner1988, Krajnovic2011, Loubser2022}. NGC 4261 was also one of the first galaxies whose central massive black hole was measured. The estimated black hole mass from ionised gas as tracer is $M_{\rm BH} = (4.9 \pm 1.00) \times 10^8 M_{\odot}$ \citep{Ferrarese1996}. There are recent measurements using molecular gas kinematics for NGC 4261 which give $M_{\rm BH} = (1.62 \pm 0.05) \times 10^9 M_{\odot}$ \citep{Boizelle2021, Ruffa2023}. We derive a stellar-based black hole mass of $(1.14^{+1.08}_{-0.95}) \times 10^9$ M$_{\odot}$ from triaxial Schwarzschild modelling, which is consistent with both gas-based results. However, given the low quality of the SINFONI data, our derived black hole mass has large uncertainties.

\subsection{NGC 4636}  

NGC 4636 has a weak radio jet originating from the galaxy's active nucleus, which was classified as LINER \citep{Ho1997}. The galaxy is a very circular-shaped core-elliptical, which is embedded in a fainter envelope containing a large number of globular clusters \citep{Dirsch2005, Lee2010}. The kinematics of these globular clusters have been analysed by \cite{Schuberth2006}, and they revised earlier claims based on X-Ray data that this galaxy contains exceptionally large amounts of dark matter \citep{Loewenstein2003}. In fact, dynamical models suggest a similar dark matter fraction as in other elliptical galaxies \citep{Poci2016}. We confirm the result by \cite{Poci2016} with our triaxial Schwarzschild models. In the isophotal map of NGC 4636, a small dust lane is visible close to the centre of the galaxy. \cite{Temi2003} suspect that this dust lane was accreted in a recent merger with a gas-rich galaxy. As the lane is very small, we automatically correct for the dust using the symmetry of our light model. As previously described (Section \ref{s:kin}), NGC4636 has very unusual stellar kinematics. While there is hardly any coherent rotation noticeable within the SINFONI FOV, large-scale kinematics reveal a KDC ($\sim10\arcsec$ in diameter), which rotates approximately around the major axis of the galaxy. The rotation on even larger scales is in the opposite sense with respect to the KDC, making NGC4636 a rare case of prograde and retrograde rotation around the major axis. The velocity dispersion peaks in the centre at just under 300 km/s, but the high $\sigma$ region is somewhat elongated. 

NGC 4636 also belongs to the \cite{Beifiori2012} sample, and the authors measured based on ionised gas kinematics an upper mass limit of $M_{\rm BH} = (3.80 \pm 2.50) \times 10^8 M_{\odot}$. 
Our black hole mass measurement from triaxial Schwarzschild modelling $M_{\rm BH} = (4.68^{+2.99}_{-4.26}) \times 10^8$ M$_{\odot}$ is larger but consistent with the gas-based mass measurement. We note that the uncertainty in the SINFONI data exceeds 15 \%, and therefore it does not contribute significantly to constraining the black hole mass, leading to large uncertainties. Our measurements from the axisymmetric Schwarzschild models are also consistent with the result from ionised gas modelling, while the spherical Jeans model disagrees by a factor of 3.

\subsection{IC 4296}

IC 4296 shows elliptical morphology and is the brightest cluster galaxy of the cluster Abell 3565. Similar to NGC 4261, HST observations have unveiled a distinctive nuclear dust disk in this galaxy \citep{Lauer2005}. The disk component envelops a strong radio source from which a double-sided jet extends to $5'$ from the nucleus. IC 4296 has furthermore been classified as a core elliptical galaxy in earlier works \citep[e.g.,][]{Kormendy2013}, even though, for similar arguments as for NGC 4261, the core classification might be the result of dust contamination. Our derived SINFONI kinematics of this galaxy show a small counterrotating component in the centre. This feature was already found in early work by \cite{Franx1989} and \cite{Saglia1993}. Unfortunately, the observations of this galaxy have the lowest S/N in our sample, and the large error bars make it difficult to identify the counter-rotation in the other moments. IC 4296 was also analysed by \cite{DallaBonta2009}, who report $M_{\rm BH} = (1.34 \pm 0.46) \times 10^9 M_{\odot}$ and $M/L_{\rm B} = 6.3\,M_{\odot}/L_{\odot}$ by modelling the kinematics of ionised gas.
We measure a black hole mass of $M_{\rm BH} = (3.51^{+3.37}_{-2.57}) \times 10^9$ M$_{\odot}$, which is almost three times larger than the previous result, and fits well to our kinematic data (Fig.~\ref{ff:sigma_comparison}). Our JAM and axisymmetric Schwarzschild models are not able to constrain the lower $M_{\rm BH}$ limits, but our derived upper limits are consistent with the measurements derived by \cite{DallaBonta2009}.

\subsection{IC 4329}

IC 4329 is a giant lenticular and the brightest cluster galaxy of the cluster Abell 3574. \cite{Laine2003} noticed a core in the surface brightness of this galaxy. While the central kinematics do not look very disturbed, due to their proximity, IC 4329 is likely interacting with its neighbour IC 4329A\footnote{https://heasarc.gsfc.nasa.gov/docs/objects/galaxies/ic4329.html}. IC 4329 has a cone-shaped shell system \citep{Carlsten2017}, which is likely related to the merger event responsible for the large-scale long-axis rotation. The stellar population properties of IC 4329 are very similar to those of NGC 3706 and NGC 3923, with an overall old population and a young subpopulation \citep{Carlsten2017}. We derive an $M_{\rm BH} = (2.47^{+1.44}_{-2.46}) \times 10^9$ M$_{\odot}$, which is the first black hole mass estimate for this galaxy. The mass using triaxial Schwarzschild modelling is roughly five times the SMBH mass of the axisymmetric Schwarzschild models and half the SMBH mass of the Jeans models.

\section{Dust correction and PSFs}

\begin{figure}
  \centering
    \includegraphics[width=0.5\textwidth]{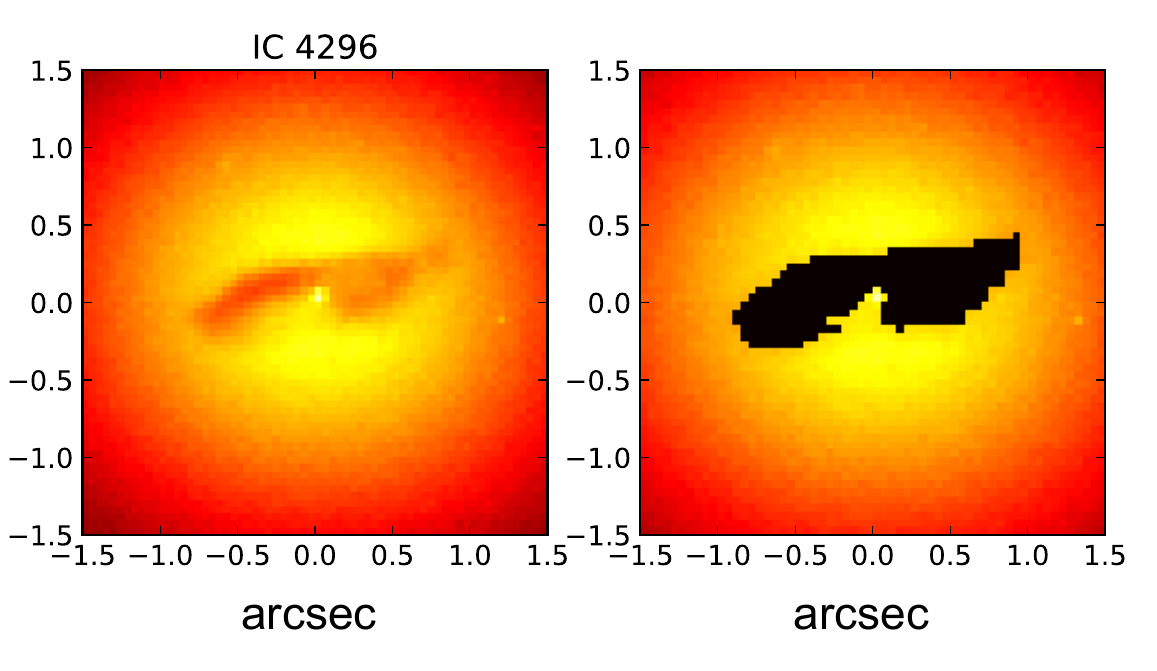}
      \caption{Dust-masked region of the HST WFPC2 image of IC 4296. The left panel shows the original HST image, the right panel shows the same image overplotted with the dust mask (black-coloured pixels).}
      \label{ff:dustmask}
\end{figure}

\subsection{Dust correction} \label{app:A.1}
Only IC 4296 required a detailed dust masking for the galaxies of our sample. We applied the same method as in \cite{Thater2019} for IC 4296. The resulting dust mask can be seen in Figure B.1. where it is compared with the original image. The dust mask works well in covering pixels that are attenuated by dust.

\subsection{HST spatial resolution}
In this section, we list the HST spatial resolution for our galaxies. The spatial resolution was obtained by fitting concentric circular Gaussians to an artificial PSF image generated by TinyTim \citep{Krist2001}. The Gaussians were afterwards normalised. Table 2 of the supplementary material lists the derived PSF parameters.

\subsection{SINFONI spatial resolution}
\label{ss:SINFONI PSF}
The SINFONI spatial resolution was derived following the approach of \cite{Krajnovic2009} and \cite{Thater2019}. We created integrated light images of the SINFONI IFU observations and compared them with HST images, which were of significantly higher spatial resolution than SINFONI. The HST images were convolved with trial PSFs until they matched the SINFONI observations. We performed the match using a Levenberg-Marquardt least-squares minimisation with mpfit \citep{Markwardt2009} within the FOV of SINFONI over 2 dimensions. The PSF was parameterised using a sum of two concentric Gaussians of different weights. Similar to \cite{Thater2022}, for NGC 4261 and IC 4296, we encountered problems with this method due to nuclear dust contamination. In addition, IC 4329 has a foreground star in the direct vicinity of the nucleus. We carefully masked the
dust-affected regions and the foreground star to improve the PSF fit. We show the best-fitting match between the convolved HST image and the reconstructed SINFONI observations in Figure S4 of the supplementary material. The PSF parametrisation of each galaxy is shown in Table 2 of the supplementary material and used in the dynamical modelling.

\subsection{MUSE spatial resolution}
\label{ss:MUSE PSF}
We followed the method described in \cite{Thater2022b} to estimate the spatial resolution of the MUSE observations. Therefore, we selected at least three well-visible stars within the MUSE FOV and fitted a 2-dimensional Gaussian function to these stars using mpfit. The derived seeing values differed over the MUSE FOV by about 10\% in $\sigma_{\rm PSF(MUSE)}$.

\section{Large-scale kinematics}
\label{app:kin}
\begin{figure*}[!htb]
  \centering
    \includegraphics[width=0.86\textwidth]{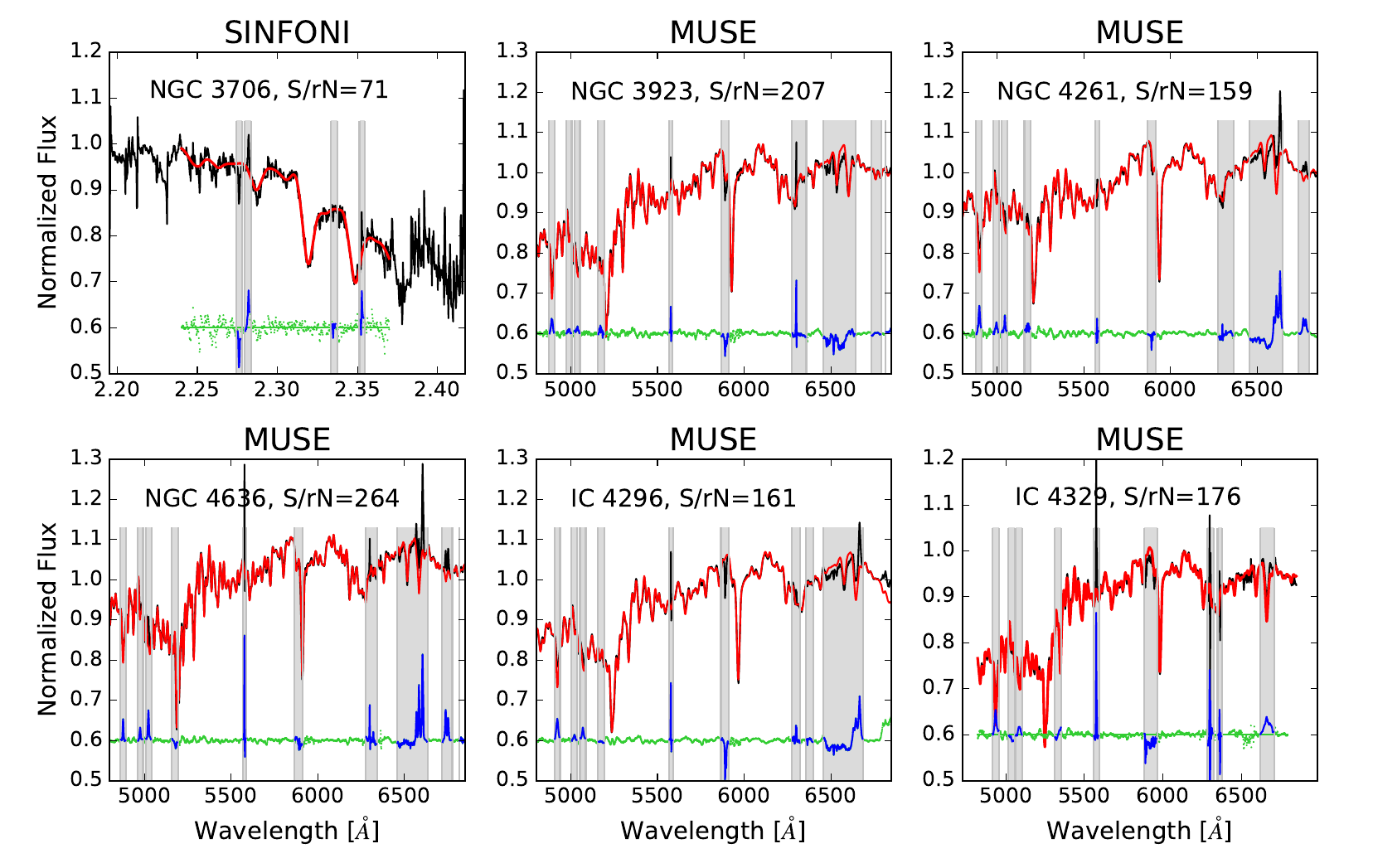}
      \caption{Integrated SINFONI and MUSE 
       spectra and \texttt{pPXF} fits of our target galaxies. The integrated spectra (black solid lines) were obtained by summing up all spectra of the IFU data cubes and fitted using the \texttt{pPXF} routine (red lines) in order to derive an optimal template. The fitting residual between the spectrum and best-fitting pPXF model is shown as green dots and is shifted up by 0.5. Regions which were masked in the fit (often due to emission lines or insufficient sky subtraction) are indicated as grey shaded regions, and their residuals are indicated in blue.}
      \label{ff:ppxf_overview_ls}
\end{figure*}

We show here the large-scale kinematics that were used for the \texttt{DYNAMITE} triaxial Schwarzschild modelling presented in the main text. For NGC 3706, we only had large-scale SINFONI observations; for the remaining galaxies, high-quality MUSE observations were used. Figure \ref{ff:ppxf_overview_ls} shows the \texttt{pPXF} fits of the integrated large-scale SINFONI and MUSE spectra. This figure is similar to Figure \ref{ff:ppxf_overview} of the main paper, and it shows the quality of the \texttt{pPXF} fits to the global spectra obtained by summing the full data cube. The quality of the MUSE observations is remarkably good, reaching high values for the signal-to-residual noise. The resulting Voronoi-binned large-scale kinematics are shown in Figure \ref{ff:kinematics_ls1}. We show a comparison between the SINFONI and MUSE observations in Figures S8 and S9 of the supplementary material.

\begin{figure*}
  \centering
    \includegraphics[width=.90\textwidth]{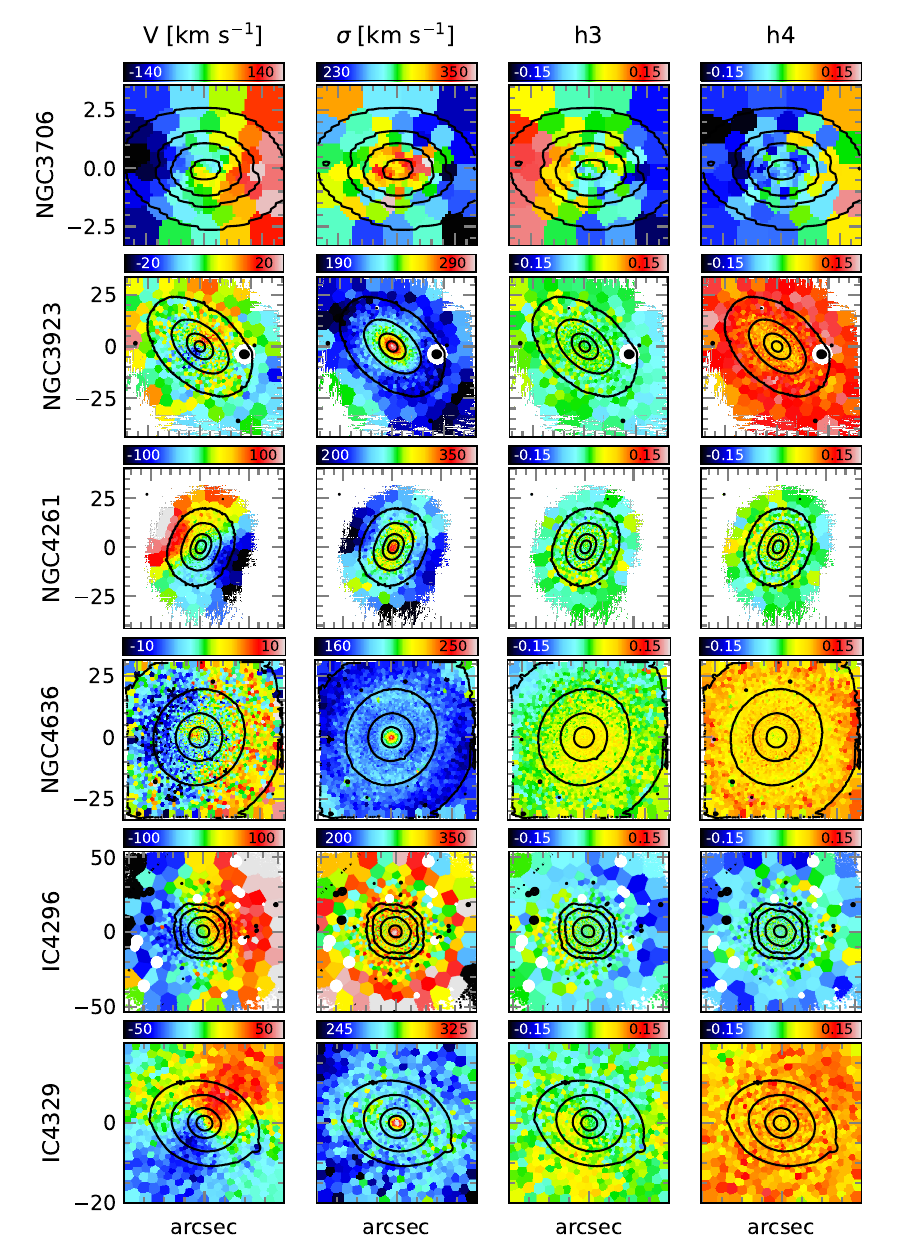}
      \caption{Large-scale stellar kinematics extracted from SINFONI (NGC 3706) and MUSE IFU observations. Shown are the mean velocity V, velocity dispersion $\sigma$ and the $h_{3}$ and $h_4$ Hermite polynomials extracted using \texttt{pPXF}. The image orientation is such that north is up and east is left.}
      \label{ff:kinematics_ls1}
\end{figure*}

\section{Triaxial models}
\label{ss:triangle}
We show the $\chi^2$ distributions of our \texttt{DYNAMITE} Schwarzschild models presented in the main text. Before running a fine grid between black hole mass and mass-to-light ratio, we had to constrain the intrinsic shape parameters (p,q,u) and the dark matter fraction of our galaxies. Figure \ref{ff:chi2} shows the corner plots of these parameters for each of our galaxies. In Figure \ref{ff:chi2_rad}, we show a comparison of the $\chi^2$ distributions calculated using different circular apertures for each galaxy. The circular apertures are chosen to be very small (R<5 arcsec), intermediate (R<10 arcsec) or the full FOV, and have only an effect on the MUSE data, as the SINFONI data have a FOV of 3 arcsec. For NGC 3923, IC 4296 and IC 4329, the $\chi^2$ contours are relatively stable for the different aperture sizes. However, strong changes are seen for NGC 4261 and NGC 4636. We have chosen to show the aperture sizes where most effects occur in the $\chi^2$. It is clear that the derived black hole mass would change significantly using different aperture sizes. For NGC 4261, an upper limit becomes a constraint on black hole mass, and for NGC 4636, the mass measurement changes by a factor of 10. NGC 4261 and NGC 4636 are the galaxies with the most complex kinematics. The strongest change in the $\chi^2$ distribution occurs roughly where we see a change from long-axis rotation to short-axis rotation in the stellar velocity field. 

\begin{figure*}
  \centering
    \includegraphics[width=0.4\textwidth]{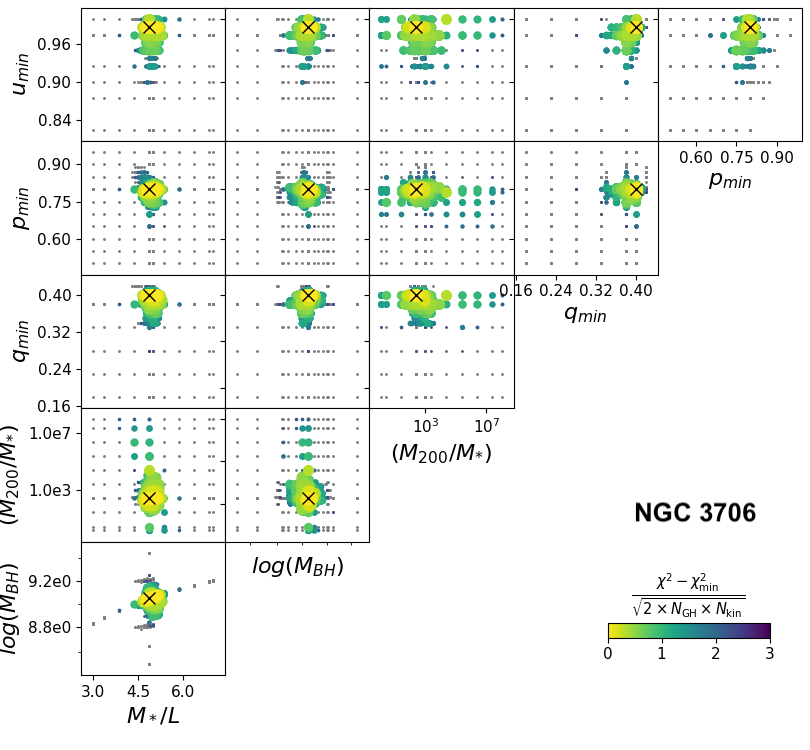}
        \includegraphics[width=0.4\textwidth]{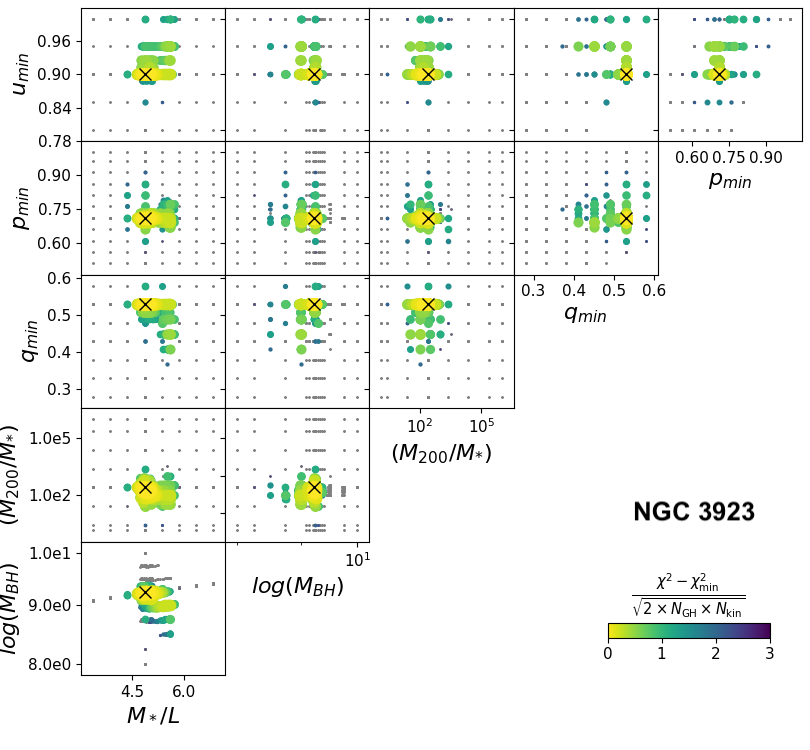}
        \includegraphics[width=0.4\textwidth]{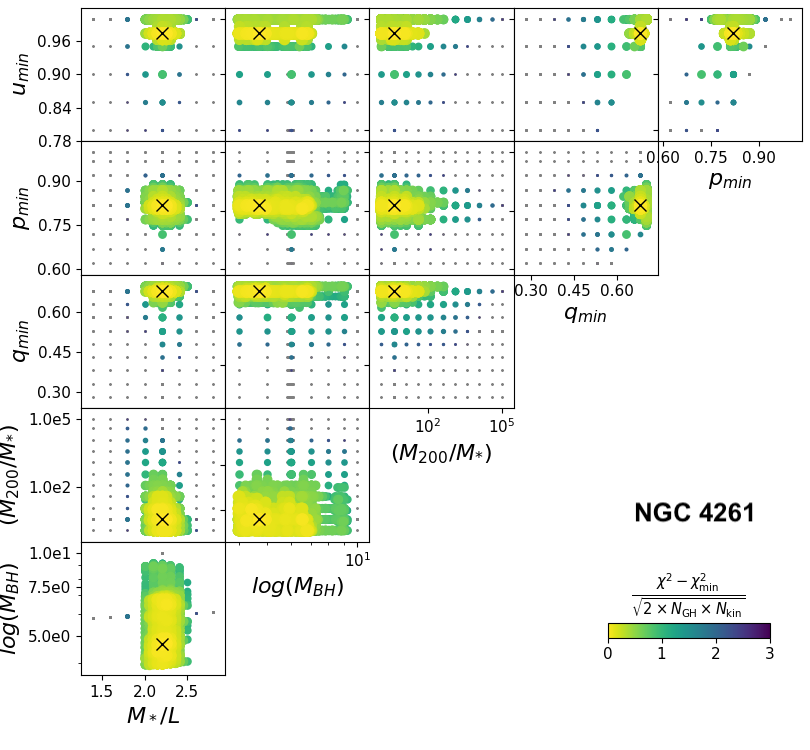}
        \includegraphics[width=0.4\textwidth]{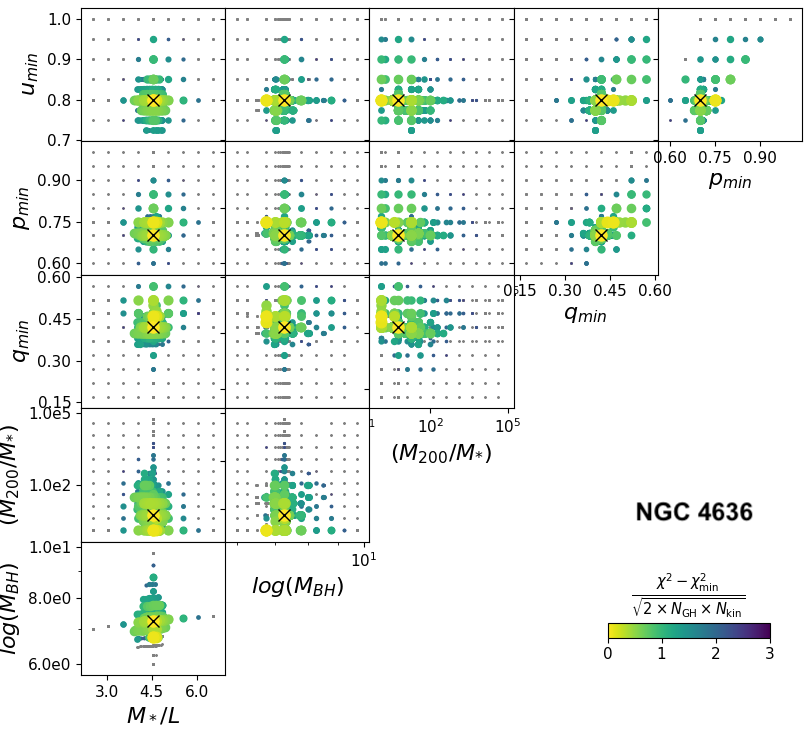}
        \includegraphics[width=0.4\textwidth]{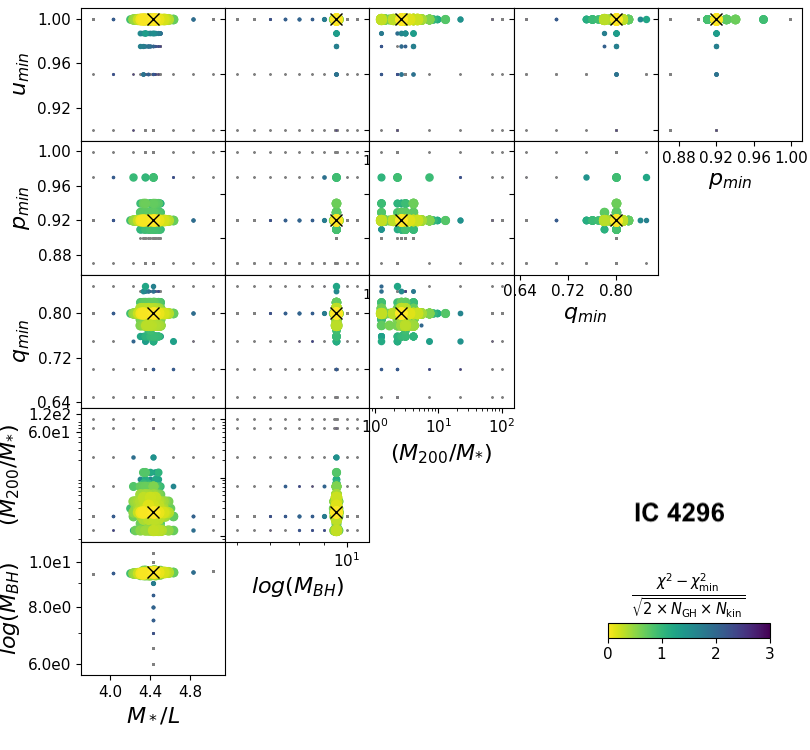}
        \includegraphics[width=0.4\textwidth]{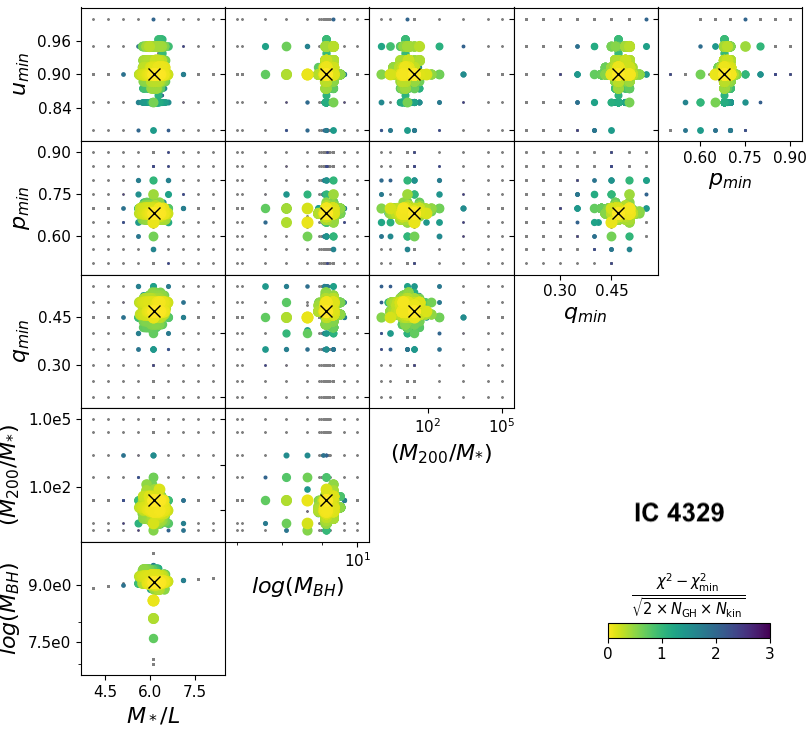}

      \caption{$\chi^2$ distribution of our \texttt{DYNAMITE} triaxial Schwarzschild models to constrain intrinsic shape and dark matter parameters. The six hyper-parameters are the stellar mass-to-light ratio $M_*/L$, the black hole mass in solar units, the dark matter halo mass in units of the stellar mass $M_{200}/M_{\rm star}$, the intrinsic minor-to-major axis ratio q, the intrinsic medium-to-major axis ratio p, and the ratio between projected and intrinsic major axis u. Each point is one model that is colour-coded according to its $\Delta\chi^2$ value as shown in the colour bar, where ($\chi^2$-$\chi^2_{\rm min}$)/$\sqrt{2 \times N_{\rm GH} \times N_{\rm kin}}  < 1$ indicates the models within $1\sigma$ confidence interval. Here, $N_{\rm GH}$ corresponds to the number of fitted Gauss-Hermite moments (4) and $N_{\rm kin}$ is the number of kinematic bins. Grey points are models outside 3$\sigma$ confidence.}
      \label{ff:chi2}
\end{figure*}

\begin{figure*}
  \centering
    \includegraphics[width=0.9\textwidth]{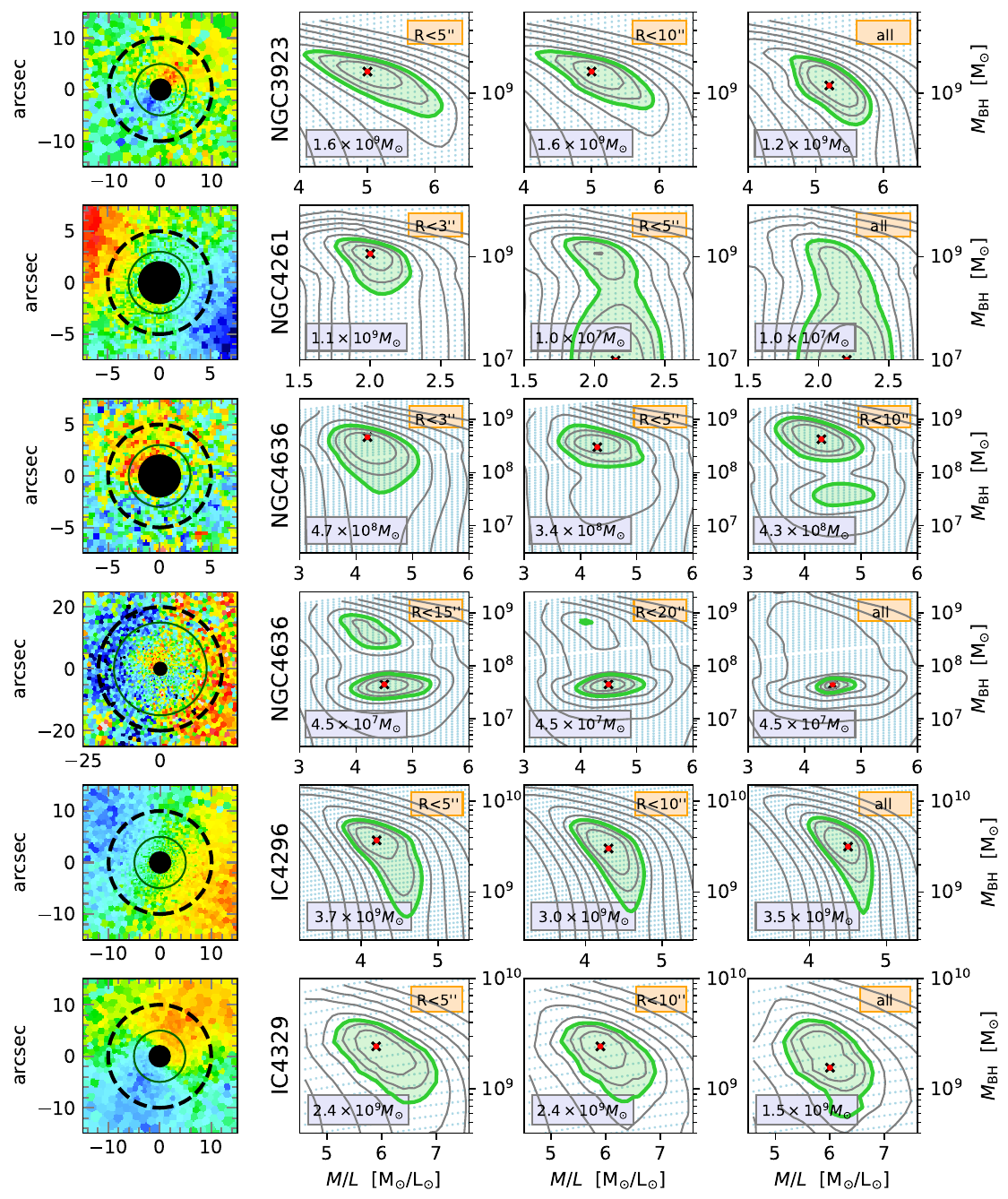}

      \caption{$\chi^2$ distribution of our \texttt{DYNAMITE} triaxial Schwarzschild models for each galaxy, calculated over different circular kinematic apertures. From left to right, the panels show a zoom-in of the stellar velocity with circles indicating the apertures used for the $\chi^2$ calculation, and three $\chi^2$ distributions calculated over increasing apertures. The orange boxes in each $\chi^2$ distribution panel show the radius of the circular aperture. The smallest aperture of each row corresponds to the solid circle in the velocity map, the intermediate aperture to the dashed circle (i.e., for NGC 3923, the solid circle has a radius of 5 arcsec and the dashed circle a radius of 10 arcsec). The black-filled circle in the velocity field corresponds to a region which was masked during the combined \texttt{DYNAMITE} run. {Within the masked region, the models are constrained by the SINFONI data. The $\chi^2$ distributions are similar to Figure 4 and show the $3\sigma$ best-fit models within the green contours. The best-fitting black hole mass of each $\chi^2$ distribution is shown in the purple box.}
      \label{ff:chi2_rad}}
\end{figure*}

\begin{figure*}[!htb]
  \centering
    \includegraphics[width=0.9\textwidth]{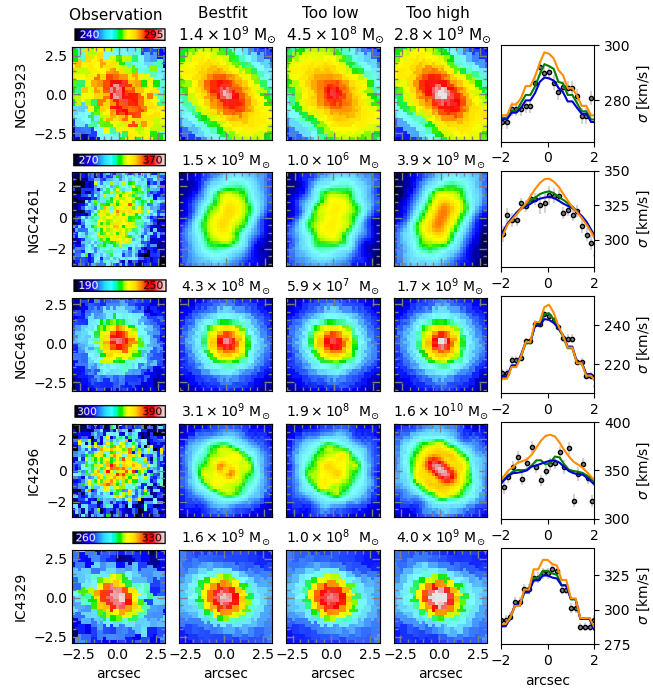}
      \caption{Comparison of the velocity dispersion maps from the MUSE observations and the \texttt{DYNAMITE} Schwarzschild models of the MUSE only run. Each row shows the maps of one galaxy, respectively. From left to right, we present the observed velocity dispersion from the nuclear MUSE data, and the velocity dispersion maps of the Schwarzschild models from the best fitting, a too low and a too high black hole mass. The last column shows the velocity dispersion profiles along the x=0 axis. The too low (blue) and too high (orange) black hole masses are chosen to be just outside of the $3\sigma$ confidence intervals. All models are shown at the respective best-fitting M/L. The high- and low-mass models are clearly ruled out for most galaxies. NGC 3706 is not included as it does not have MUSE data.}
      \label{ff:sigma_comparison_muse}
\end{figure*}

\section{Jeans models}
\label{ss:jeans}

In an alternative dynamical modelling approach, we used the Jeans Anisotropic Modelling method \citep[JAM;][]{Cappellari2008, Cappellari2020} to fit the derived SINFONI kinematics only and derive the supermassive black hole masses for our sample galaxies. JAM solves the Jeans equations and calculates the second velocity moment projected along the line-of-sight, which compares well with the observable root mean square velocity of the observations $V_{\rm rms}=\sqrt{V^2+\sigma^2}$. We used two different implementations of the JAM code: For NGC 3706, which is the only fast-rotator, we assumed an axisymmetric potential (with free inclination), while for the other galaxies, all slow-rotators, we assumed a spherical potential. Although slow-rotators are often triaxial and have twists in their kinematics \citep[e.g.,][]{Krajnovic2011}, a triaxial approach with three independent axes of the velocity ellipsoid would be a more appropriate modelling method \citep{vandenBosch2008}. However, on a first-order approximation, a spherical JAM solution is a simple approach to fit the often roundish-shaped slow-rotating galaxies. In contrast to \cite{Thater2019}, we allow for a varying spherical anisotropy $\beta = 1- \sigma^2_{\theta} /\sigma^2_{r}$ when constructing spherical JAM models to distinguish between the core and the outer profile (see also \citep{Cappellari2015, Drehmer2015, Ene2019, Chae2019, Simon2023}. Therefore, we use the logistic function implemented in JAM, which parametrises the anisotropy as 
\begin{equation}
    \beta(R) = \beta_{0} + (\beta_{\infty} - \beta_{0})/[1 + (R_{\rm a}/R)^\alpha]
\end{equation}
where $\beta_{0}$ is the anisotropy parameter in the galaxy centre ($R=0$), $\beta_{\infty}$ is the anisotropy parameter at $R=\infty$, $R_{\rm a}$ represents the transition radius and $\alpha$ controls the sharpness of the transition. We fixed the anisotropy transition radius to $1.5''$, which is where the slope of the surface brightness profiles changes between the central flat and outer steep profiles. $\alpha$ was fixed to 1.2 based on the results of \cite{Simon2023} for M87. In addition, we assumed physically motivated priors for $\beta_{0}$ and $\beta_{\infty}$ to break the mass-anisotropy degeneracy. We assumed tangential anisotropy in the galaxy centres ($\beta_{0}$<0) and limited the outer anisotropy parameter to $\beta_{\infty}$<0.3 based on the typical outer anisotropy parameters for slow rotators (Fig. 10 by \citealt{Cappellari2025}). 

Consequently, our spherical JAM models are specified by four parameters: the black hole mass ($M_{\rm BH}$), the inner anisotropy ($\beta_{0}$) and outer anisotropy ($\beta_{\infty}$) and the mass-to-light ratio ($M/L$). In order to find the best-fitting model, we used the Bayesian inference method as explained in \cite{Thater2019}. The best-fit values were calculated as the median of the posterior probability of each parameter, as shown in Figure S6 of the supplementary material. We show a comparison between our observed $V_{\rm rms}$ and the best-fitting JAM models in Figure S7. For each of our galaxies, the JAM models reproduce the central $V_{\rm rms}$ peak of the observations, while the outer kinematics suffer from large scatter and large uncertainties. Using the priors on the anisotropy parameters strongly aided in constraining the black hole mass $M_{\rm BH}$ and mass-to-light ratio $M/L$. The derived JAM values for $M_{\rm BH}$ are very consistent with the results from the triaxial modelling, while the JAM $M/L$ are often a bit lower. We summarise our respective JAM best-fit parameters in Table~\ref{t:results} of the main text.

\section{Axisymmetric Schwarzschild models}
\label{ss:comparison_tri_axi}

In our third dynamical modelling approach, we used axisymmetric Schwarzschild models to fit the full line-of-sight velocity distribution (V, $\sigma$, h$_3$, h$_4$) derived from both the high-resolution SINFONI and the large-scale data. This method is based on the numerical orbit superposition method by \cite{Schwarzschild1979} and was described and applied in \cite{Cappellari2006}. The slow rotation and isophotal profile of our galaxies indicate a triaxial shape, meaning an axisymmetric approach can impose significant uncertainties on our results, as only orbits allowed in the axisymmetric potential are used to fit the data. Nevertheless, similar galaxies have often been studied based on axisymmetric models only in the past \cite[e.g.,][]{Shen2010, Rusli2013, Mazzalay2016, Saglia2016}. Therefore, we used the axisymmetric Schwarzschild models to test the applicability of this method to our galaxies. Schwarzschild modelling has been extensively described in the main part of this paper, and here we will provide only a summary of the axisymmetric method. 

In the axisymmetric Schwarzschild modelling method, assuming a stationary gravitational potential from the derived galaxy mass density (including a trial mass for the black hole), a library of orbits is created which covers the phase space of the three integrals of motion ($E$, $L_{z}$, $I_{3}$). This orbit library is then fitted to the observed kinematics to find the best-fitting superposition of orbits for the respective gravitational potential. Schwarzschild models for a grid of different black hole masses are computed, and the best-fitting model is determined based on a $\chi^2$ comparison. The best-fitting model then indicates the best-fitting gravitational potential and thus black hole mass.  We created our orbit library on 21 equipotential shells with logarithmically spaced radii covering the limits of the MGE model, and at each energy, we sampled 8 angular and 7 radial values. The initial set of orbits is doubled, taking into account prograde and retrograde orbital motion, and is additionally dithered by $6^3$ individual orbits. The resulting 508 032 orbits were then integrated in the galaxy's potential. A non-negative least squares method is used to assign weights to the orbits and fit the stellar kinematics on both small and large scales while accounting for PSF effects and aperture binning. We highlight here that the large-scale kinematics used to constrain the dynamical models is different from that used in triaxial models. Here we use the archival VIMOS data and the ATLAS$^{\rm 3D}$ Survey SAURON data \citep{Cappellari2011, Krajnovic2011}, as these models were already presented in the PhD thesis of the first author (\citealt{Thater2019}). VIMOS and SAURON data are consistent with the MUSE data used in the main paper, but are somewhat inferior. The main differences are: lower S/N, smaller FOV and lower spatial resolution. Furthermore, as the axisymmetric models require a symmetry around the minor axis (e.g. no long-axis rotation, or kinematic twists), both the small and large scale kinematics was bi-(anti)-symmetrized by averaging the kinematics of the positions ((x,y), (x,-y), (-x,y), (-x,-y)).

We fixed the inclination angles to the values shown in Table \ref{t:properties} and only varied the mass-to-light ratio $M_*/L$ and the SMBH mass $M_{\rm BH}$. We first ran coarse grids of the two parameters centred on the best-fit results of JAM to find the global minimum of the $\chi^2$ distribution. In a second run, we refined the grids around the minimum of the coarse grid to find the best-fitting values of $M_{\rm BH}$ and $M_*/L$. We present our Schwarzschild model grids in Figure~\ref{ff:schwarzschild_grid3} and provide a visual comparison between the best-fit model and the stellar kinematics in Figures S10 and S11 of the supplementary material. 

For half of our galaxies (NGC 4261, IC 4296, IC 4329), the models are not able to constrain the black hole mass. 
A visual comparison at NGC 4261 shows that the velocity dispersion is not well reproduced. The observed velocity dispersion suffers from large errors, which could cause problems for this galaxy. IC 4296 and IC 4329 have $\chi^2$ minima at slightly too low black hole masses, which could be related to the large error bars, at least for IC 4296. The velocity dispersion of NGC 3923 reveals a slight $M_{\rm BH}$ overestimation and an $M_{\rm BH}$ for NGC 3706. Altogether, the axisymmetric Schwarzschild models did not reproduce the derived stellar kinematics very well. This could be due to the overall limited data quality, but it is probably connected to the triaxial structure of galaxies and the inability of the axisymmetric models to produce the observed kinematics.

\end{appendix}
\label{lastpage}
\end{document}